\documentclass[final]{ar-astro2e}

\usepackage{ulem}  % For Copy Editors

\usepackage{ARAstroBib}
\newcommand{\arcsec}{\hbox{$^{\prime\prime}$}}
\newcommand{\arcmin}{\hbox{$^{\prime}$}}
\def\spose#1{\hbox to 0pt{#1\hss}}
\def\simlt{\mathrel{\spose{\lower 3pt\hbox{$\mathchar"218$}}
     \raise 2.0pt\hbox{$\mathchar"13C$}}}
\def\simgt{\mathrel{\spose{\lower 3pt\hbox{$\mathchar"218$}}
     \raise 2.0pt\hbox{$\mathchar"13E$}}}

\begin{document}
\input epsf.tex    %<-If you need EPS figures to be
                   %  called in {figure} environment for PC
\input epsf.def   %<-If you need EPS figures to be
                   %  called in {figure} environment for Macintosh

\input psfig.sty

\jname{Annu. Rev. Astron. Astrophys.}
\jyear{2012}
\jvol{}
\ARinfo{1056-8700/97/0610-00}

\title{Adaptive Optics for Astronomy}

\markboth{Davies \& Kasper}{Adaptive Optics for Astronomy}

\author{Richard Davies
\affiliation{Max-Planck-Institut f\"ur extraterrestrische Physik, \\Postfach 1312, Giessenbachstr., 85741 Garching, Germany}
Markus Kasper
\affiliation{European Southern Observatory, \\Karl-Schwarzshild-Str. 2, 85748 Garching, Germany}}


\begin{abstract}
Adaptive Optics is a prime example of how progress in observational astronomy can be driven by technological developments. At many observatories it is now considered to be part of a standard instrumentation suite, enabling ground-based telescopes to reach the diffraction limit and thus providing spatial resolution superior to that achievable from space with current or planned satellites. In this review we consider adaptive optics from the astrophysical perspective.
We show that adaptive optics has led to important advances in our understanding of a multitude of astrophysical processes, and describe how the requirements from science applications are now driving the development of the next generation of novel adaptive optics techniques.
\end{abstract}

\maketitle

%%%%%%%%%%%%%%%%%%%%%%%%%%%%%%%%%%%%%%%%%%%%%%%%%%
\section{Introduction}
\label{sec:intro}

\begin{figure}
\psfig{file=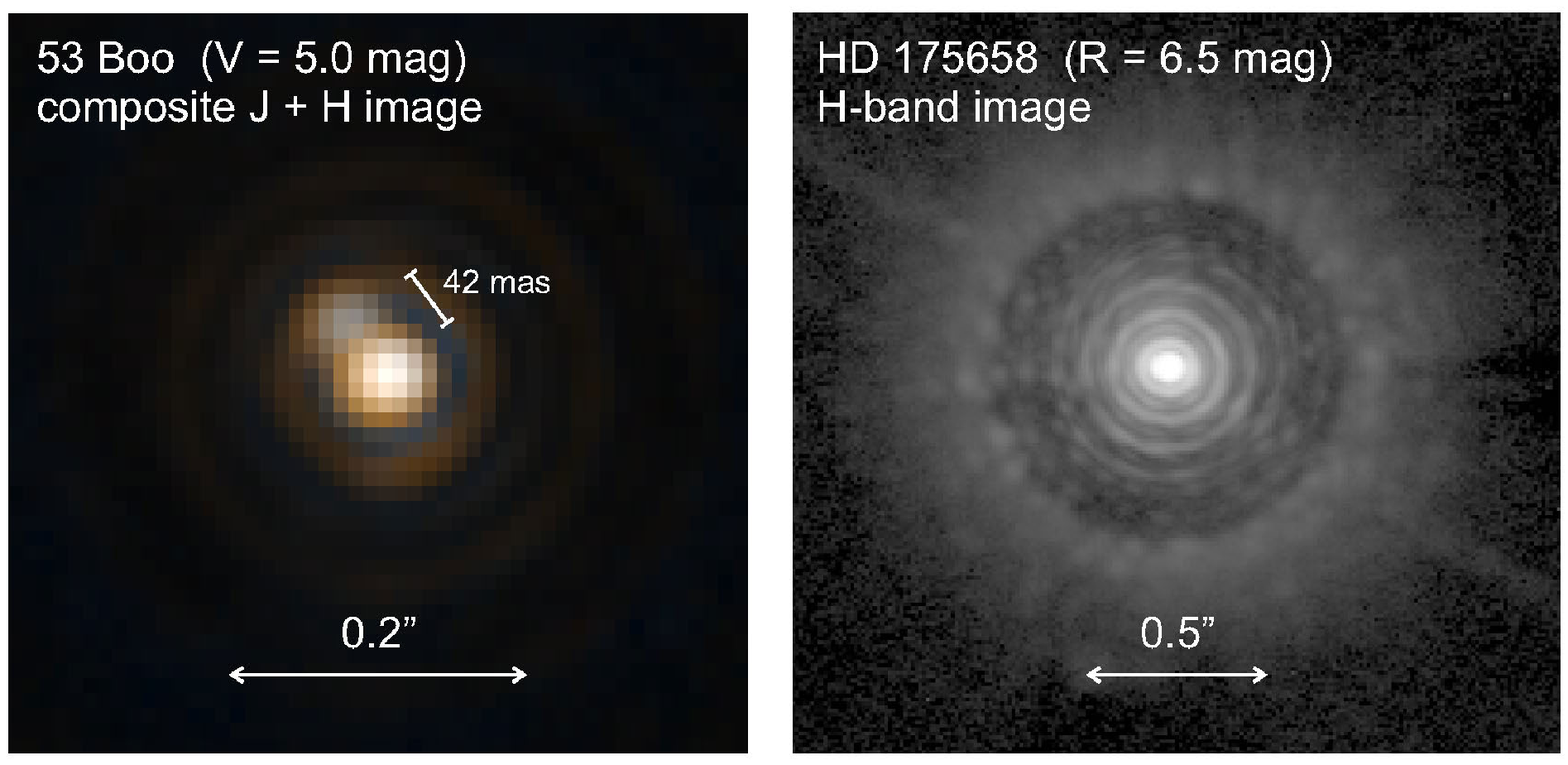,width=10cm}
\caption{Images from commissioning of the LBT adaptive secondary AO system that demonstrate the capabilities of modern adaptive optics.
Left: the double star 53\,Bootes is well resolved even though the separation is only 42\,mas.
Right panel: the H-band correction on bright stars is so good that one can count up to 10 diffraction rings, some of which are slightly fragmented by residual uncorrected aberrations. Beyond this ``dark hole'' (see Sec.~\ref{sec:estperf}), the PSF brightens because the very high orders remain uncorrected.
Adapted from \cite{esp10} (courtesy of S.~Esposito).}
\label{fig:lbt}
\end{figure}

The first successful on-sky test of an astronomical adaptive optics (AO) system was reported with the words ``An old dream of ground-based astronomers has finally come true'' \citep{mer89}.
This jubilant mood resulted from successfully reaching the near-infrared diffraction limit of a 1.5-m telescope.
Since then, both the technology and expectations of AO systems have advanced considerably.
The current state of the art is shown in Fig.~\ref{fig:lbt}.
Here, the adaptive secondary AO system of the Large Binocular Telescope (LBT), which has 8.4-m primary mirrors, recorded a phenomenal 85\% Strehl ratio in the H-band (1.65\,$\mu$m) \citep{esp10}.
In parallel to improving the performance at near-infrared wavelengths, there is an effort to reach the diffraction limit in the optical. 
Using AO with on-the-fly image selection, \cite{law09a} achieved a resolution of 35\,milliarcsec~(mas), the 700\,nm diffraction limit of the 5-m telescope at Palomar Observatory.
This is close to the highest resolution direct optical image, which had a FWHM of 22\,mas and was achieved at 850\,nm on the 10.2-m Keck~II telescope in good atmospheric conditions \citep{wiz00}.

These technical demonstrations illustrate the outstanding performance of current AO systems.
As such, they prompt the question that has prevailed through the early development of AO systems, ``Here's what AO does, can we find a use for it?''.
In contrast, the real utility of such a capability is only demonstrated by its successful scientific applications -- and AO observations are now leading to major new insights and advances in our understanding of the physical mechanisms at work in the universe.
It is this astrophysical perspective, which is now driving the development of future and novel adaptive optics techniques, that we wish to emphasize in this review.
By doing so, our aim is to motivate astronomers to ask instead, ``Here's my science, how can AO enhance it?''

To set the context, we begin in Sec.~\ref{sec:basicao} with a brief description of a simple AO system, highlighting key issues related to the current generation of AO systems.
We then turn to the astrophysical applications in Sec.~\ref{sec:science}, discussing a few examples drawn from a wide variety of fields.
A common conceptual limitation of AO concerns the point spread function (PSF), about which some knowledge is required for most science applications.
Sec.~\ref{sec:psf} discusses the different ways in which the PSF can be used for scientific analysis, the level of detail to which it must be known, and how it might be measured.
While simple adaptive optics systems are already in use at many observatories, considerable effort is being made to demonstrate novel techniques that will vastly broaden their potential. 
Sec.~\ref{sec:novel} describes how these AO systems are being tailored to meet the specific requirements of diverse science cases.
We finish in Sec.~\ref{sec:future} by asking whether there are lessons to be learned that might guide future AO development, to further augment and broaden its scientific impact.

%%%%%%%%%%%%%%%%%%%%%%%%%%%%%%%%%%%%%%%%%%%%%%%%%%
\section{Basic Adaptive Optics}
\label{sec:basicao}
%History

\begin{figure}
\psfig{file=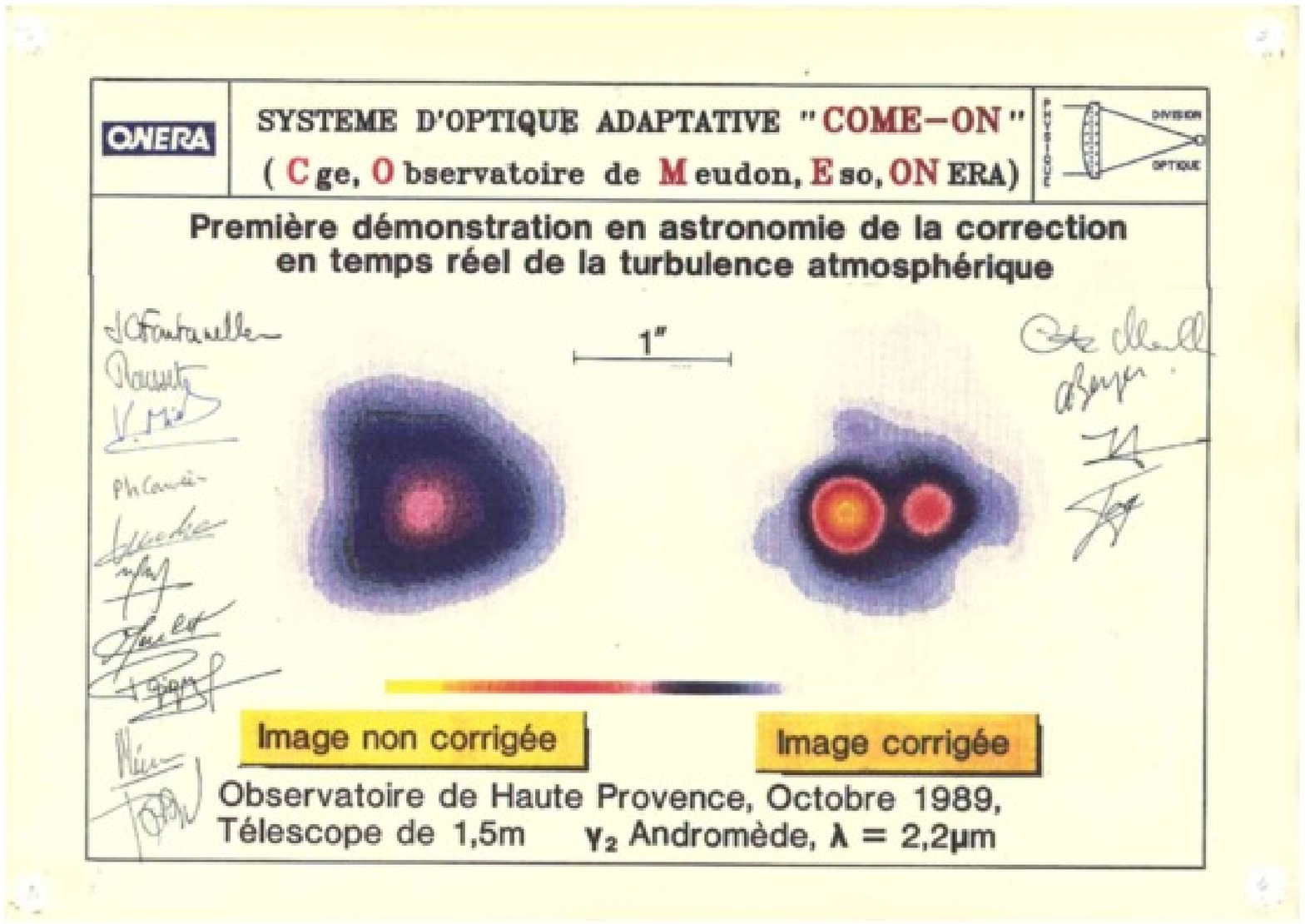,width=\textwidth}
\caption{First astronomical AO corrected image obtained with COME-ON \citep{rou90}.}
\label{fig:comeOn89}
\end{figure}

For centuries, astronomers have lived with blurry images of their targets when looking through ground-based telescopes. The images were blurred by the astronomical seeing that originates from the light passing through the turbulent refractive atmosphere before reaching the Earth's surface. Similar to the effect when looking across a camp-fire, air cells of different temperature, density and hence refractive index introduce spatial and temporal variations of the optical path length along the line of sight. Astronomers try to mitigate these detrimental effects of the atmosphere by building observatories on mountain tops, and ultimately launching them into space. 
However, in 1953 the Mount Wilson and Palomar astronomer Horace W. Babock had a ground-breaking idea: ``If we had the means of continually measuring the deviation of rays from all parts of the mirror, and of amplifying and feeding back this information so as to correct locally the figure of the mirror in response to the schlieren pattern, we could expect to compensate both for the seeing and for any inherent imperfection of the optical figure.'' \citep{bab53}. It still took more than 30 years until technology had matured to a level supporting a practical implementation of Babcock's concept, now called adaptive optics (AO). While the US military started to invest in AO in the 1970s and had commissioned the first practical adaptive optical system, called the Compensated Imaging System, on the 1.6-m telescope on top of Haleakala on the island of Maui in 1982,  it was in the late 1980s that the first astronomical AO instrument COME-ON was tested at the 1.52-m telescope \citep[][Fig.~\ref{fig:comeOn89}]{mer89,rou90} of the Observatoire de Haute-Provence and later installed at ESO's 3.6-m telescope on La Silla in Chile \citep{rig91}.  A thorough summary of the history of Adaptive Optics was presented by \citet{bec93}.

\subsection{The Principle of AO}

Atmospheric turbulence introduces spatial and temporal variations in the refractive index and the optical path length along the line of sight. Those variations mostly originate in the troposphere which typically extends up to a height of about 15\,km and contains 80\% of the atmosphere's mass. Large scale temperature variations produce pressure gradients and winds which lead to turbulent mixing of air with variable density and hence refractive index. These refractive index variations show a very small dispersion between visible and mid-infrared wavelengths \citep{cox01}, so \textit{optical path variations are approximately achromatic over this spectral range}.

According to the Kolmogorov model, energy is injected and mostly contained at large scales and then dissipated to smaller scales which carry less and less energy. Hence, also the \textit{optical path variations carry most energy at small spatial frequencies, i.e., at large scales}.

The vertical distribution of the strength of refractive index variations through atmospheric turbulence as a function of height $z$ is described by the continuous $C_n^2(z)$ profile. It is typically modeled by a number of thin turbulent layers of variable strength and height which move at different speeds in different directions (frozen flow hypothesis). In the near-field approximation the optical path variations introduced by each layer simply sum up. Amplitude or scintillation effects are neglected, which is a good approximation for most astronomical AO applications.
There are three main atmospheric parameters that drive the design and performance of AO systems on telescopes up to 8--10-m class:
\begin{itemize}
	\item The \textit{Fried parameter}, $r_0 \propto [\lambda^{-2} (\cos\gamma)^{-1} \int{C_n^2(z) dz}]^{-3/5}$, gives the aperture over which there is on average one radian of root mean square (rms) phase aberration. If an AO system aims at achieving a moderate Strehl ratio of around 40\%, it needs to correct on spatial scales of $r_0$. Since $r_0 \propto \lambda^{6/5}$, a good correction at longer wavelengths requires coarser sampling of the telescope pupil. Obviously, $r_0$ shrinks at larger zenith angles $\gamma$ and with increasing strength of refractive index variations. $r_0$ also happens to be the aperture which has about the same full-width at half maximum (FWHM) image resolution as a diffraction limited aperture in the absence of turbulence.
This gives rise to astronomical seeing, for which FWHM $\sim \lambda / r_0$. $r_0$ is of the order 10\,cm at visible wavelengths in 1\arcsec\ seeing.
	\item The \textit{isoplanatic angle}, $\theta_0 \propto (\cos\gamma) r_0 / h$, with $h$ denoting the characteristic height of the turbulence, describes the angle out to which optical path variations deviate by less than one radian rms phase aberration from each other. Given a certain correction direction, $\theta_0$ provides the maximum angular radius from this direction at which reasonably good correction is achieved. $\theta_0$ is typically of order a few arcsec at visible wavelengths and strongly depends on the height distribution of the turbulent layers.
	\item The \textit{coherence time}, $\tau_0 \propto r_0/v$, with $v$ denoting the average wind speed, describes the time interval up to which optical path variations deviate by less than one radian rms phase aberration from each other. $\tau_0$ therefore defines the required AO temporal correction bandwidth, which is typically a few milliseconds at visible wavelengths.
\end{itemize}

A fourth parameter that becomes increasingly important for larger telescopes -- notably the future 30--40-m class extremely large telescopes (ELTs) -- is the \textit{outer scale} $L_0$, which is typically a few tens of metres although it can be larger.
A wavefront that has propagated through the atmosphere does not de-correlate any further on size scales greater than $L_0$.
This directly influences the performance of an AO system, most dramatically when $L_0$ is comparable to the size of the telescope aperture.
Thus, for the ELTs, $L_0$ may become as important as $r_0$, $\theta_0$, and $\tau_0$ for deciding which AO science programmes can be observed on a given night.

\begin{figure}
\psfig{file=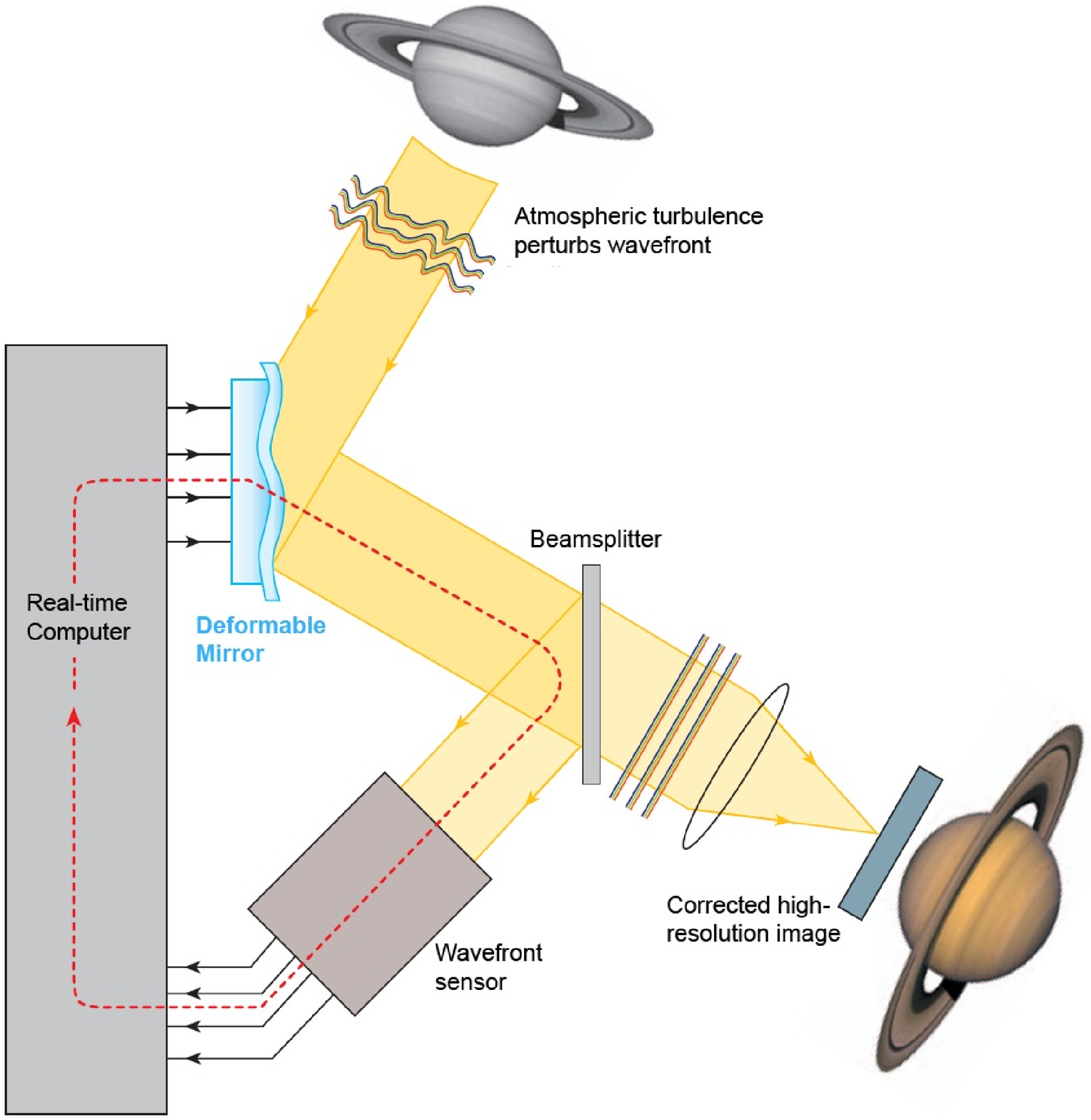,width=\textwidth}
\caption{Adaptive optics working principle (courtesy of S. Hippler).}
\label{fig:ao_prince}
\end{figure}

The principle of AO is depicted in Fig.~\ref{fig:ao_prince}. The AO system tries to regulate the optical path variations (wavefront) by measuring the deviations using a wavefront sensor (WFS),  calculating an appropriate correction, and applying this correction to a deformable mirror (DM). This feedback loop is carried out several hundred times a second in order to comply with the temporal bandwidth requirement set by $\tau_0$. The size of the resolution elements of the wavefront sensor (subapertures) and the deformable mirror (actuators) projected on the telescope entrance aperture should approximately match with $r_0$. 
The depicted setup uses WFS measurements of a single guide star to correct the wavefront in its direction. Such a setup is most simple, widely used, and called Single-Conjugated Adaptive Optics (SCAO). SCAO suffers from image degradation over the field of view set by $\theta_0$. There is a wealth of other concepts involving multiple-guide stars and/or DMs as well as more complex control strategies which will be introduced in Sec.~\ref{sec:novel}.

Since AO wavefront sensing requires a light-source above the atmosphere and near to the astronomical object, it is very often photon starved, and a good sensitivity of the wavefront sensor is essential. In many cases, a visually bright enough natural guide star (NGS) is not available. While astronomical targets that are cold or obscured by dust may still be bright enough for wavefront sensing in the near-infrared (near-IR, 1--2.5\,$\mu$m), the ultimate way to achieve a decent sky coverage is to create one's own guide star where needed. These laser guide stars (LGS) are introduced in Sec.~\ref{sec:lgs}.

The number of resolution elements for current AO systems at 8--10-m class telescopes correcting in the near-IR is typically a few hundred. In the standard approach, the complexity of a single wavefront calculation scales with this number squared, and the number of calculations per second is inversely proportional to $\tau_0$. Hence, the complexity of an AO system roughly scales with $r_0^{-3}$ or $\lambda^{-18/5}$, so it is much easier to achieve a good correction at longer wavelengths.

\subsection{Key Components of an AO system}

\subsubsection{Wavefront Sensing}
The objective of the WFS is to provide a signal with which the shape of the wavefront can be estimated with sufficient accuracy. It generally incorporates a phase-sensitive optical device or sensing scheme and a low noise, high quantum efficiency, photon detector. Since the wavefront is nearly achromatic, wavefront sensing is typically done at visible wavelengths where detector technology -- Charge-Coupled Devices (CCDs) or Avalanche Photo Diodes (APDs) -- is most mature and state-of-the-art detectors have quantum efficiency near one and read-noise near zero. 

\begin{figure}
\psfig{file=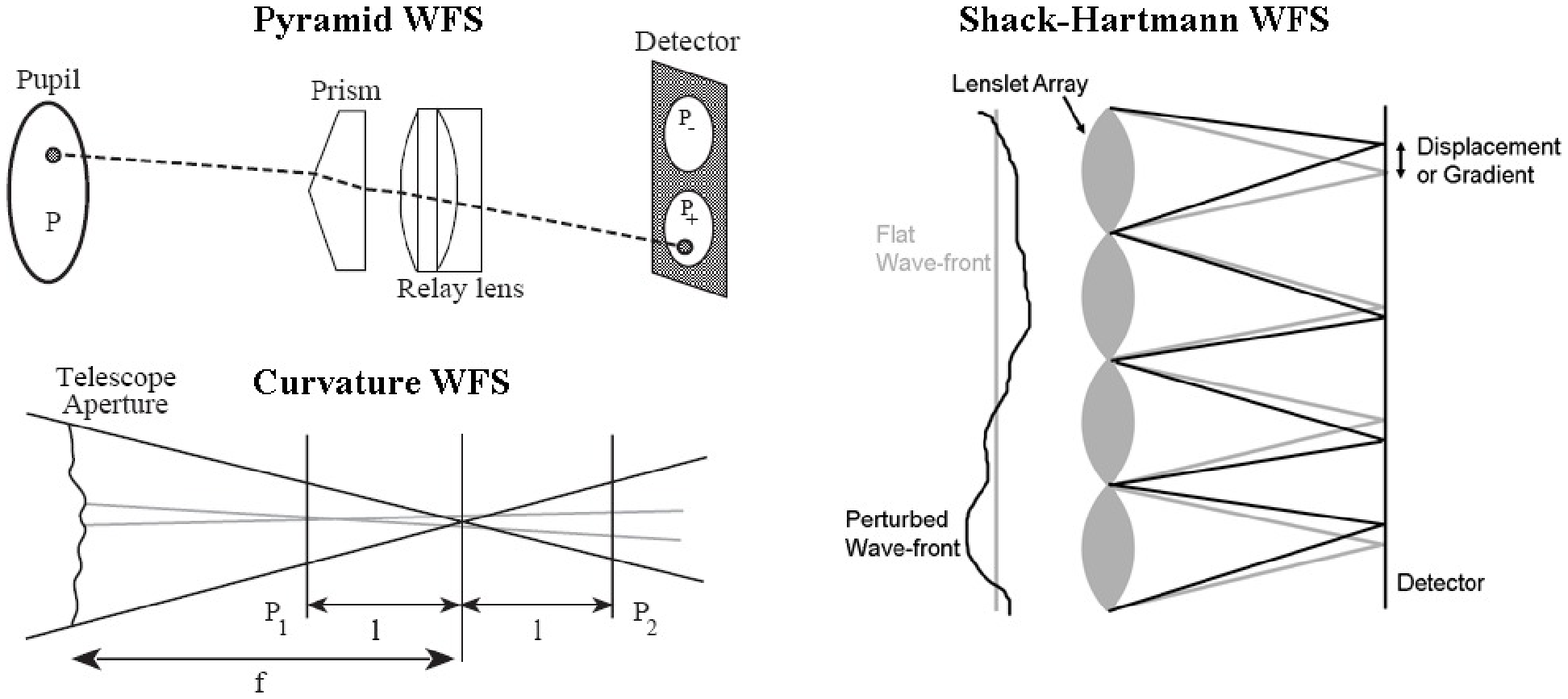,width=\textwidth}
\caption{Schematic drawings of the three main WFS working principles.}
\label{fig:wfs_princ}
\end{figure}

Three flavours of WFS are currently used in AO: the Pyramid WFS, Shack-Hartmann WFS and the curvature WFS. They all work with broad-band light, but differ in dynamic range and sensitivity. While the dynamic range is less important for closed-loop systems which are supposed to operate on small residual errors, the sensitivity (i.e. noise propagation properties) is the parameter that is to be traded against practical issues like technological feasibility.
Fig.~\ref{fig:wfs_princ} shows the working principle of the three WFS mentioned above. 

The \textit{Shack-Hartmann WFS} employs an array of lenslets across the aperture which produce an array of spots corresponding to the local wavefront. The positions of these spots represent the average wavefront slope or gradient over the subaperture. 

The \textit{Pyramid WFS} \citep{rag96} very much represents the Shack-Hartmann WFS when the pyramid (or prism in Fig.~\ref{fig:wfs_princ}) is modulated. When an aberrated ray hits the prism on either side of its tip, it appears in only one of the multiple pupils as displayed by the upper left panel of Fig.~\ref{fig:wfs_princ}. The intensity distributions in the multiple pupil images are therefore a measure for the sign of the ray's slope. If the prism modulates, the ray will appear in either of the pupil images depending on the modulus of the local slope. Thus, the intensity distribution integrated over a couple of modulations also measures wavefront slopes in the pupil. The Pyramid WFS, however, offers the flexibility to adjust the modulation amplitude and hence the sensitivity of the sensor to the observing conditions. In the extreme case of no modulation at all, the Pyramid WFS very much behaves like an interferometer and measures the phase directly rather than its slope \citep{ver05}.

The \textit{curvature WFS} measures intensity distributions in two different planes on either side of the focus, corresponding to the wavefront's curvature or 2nd derivative \citep{rod88}. Wavefront shapes with zero curvature (like tip, tilt and astigmatism) are measured through the beam circumference only. The beauty of the curvature WFS is its simplicity and ease of use. Using an oscillating membrane in the focal plane, a truly differential measurement can be performed by a single pixel detector per subaperture, well suited for APDs, which have the same quantum efficiency as a CCD but virtually zero read-out noise and delay.

\citet{guy05} carried out a theoretical study of noise propagation properties of the different WFS for high-contrast imaging applications. While a phase sensor like the non-modulated Pyramid WFS has equal sensitivity at all spatial frequencies, the sensitivity of a slope or 1st derivative sensor like the modulated Pyramid WFS or the Shack-Hartmann WFS degrades towards low spatial frequencies. This degradation is even more prominent for the curvature WFS where low spatial frequencies are only poorly sensed. The curvature WFS, however, has a better sensitivity than the Shack-Hartmann WFS at high spatial frequencies, and there are ways to cope with the poor low spatial frequency performance \citep{guy07}. Nevertheless, the high sensitivity and flexibility advocate the Pyramid WFS as the best choice for modern AO systems, a conclusion which is supported by the impressive recent on-sky results shown in Fig.~\ref{fig:lbt} \citep{esp10}.

\subsubsection{Wavefront Reconstruction}

The topic of wavefront reconstruction addresses the calculation of an appropriate correction vector $\textbf{v}$ (containing the voltages sent to the DM) from the WFS measurement vector $\textbf{s}$ (containing, for example, all the slopes measured by a Shack-Hartmann sensor).
Since the WFS can be assumed to be operated in an approximately linear regime when in closed loop, wavefront reconstruction is described by the linear system
\[ \textbf{D} \textbf{v} = \textbf{s} + \textbf{n},
\]
with $\textbf{n}$ denoting the measurement noise usually assumed to be Gaussian and uncorrelated, and \textbf{D} denoting the interaction matrix between DM and WFS.

In order to solve for $\textbf{v}$, current AO systems derive a reconstruction matrix $\textbf{R}$ and multiply it with $\textbf{s}$. In the most simple case, $\textbf{R}$ is the inverse of \textbf{D} -- or strictly, the Moore-Penrose pseudoinverse, since \textbf{D} is usually degenerate and not square and so not directly invertible. However, this simple approach normally leads to large noise amplification, so some kind of modal decomposition, filtering and weighting is involved in the calculation of $\textbf{R}$ \citep[e.g.][]{wal83,gend94,ver01}.

Unfortunately such vector-matrix-multiply reconstructors scale in complexity with $O(n^2)$, where $n$ is the number of degrees of freedom of the system. Since a reconstruction must be carried out at each time step ($\sim 1$\,ms) of the control loop, and the delay between measurement and corrective action must be very small ($\sim$ one time step), the computational load quickly becomes very demanding for the next generation of very high-order AO systems. Several approaches exist in order to reduce this complexity.
For example, the FFT-based reconstructor \citep{poy02} scales with $O(n \log n)$ and has successfully been tested in the laboratory \citep{poy08}; the fractal iterative method (FRIM) \citep{thi10} scales with $O(n)$, but needs a few iterations of a conjugate gradients search algorithm to converge; the cumulative reconstructor (CURE) \citep{rose11} scales with $O(n)$ in a single step.

Wavefront reconstruction and control can further be improved by predicting the system's state including the residual wavefront error in a linear-quadratic-Gaussian (LQG) or Kalman filter based control approach \citep{ler04, poy07}. In such a scheme, telescope vibrations can also be incorporated in the state vector and efficiently corrected \citep{pet08}. The drawback of such advanced control methods is again the added computational complexity, which can be mitigated by applying it to the most critical modes only, i.e. just to tip \& tilt in the case of vibration filtering.

\subsubsection{Deformable Mirrors}
\label{sec:dm}

The objective of the DM is to correct for the optical path differences introduced by the turbulent atmosphere. It usually consists of an array of actuators which are connected to a thin optical surface that deforms under the expansion of the actuators. The most important parameters for a DM are stroke, response time, spacing and number of actuators. Spacing (projected onto the telescope's aperture) and response time should agree with the requirements set by $r_0$ and $\tau_0$, while stroke and number of actuators scale with aperture diameter. Assuming an infinite outer scale of turbulence $L_0$, the stroke increases with aperture diameter as $d^{5/6}$ according to the structure function of the Kolmogorov model. In reality, the structure function flattens out at an outer scale of order 20--50\,m, and stroke requirements are of the order of tens of microns optical path difference. While the largest DMs nowadays have some thousand actuators, AO at visible wavelengths at a 40-m telescope would require a DM with several ten thousands of actuators.

\begin{figure}
\psfig{file=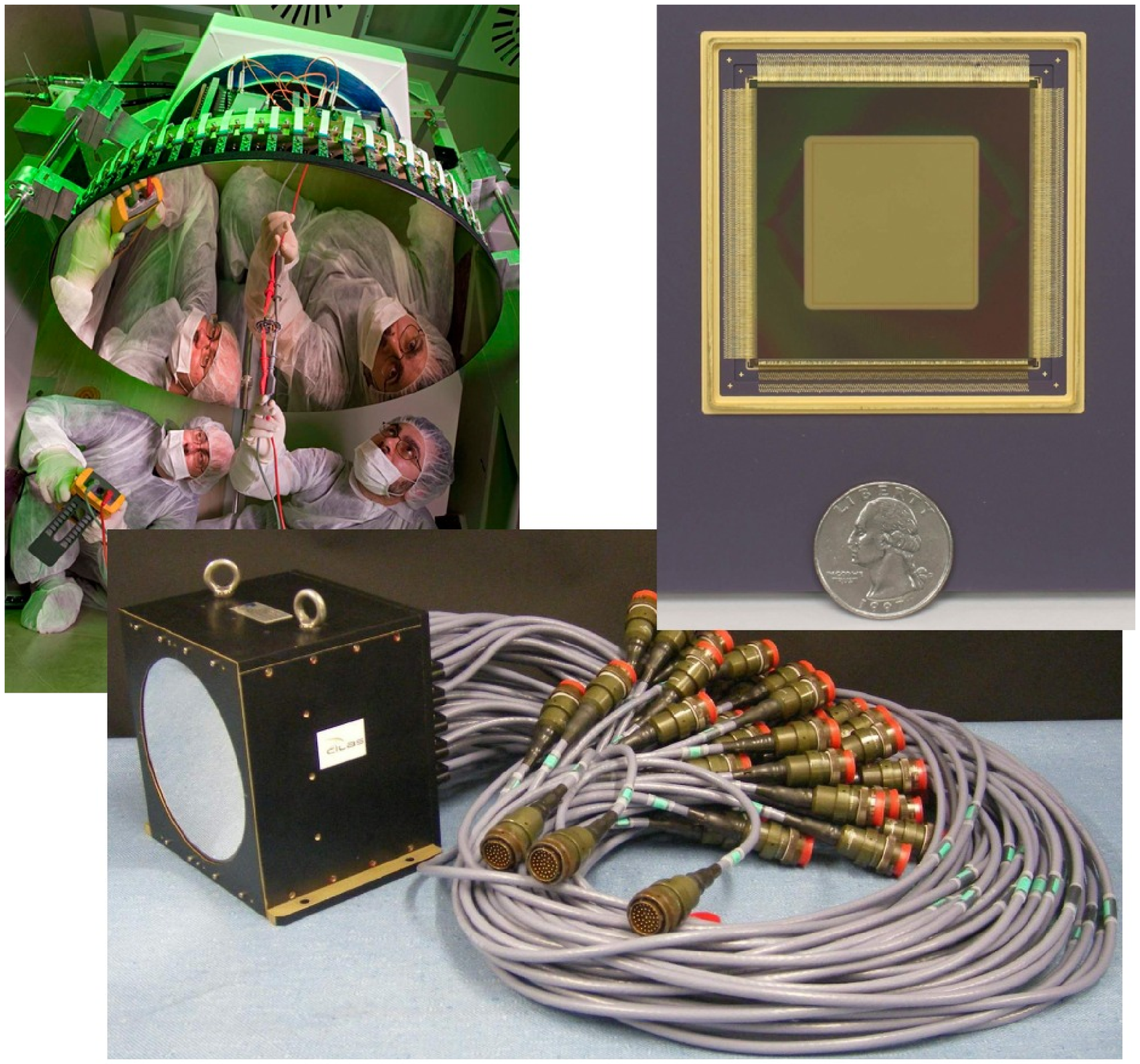,width=\textwidth}
\caption{Main technologies used to produce AO DMs. Upper left: Lab testing of the LBT adaptive secondary mirror system with 672 actuators (Photograph by R.~Cerisola). Upper right: A 4096-actuator micro-optical-electrical-mechanical-system (MOEMS) deformable mirror (Photograph courtesy of S.~Cornelissen, Boston Micromachines Corp.). Bottom: Piezo high-order 1377 actuator DM for SPHERE \citep{sin08}.}
\label{fig:mirrors}
\end{figure}

Three main technologies are used to produce AO DMs (see Fig.~\ref{fig:mirrors}). The largest are the {\it adaptive secondary} or \textit{deformable secondary mirrors} (DSM). Replacing the static secondary mirror of the telescope, they provide AO correction while maintaining high transmission and low thermal emissivity by avoiding extra relay optics. The actuators can be voice coils, which are locally positioned by an internal control loop. They are typically separated by a few~cm and attached to an optical shell, about 1\,m in diameter but only 1--2\,mm thick. Especially the polishing and handling of this shell is one of the great challenges in producing DSMs. The first such mirror was installed in 2003 at the 6.5-m Multiple Mirror Telescope (MMT) \citep{bru03}. Another one is in operation at the LBT \citep{esp10}; one is undergoing laboratory tests for the Magellan Telescope \citep{clo10b}; and one more is in construction for the Very Large Telescope \citep[VLT;][]{ars10}.

More common are medium size piezo DMs with an actuator spacing of several millimeters. They have less stroke than DSMs, but at about 10\,$\mu$m peak-to-valley it is still sufficient for 8--10-m class telescopes. In addition, piezo DMs have a significantly reduced response time of order a hundred microseconds. Since piezo DMs usually do not incorporate local position control and are affected by hysteresis and thermal expansion, they need to be controlled by an AO system to provide a precise, stable wavefront quality.

Quite recently, micro-optical-electrical-mechanical systems (MOEMS) have emerged as an alternative. They use electro-static or voice-coil actuation mechanisms and are produced using standard semiconductor fabrication technologies. With typical interactuator spacings of a few hundred microns, MOEMS are significantly smaller than other DMs. They have almost instantaneous response times, do not suffer from hysteresis, and are relatively inexpensive to produce with many actuators. However, with a very large number of actuators as required by the next generation ELTs, channel routing and high actuator yield become a technological challenge. Some MOEMS as well as small spaced piezo DMs only have a small actuator stroke of the order $\sim2$\,$\mu$m. AO systems using such mirrors need to be operated with a second large stroke DM with fewer actuators acting as a woofer that pre-flattens the wavefront to the required level \citep{ham06}.

\subsection{Estimating Performance}
\label{sec:estperf}

The performance of an AO system can be evaluated using a number of different criteria or metrics. These metrics have been chosen to be relevant for particular science cases. A ``good'' AO system for hunting exoplanets may not be useful for extragalactic science, and trade-off studies during a conceptual design must take into account the desired performance for the intended applications.
Probably the most common performance metric for AO is the \textit{residual wavefront error variance} $\sigma_{WFE}^2$ over the telescope aperture. Another very common performance metric is the \textit{Strehl ratio} $S$. These two metrics are, however, somewhat redundant since they are linked through the Mar\'echal approximation:
\[
	S \sim e^{-\sigma_{WFE}^2}.
\]
\citet{ros09} showed that this approximation holds exactly in the common AO case of Gaussian phase residuals, and it is still a very good approximation for many other distribution laws.
The residual wavefront error variance can be split up into individual contributors assuming that those are uncorrelated:
\[
	\sigma_{WFE}^2 = \sigma_{fit}^2 + \sigma_{rec}^2 + \sigma_{bw}^2 + other.
\]
The main error contributors are i) the fitting error $\sigma_{fit}^2$ which gives the residual high spatial frequency wavefront error which the DM, with its finite number of actuators, cannot fit anymore; ii) the reconstruction error $\sigma_{rec}^2$ which combines all effects that reduce wavefront reconstruction accuracy (measurement noise, calibration noise, sampling errors, aliasing, chromaticity, etc.); and iii) the temporal bandwidth error $\sigma_{bw}^2$ which accounts for the finite response time of the AO system to the dynamic turbulence. The latter two terms are within the control space of the AO system. A good AO concept properly balances the error terms in order to avoid excessive specifications of some components that do not present a benefit to the overall performance.
Depending on the application, there may be other error sources in addition to those given above. For example, observation of an object that is not coincident with the AO guide star will introduce angular anisoplanatic error, and using a single laser guide star (see Sec.~\ref{sec:lgs}) will lead to the cone-effect or focal anisoplanatic error.

High spatial frequency wavefront errors degrade many types of science more than lower frequency ones, because they scatter the light far away from the image center. Low spatial frequency errors instead  leave most light concentrated in the vicinity of the center, and the corresponding loss could be compensated with slightly bigger pixels without a large penalty on sensitivity and resolution. This reasoning led to the \textit{encircled energy} metric which defines the radius in which a certain fraction (typically 50\% or 80\%) of the total light is concentrated. The overall budgeting of multiple error sources is however difficult since the encircled energy cannot easily be written as the sum or product of individual terms like the residual wavefront variance above.

The missing budgeting capability of the encircled energy led \citet{seo09} to propose the \textit{normalized point source sensitivity} (PSSN) metric. The PSSN for AO characterization is defined as the ratio of the sum of an observed point source image intensity $I_{\textrm{PSF}}$ squared over that of a perfect imaging system working at the diffraction limit: 
	\[
	\textrm{PSSN} = \frac{\sum_{pix} I_{\textrm{PSF, obs}}^2}{\sum_{pix} I_{\textrm{PSF, perf}}^2}.
\]
Similar to the Strehl ratio, the PSSN is equal to one for perfect AO performance and approaches zero when residual errors are large. It is also approximately multiplicative for multiple error terms as long as those are low spatial frequency and weak, i.e., the total PSSN is the product of the individual PSSNs of each error term.

A somewhat special case is presented by the high-contrast imaging of faint companions and exoplanets around bright stars. Here, the objective of the AO and wavefront control system is to produce a very good correction performance or residual wavefront variance as a function of spatial frequency. This is because the residual image intensity is proportional to the residual spatial power spectrum of the wavefront \citep[e.g.][]{per03, guy05} when aberrations are small. A DM can suppress aberrations up to its correction radius $\theta_{AO} = \lambda/(2d)$, where $\lambda$ is the wavelength of the observation and $d$ is the interactuator spacing projected back to the telescope aperture.
As such, it must create a ``dark hole'' in this region (for example, as seen in the right panel of Fig.~\ref{fig:lbt}) that is void of scattered stellar light, and in which one could search for faint companions.

\subsection{Laser Guide Stars}
\label{sec:lgs}

For a reasonable correction performance in the near-IR, AO systems need sufficiently bright ($\sim 15$\,mag) guide stars within $\theta_0$ of the astronomical target. This requirement strongly limits the sky coverage that can be attained using natural guide stars (NGS); only about 10\% of the sky can be observed with AO on average while there is a large difference between the galactic plane (several tens of percent) and the galactic pole (a few tenths of a percent) \citep{ell98}. Especially the very low sky coverage at high galactic latitudes hinders extragalactic astronomy from using AO efficiently.

It was the US military, in a classified programme, that first proposed to overcome this problem by using laser beacons to create artificial sources (laser guide star, LGS) \cite[see][]{bec93}.
A few years later the concept was independently proposed in an astronomical context by \citet{foy85}.
The two mechanisms considered so far are i) Rayleigh scattering in the dense regions of atmosphere up to altitudes of about 30\,km above the ground, and ii) resonance fluorescence of sodium atoms which are concentrated in a layer at about 90\,km height. The first successful tests of a Rayleigh LGS were performed at the Starfire Optical Range 1.5-m telescope \citep{fug92}; the first astronomical LGS AO systems were installed at the Lick \citep{max97} and Calar Alto \citep{eck00} observatories in the mid 1990s. Still it took another ten years until the technology had matured enough to be installed at 8--10-m class telescopes such as Keck~II \citep{wiz06}, VLT \citep{bon06}, Gemini North \citep{boc06}, and Subaru \citep{hay10}. 
%Science output has steadily been increasing since then \citep{liu08}.

\begin{figure}
\psfig{file=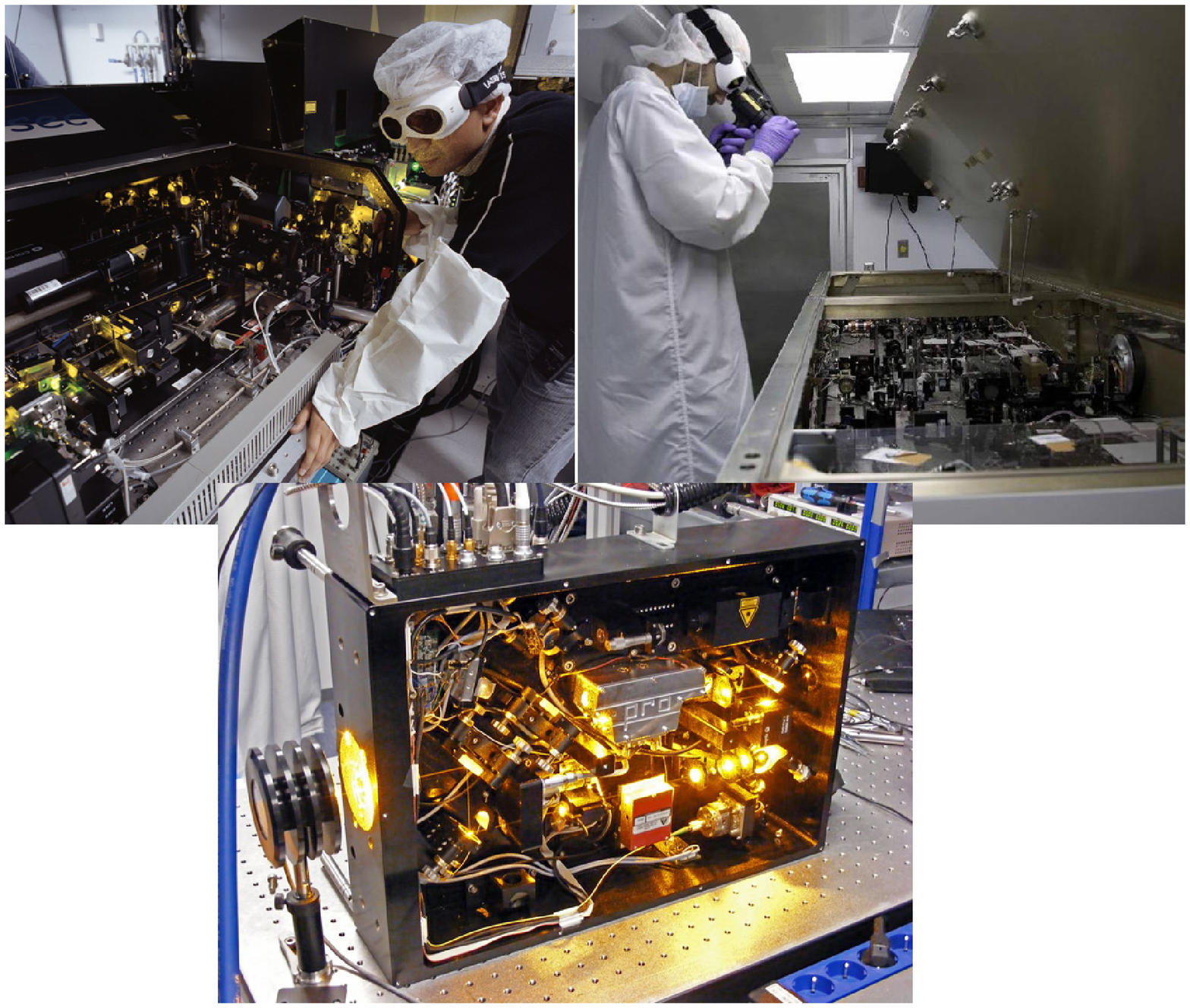,width=\textwidth}
\caption{Left: VLT LGSF dye laser PARSEC (\citealt{rab02}, picture credit ESO/H. Zodet) with $>10$\,W output power. Right: Solid state sum-frequency laser for the Gemini South MCAO system (courtesy of C.~D'Orgeville). Bottom: Raman fiber laser prototype (picture credit ESO) with 20\,W output power.}
\label{fig:lasers}
\end{figure}

Most of the LGS AO systems mentioned above use sodium laser guide stars generated by lasers emitting light of 589\,nm wavelength to excite sodium D$_2$ lines. One of the greatest technological challenges has been the generation of the laser light itself.
Dye lasers were a common choice for early systems, but these are bulky (see top left of Fig.~\ref{fig:lasers}), inefficient and sensitive to environmental conditions. Dye LGS therefore require high maintenance and preparation time before an observing night and produce significant operation overheads.
Solid state sum-frequency lasers, which may be the best option for a pulsed laser beam, are used at some observatories and offer a significant improvement to this situation.
They are also currently the most powerful sodium line lasers available.
The Gemini South MCAO system, for example, uses a single 50\,W laser to produce its 5 LGS \citep{org11}.
The maintenance issues of both dye and sum-frequency lasers now appear to be solved by the ruggedized and compact Raman fiber laser technology \citep{bona10} which has demonstrated a 20\,W output power.
It will generate the LGSs for ESO's adaptive optics facility \citep{ars10}, and is planned to be used for the European ELT.

LGS also possess considerable drawbacks with respect to NGS. 
First, due to the finite distance between telescope and LGS, the backscattered beam does not sample the full aperture at the height of the turbulent layers. This \textit{cone effect} or focal anisoplanatism is more severe for larger apertures and higher turbulent layers. Second, an LGS cannot be used to measure tip and tilt of the wavefront since the contribution of the upward projection jitter cannot be disentangled from the measurement. Hence, an NGS is still required, but the reduced demands on its brightness and distance to the target lead to a ten-fold increase in sky coverage \citep{ell98}.

ELTs offer an interesting alternative to using LGSs, if the AO system operates in open loop.
This is a very different mode of operation from typical closed loop AO systems, where the deformable mirror is in the optical path between the incoming light and the wavefront sensor so that one is only measuring the residual wavefront error.
In these systems, if the AO is functioning well, the measured residual wavefront will be close to zero.
In contrast, with open loop control, the full wavefront error is measured in each cycle and applied to a DM that corrects the path over a restricted field in one or more specific directions.
There is thus no direct feedback about how well the wavefront in the science field is being corrected.
With increasing telescope diameter, the beam overlap in the highest turbulent layers also increases and the angular separation of stars to sample the atmosphere can be larger. 
In an application of the multi-object AO technique (see Sec.~\ref{sec:moao}), \citet{rag11} argues that for 40-m aperture and 10\arcmin{} search area, the probability to find at least three stars which are bright enough for wavefront sensing (V$\approx$15\,mag) and reasonably well sample the highest layers over a 2\arcmin{} scientific field approaches 100\%. 
Tomographic reconstruction and open loop control would then allow for nearly all-sky coverage using only NGS.

%%%%%%%%%%%%%%%%%%%%%%%%%%%%%%%%%%%%%%%%%%%%%%%%%%
\section{Astrophysical Applications of Adaptive Optics}
\label{sec:science}

During the last decade, simple SCAO using natural and laser guide stars, has emerged from being a niche technology and is beginning to have an impact in mainstream astrophysics.
In this section we focus on a few examples drawn from what is now a vast AO literature, to illustrate how it has enabled a deeper understanding of the physical processes at work across a vast range of scales and epochs throughout the universe.

\subsection{Sun and Solar System}
\label{sec:solarsystem}

\subsubsection{The Sun}

One of the biggest successes of adaptive optics has been in driving forward our understanding of the sun's photosphere.
This complex layer is the interface between deep regions where turbulent convection dominates the dynamics, and the magnetically controlled chromosphere and corona above.
It is characterised by granulation from hot upflowing convective cells which sweep the uplifted magnetic field into the dark cooler downflowing lanes between them.
As pointed out by \cite{rim11} in their review, AO systems, where they have been installed on the current generation of 1-m class solar telescopes, are used for the vast majority of observations.
The scientific impetus for developing the next generation of 4-m class solar telescopes is entirely due to the enhanced performance afforded by AO.
Solar adaptive optics faces different challenges to those of night-time astronomy.
Observations are, by definition, performed during the day, and often at high airmass.
In addition, the wavelengths of scientific interest are typically in the optical regime (e.g. the {\it G}-band at 430\,nm).
All these make good AO performance hard to achieve.
Yet the wavefront sensing is accomplished using the low contrast, time varying, granulation of the sun's photosphere.
As such, the spots needed for a standard Shack-Hartmann sensor are created using a cross-correlation technique \citep{rim00}.
The reason that adaptive optics is able to have such an impact on solar physics is because the large scale structures are intimately linked to the small scale dynamics.
The two scale lengths that determine the structure of the solar atmosphere, the pressure scale height and the photon mean free path, are both of order 70\,km, corresponding to about 0.1\arcsec\ \citep{san08}, the optical diffraction limit of 1-m class telescopes.
Amongst the largest solar telescopes currently operational is the 1.6-m clear aperture telescope at Big Bear Solar Observatory.
The first results with its 76-aperture AO system reached a resolution of 0.12\arcsec\ at 706\,nm \citep{goo10}.
The data show that, in contrast to expectations and previous lower resolution observations, the smallest scale magnetic fields along the inter-granular lanes appear to be isolated points.
Their long lifetimes of several minutes suggest they are anchored deep beneath the photosphere.
Thus, as the granules on either side collide, they are forced into cyclonic flows. 
The resulting twisting of the magnetic field may be an important way to transfer energy upwards from a significant depth.

\subsubsection{Asteroids}

At the opposite end of the solar system's mass range, AO is showing that massive asteroids are surprisingly porous, with implications on their formation mechanism.
The best method to measure the mass, and hence bulk density, of an asteroid is to observe the orbit of a `moon' around it.
Today, about 150 main belt binary asteroids are known, and AO -- using the asteroids themselves as the wavefront references -- is a key diagnostic for studying these systems.
%few have been directly imaged at high resolution.
The first was discovered using the PUEO AO system on the Canada-France-Hawai`i Telescope (CFHT): 
the 4.7\,day orbit of Petit\,Prince around 45\,Eugenia revealed that the density of Eugenia is only $\sim20$\% greater than that of water \citep{mer99}.
%Spatially resolving main belt asteroids is complementary to light curve measurements or radar techniques:
AO on the VLT led to the confirmation that some asteroids are multiple systems, with the discovery that 87\,Silvia has 2 moons \citep{mar05}.
In a larger survey using AO on Keck~II, \cite{mar06b} imaged 33 $V\simlt13$\,mag main belt asteroids, achieving 2\,$\mu$m resolutions of 50--60\,mas, corresponding to $<100$\,km at a typical distance of 2\,AU.
One remarkable system, 216~Kleopatra, is highly elongated in a `dog-bone' shape, and also has 2 moonlets about 6\,mag fainter \citep{des11}.
Its bulk density of 3.6\,g\,cm$^{-3}$ is a factor of two lower than that of pure iron, and of the spectroscopically estimated surface grain density of its meteoritic analogue, implying that the asteroid is rather porous. 216~Kleopatra may therefore have formed as matter re-accumulated after a catastrophic impact disrupted a parent asteroid; its loosely packed material means that as it spun up, it became elongated, analogously to a spinning liquid mass; and that the satellites result from shedding of some of this material.
These authors have analysed a number of other weakly bound aggregates in similar detail using adaptive optics also on the VLT and Gemini North.
Among these is 617~Patroclus, which comprises 2 similar components separated by 680\,km with a bulk density of only 0.8\,g\,cm$^{-3}$.
Its angular momentum is high enough to not only have elongated the body but to split it into the observed binary \citep{mer01,mar06a}.
These results, derived from applying AO imaging at multiple epochs, have led \cite{des08} to propose such `rubble-pile' models as a general evolutionary scenario for these asteroids.

\subsubsection{Planets and their Satellites}

\begin{figure}
\psfig{file=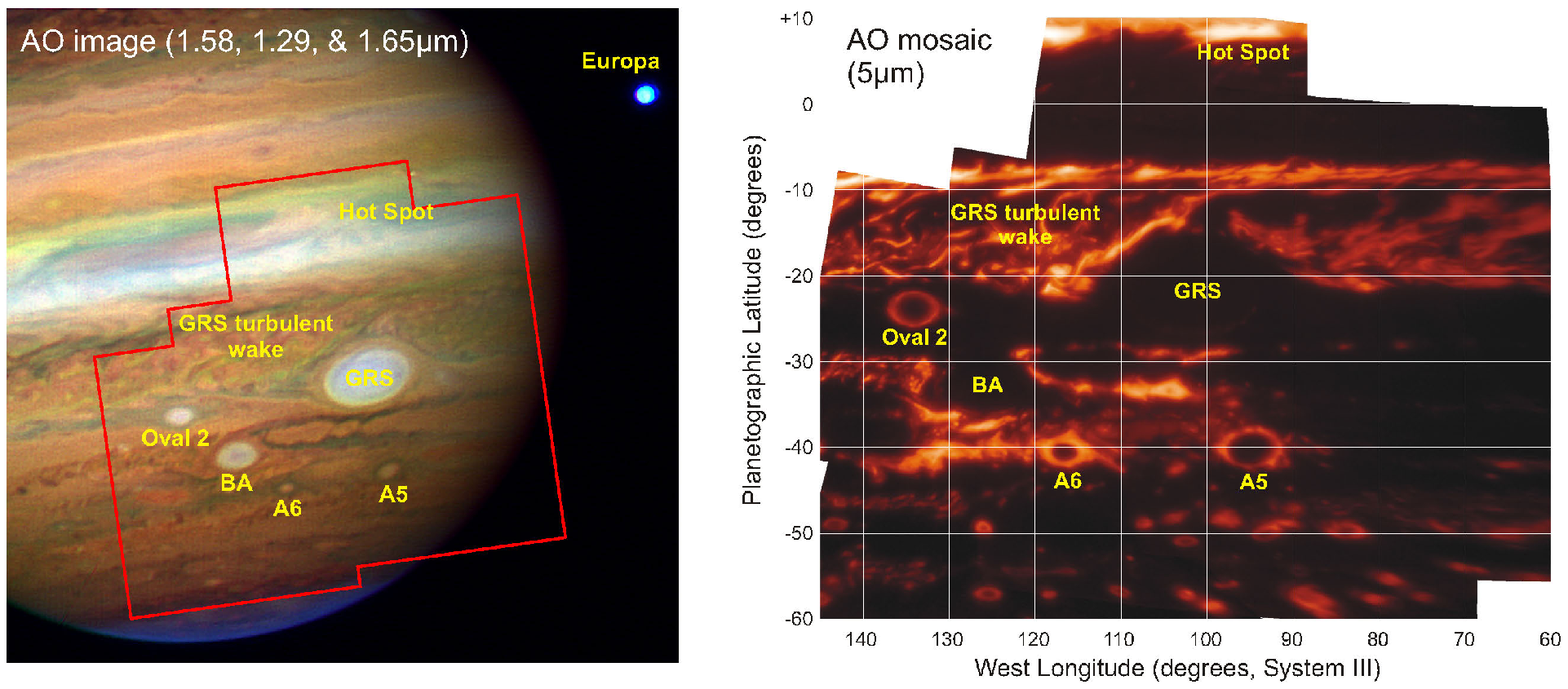,width=\textwidth}
\caption{Multi-wavelength imaging using AO on 8--10-m class telescopes (often together with HST optical imaging) is central to many science analyses.
Here this complementarity is applied to Jupiter, using AO observations from 1\,$\mu$m to 5\,$\mu$m.
The left panel shows a number of southern ovals at 1.3--1.6\,$mu$m wavelengths.
The right panel shows that at 5\,$\mu$m the smaller ovals are surrounded by rings.
Adapted from \cite{pat10} (courtesy of I.~de~Pater).}
\label{fig:jupiter}
\end{figure}

The planets, with angular diameters of a few arcsec to an arcmin, are obvious targets for adaptive optics.
Ground-based observations of Jupiter and Saturn \citep{gle97} using LGS at the USAF Starfire Optical Range, and of the three ring arcs around Neptune \citep{sic99} using PUEO at the CFHT, were among early successful applications to planetary science.
AO is particularly suited to studying the atmospheres of planets and their satellites, since different spectral bands are able to probe changes in the distribution and abundance of a variety of molecules and ices to different depths.
One frequent target is Titan, the only satellite to have a dense atmosphere.
It has now been observed for more than a decade with a variety of AO systems, spatially resolving its 0.8\arcsec\ diameter face and tracking seasonal and diurnal changes in its atmosphere \citep{har04,hir06,ada07}.
%It has also been studied using the fortuitous occultation of a binary star, for which AO revealed one component traversing each hemisphere \citep{bou03}.

Affording a special technical challenge to AO are observations of a planet or satellite while using another as the guiding reference, because of their different non-sidereal motions.
One example is Io, which has been observed in Jupiter's shadow to highlight its volcanic activity, with Ganymede as the AO reference \citep{pat04}.
Jupiter itself is another example.
Although such observations are complex, they enable one to obtain data across a vast wavelength range at similar spatial resolution, from optical space-based imaging with the Hubble Space Telescope (HST) to near and mid-infrared ground-based AO on 8--10-m class telescopes.
This provides a view of the reflected sunlight as well as Jupiter's own thermal emission, enabling one to probe to greater depths through the atmosphere.
One remarkable result, illustrated in Fig.~\ref{fig:jupiter}, is that the oval storm systems are typically surrounded by bright rings at 5\,$\mu$m \citep{pat10}.
The 225--250\,K brightness temperature of these rings suggests they are cloud-free down to a depth at which the pressure reaches at least 4\,bar.
Radiative transfer modelling of the 1--2$\mu$m data shows that the ovals have similar vertical structure, the main differences between them being the 
particle densities in the tropo- to stratospheric hazes.
These observations have led to a model in which the ovals are anticyclones where air is rising at the centre up to the tropopause at a pressure of $\sim0.1$\,bar, and descending around the periphery \citep{pat10}.
Importantly, the downflows cannot exist at radii greater than 1--2 times the Rossby radius, where rotational effects become as important as buoyancy effects (i.e. where the coriolis force has turned the velocity vector by 90$^\circ$), about 2300\,km at the latitude of the observations.
This is why the 12000\,km diameter Oval~BA does not exhibit a 5\,$\mu$m ring: instead the downflow occurs in the outer red part of the oval.
For the much larger Great Red Spot, the size of the system means that the Rossby radius, and hence the downflow, is located more towards the centre.
%, where the optical colour of the feature is also redder.

\subsection{Star Formation}
\label{sec:starform}

\subsubsection{Stellar Multiplicity}
\label{sec:browndwarfs}

Using adaptive optics to assess the multiplicity of stars is an obvious science goal, but one that is difficult to achieve: it requires a large well selected target sample, and very careful bias and sensitivity corrections that depend on the performance of the AO system as well as the magnitude and separation of potential companions.
AO offers clear advantages over the earlier speckle imaging \citep{ghe93} by enabling one to study higher contrast systems as well as fainter objects.
In the former case, AO has been cardinal in searches for low mass companions in the vicinity of higher mass stars, from an early survey of OB stars using ADONIS \citep{sha02} to observations around solar analogues \citep{met09}.
%;
%see \cite{duc12} for a discussion of this and other work.
In the latter case, a key role for AO has been to resolve very low mass binaries ($M_{tot}<0.185M_\odot$) in the near-IR where these cool objects emit most of their luminosity, and where AO works well.
Multi-year studies of their orbits have allowed accurate masses and luminosities to be measured; which in turn has led to calibration and verification of their evolutionary mass, luminosity, and age tracks \citep{zap04,clo07b}.
Intriguingly, while theoretical tracks \citep{cha00} are a reasonable match to the data for low mass stars, applying the same atmospheric models to even lower masses ($<13\,M_J$) fails to predict the very red low luminosity objects now known as giant exoplanets (see Sec.~\ref{sec:extrasolar}).

One of the largest AO surveys of very low mass stars was carried out by \cite{clo03} and \cite{sie03}, who used the Hokupa`a AO system on Gemini North to survey 69 stars of spectral type M6.0 to L0.5.
They found 12 systems with very low mass or brown dwarf companions, yielding
a sensitivity corrected binary fraction of order 10\%.
The pairs in each binary have similar masses, and their separations are only a few AU (with none beyond 15\,AU).
Remarkably, these characteristics differ significantly from the slightly more massive G dwarfs for which the binary fraction is around 50\%, and that exhibit a wide distribution of separations centered around 30\,AU.
An additional discrepancy between the populations was pointed out by \cite{dup11} and \cite{kon10} who, both using AO on Keck~II, compiled orbital data for 16 and 15 very low mass and brown dwarf binaries respectively.
Both studies highlighted a preponderance of almost circular orbits, and find at best only a marginal correlation between eccentricity and period.
The favoured model to explain these population differences, for which recent hydrodynamical simulations are reported by \cite{bat09}, suggests
that older ($\sim$Gyr) field brown dwarfs are systems that have been ejected at speeds of a few~km\,s$^{-1}$ from the cluster in which they formed.
For very low mass stars, only the most tightly bound systems survive such a velocity kick, resulting in both low multiplicity and small separations, and favouring low eccentricities.
While some discrepancies between observations and simulations remain, the global view is compelling.
In a new twist to this picture, \cite{clo07a} and \cite{bil11} find that young ($<10$\,Myr) very low mass and brown dwarf binaries can have much wider separations, beyond even 100\,AU.
The proposed explanation is that the survival time of a binary system depends on how tightly bound it is, as well as the stellar density in its local environment.
Wide binaries seen at young ages are likely to be disrupted by stellar interactions, and so are absent from the population of older field binaries.
Confirmation that a high multiplicity is established early on comes from AO observations showing that 30--50\% of embedded protostars are binaries with separations up to $\sim1000$\,AU \citep{duc07} and that many close protostar pairs may have formed via ejection of a third star \citep{con09}.

\subsubsection{Circumstellar Disks}

\begin{figure}
\psfig{file=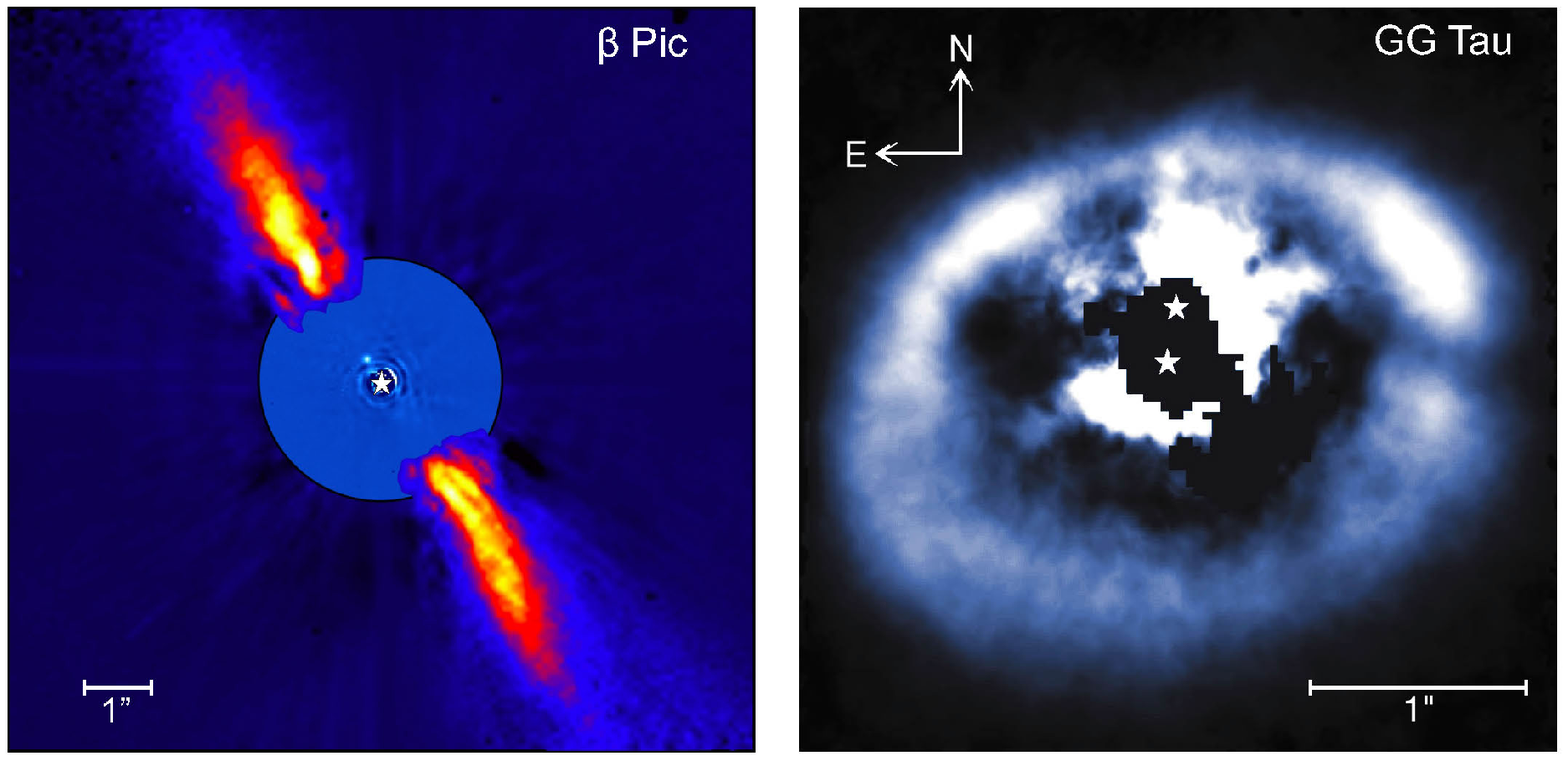,width=\textwidth}
\caption{AO images of circumstellar disks.
Left: composite image of $\beta$~Pic. The outer part shows the dust disk observed in the K-band with ADONIS; the central part is an L-band image from NaCO on the VLT which shows a companion 0.4\arcsec\ (8\,AU) to the north-west of the star (image credit: ESO/A.-M. Lagrange et al).
Right: H-band image of light reflected by the disk around GG~Tau, obtained with Hokupa`a on Gemini North, confirming the presence of a gap in the disk at a position angle of 270$^\circ$. The locations of the binary stars are indicated. (image credit: Daniel Potter/University of Hawaii Adaptive Optics Group/Gemini Observatory/AURA/NSF).}
\label{fig:circum}
\end{figure}

It is the disks around young stars that can shed light on how binary stars and planetary systems form.
The interest and importance of this topic is reflected in the huge activity over the last decade in studying disks around stars of all mass ranges, stimulated by the advent of high resolution imaging and interferometry.
%, notably millimeter interferometry of molecular lines \citep{dut07} and various imaging techniques \citep{wat07}.
While AO has not been the key driver in this field, it has added a new aspect by providing near- and mid-infrared data at the same resolution as optical HST images. 
The resulting multi-colour data make it possible to probe the size and structure of the dust grains and hence infer how the disk evolves.
The first circumstellar disk observed with AO was $\beta$~Pic \citep{gol93}, and numerous subsequent AO observations have revealed warps and sub-structure in this disk on 1\,AU scales (Fig.~\ref{fig:circum} left).
The binary T~Tauri system GG~Tau~A-B (Fig.~\ref{fig:circum} right) was also the focus of early studies \citep{rod96}.
AO imaging with Keck~II of its 3.8\,$\mu$m scattered light, combined with optical HST data, has indicated that the dust in the ring may be stratified, with the larger grains either settling towards the midplane or preferentially growing there faster \citep{duc04}.
However, it is not clear how universal these processes are since AO observations of HV~Tau~C \citep{duc10} and HK~Tau~B \citep{mcc11} show that, despite their similar ages, these 3 disks exhibit differing dust properties, which implies differing growth and/or settling times for their constituent grains.
Complementary to work based on the scattering phase function of grains are direct measurements of absorption from specific molecules.
Using the MMT adaptive secondary AO system to obtain spatially resolved spectroscopy of the 10\,$\mu$m silicate feature in each component of nine T~Tauri systems, \cite{ske11b} pointed out that, despite the large variation among different systems, shared properties within a system likely play an important role in dust grain evolution.
In a more massive system, detection of water ice via its 3.08\,$\mu$m absorption in the disk around the Herbig Ae star HD\,142527, was achieved using AO coronagraphy on Subaru \citep{hon09}.
While water ice is believed to promote the formation of cores in protoplanetary disks, this observation is perhaps the most direct evidence confirming its existence.

Returning to GG~Tau, there are intriguing hints that its disk may be tilted with respect to the orbit of its system.
Astrometric data obtained with AO on the VLT shows that the abrupt inner edge of the disk cannot be due to the stars unless their orbit is tilted by $\sim25^\circ$ with respect to the disk \citep{koh11}.
The only way to avoid this conclusion is to hypothesize the existence of a companion.
%beta pic?
Many circumstellar debris disks do show significant sub-structure, which is believed to be driven by orbiting companions, and can provide insights into the mechanisms of planet formation \citep{wya08}.
Indeed AO, together with non-redundant aperture masking for which the interferometric analysis leads to superior PSF calibration and speckle suppression, is now providing evidence that the central holes in the disks are due to the presence of a binary star \citep{kra11a} or are carved by giant planets \citep{hue11,kra11b}.

%However, while it is only a small conceptual step to directly imaging exoplanets, the technological hurdle is substantial.

\subsubsection{Extrasolar Planets}
\label{sec:extrasolar}

Direct imaging of exoplanets is a topical high priority science goal and one of the drivers of future ELTs.
It requires extremely high contrast ($>10^{-9}$), long exposure, coronagraphic imaging, combined with very careful control and characterisation of the residual speckle pattern \citep{opp09}.
These lead to demanding requirements for the AO, as well as for the associated instrumentation and post-processing techniques (see Sec.~\ref{sec:speckle} and~\ref{sec:extremeao}).
%The resulting designs have been incorporated into two instruments, SPHERE \citep{beu06} and GPI \citep{mac08}, which are close to commissioning at the VLT and Gemini observatories.
%It is predicted that such instruments will find large numbers of giant planets 
%with masses $<13.6$\,M$_J$ (the dividing line between planets and brown dwarfs) 
%in 5--50\,AU orbits, which are inaccessible to current Doppler techniques \citep{mac07}.
%There are far fewer systems amenable to current instrumentation, as evidenced
Although a few exoplanets have now been imaged directly, a number of surveys -- e.g. \cite{bil07} (45 targets), \cite{laf07} (85 targets), and \cite{kas07} (22 targets) -- have failed to detect any planets down to limits of a few~$M_J$.
%A meta-analysis of 118 such stars of $\simlt1$\,M$_\odot$ finds that 4$\,M_J$ planets should be very rare at separations beyond 22\,AU \citep{nie10}.

The first unambiguous confirmation by proper motion analysis of a multi-planet system was reported by \cite{mar08}.
HR~8799 is a 1.5\,M$_\odot$ A5V star at a distance of 39.4\,pc with an age of 30--160\,Myr.
It is now known to have 4 planets at distances of 14--68\,AU and masses in the range 7--10\,M$_J$, which all orbit in the same direction \citep{mar10}.
But the existence of giant planets at such a wide range of distances, as well as their low luminosities, is a puzzle \citep{clo10}.
While the outer 3 could have been formed via the fragmentation of a disk of gas and dust, at the distance of the innermost planet the disk would have been neither cold enough nor rotating slowly enough for this mechanism to work.
The alternative process is via the agglomeration of grains. 
This has been suggested for the $\sim9$\,M$_J$ planet $\beta$~Pic~b, also discovered by AO imaging, which orbits a 1.8\,M$_\odot$ star at a distance of 8--15\,AU \citep{lag10} (Fig.~\ref{fig:circum}).
However, for HR~8799~d, it is barely fast enough to form the planet before the matter is accreted onto the star itself, and certainly could not have formed the outer planets.
A more rapid growth rate could be achieved if the surrounding disk were dynamically cool, which would require it to consist of Pluto-mass planetesimals \citep{cur11}.
While this is not impossible, it is certainly considered an infrequent phenomenon.
%Since it also seems unlikely for the planets to have formed at larger or smaller scales and then migrated to their current separations, the formation of this planetary system is still the subject of debate.
In terms of their low luminosities, the favoured explanation -- which was successfully applied to the first exoplanet directly imaged, 2MASS~1207~b \citep{cha05} -- is that very thick high latitude cloud bands absorb and scatter light when the planet is viewed pole-on \citep{ske11a}.
Observational confirmation of this conjecture for HR~8799 is still at an early stage but it is a likely scenario if the clouds are thicker than found in brown dwarf atmospheres \citep{cur11}.
Given such a status quo, the next generation of planet imagers is eagerly awaited.

\subsection{Resolved Stellar Populations}
\label{sec:stellarpops}

Another major goal of ELTs is to spatially resolve stellar populations in nearby galaxies, in order to trace their star formation histories (ages, metallicities) by mapping the loci of individual stars on colour magnitude diagrams (CMD) \citep{ols03,dee11}.
Adaptive optics is a natural technical solution to the problem of crowding in these exceptionally dense stellar fields, the limiting factor in such work.
However, the better diagnostic power of CMDs at shorter (optical) wavelengths  conflicts with the increasing performance of AO systems at longer (near-infrared) wavelengths.
This has led to the requirement to enhance AO at shorter (0.7--0.8\,$\mu$m) wavelengths as well as to develop new techniques for estimating the ages of stellar populations \citep{bon10,dee11}.
These obstacles mean that AO on 8--10-m class telescopes currently struggles to be competitive with optical space-based work. 
Despite this, some progress has been made probing the stellar ages and star formation rates in the bulge and disk of M\,31 \citep{dav05,ols06} and the nearest dwarf galaxies \citep{mel10}.

The technique is highly effective when applied to galactic star clusters, notably yielding insights on the hotly debated topic of whether or not there is a universal initial mass function (IMF, defined as a power-law with slope $\Gamma$: $dN/dlogM \propto M^{-\Gamma}$, with a standard Salpeter value of $\Gamma = 1.35$) \citep{bas10}.
NGC\,3603 plays a role here because it is one of the most massive and densest star forming clusters in the Galaxy, and is often considered to be a local template for the massive star clusters found in starbursts and external galaxies.
Using AO to mitigate the crowding problems, \cite{eis98} and \cite{har08} probed the stellar population to sub-solar masses, far lower than achieved previously.
Their intriguing result is that, although at higher masses the IMF slope is consistent with the standard value, at $M<15$\,M$_\odot$ the slope has $\Gamma = 0.74$ indicative of a top-heavy IMF.
Although the mass segregation in NGC\,3603 -- evident through a flattening of the IMF at radii $<5$\arcsec\ (0.15\,pc) -- may complicate the picture, \cite{har08} found no evidence for a steepening of the IMF slope above $\Gamma \sim 0.9$ at any radius out to 110\arcsec\ (3\,pc).
However, while the IMF is distinguishably flatter than Salpeter in NGC\,3603 (and also in the Galactic Centre clusters), there are other massive star forming clusters which do not exhibit such an effect \citep{bas10}.
Thus, the cause of such variations, and the fundamental question of whether the IMF does systematically vary with environment, is still open.

The ability to image stellar fields at high angular resolution leads directly to a second scientific application: measuring the proper motions of stars, and hence deriving the internal kinematics of clusters or galaxies, as well as their global motion.
If the sources of error can be controlled to a sufficient level, then AO astrometry will become a major capability of ELTs, yielding a proper motion precision of 10\,$\mu$as\,yr$^{-1}$ after only 3--4\,yrs, equivalent to 5\,km\,s$^{-1}$ at 100\,kpc \citep{tri10}. 
Such techniques have already been applied to the Arches Cluster, which is only 30\,pc from the Galactic Center, using data from NaCo on the VLT \citep{sto08}.
Observations over a 4.3\,yr baseline show that this young cluster has moved by $24.0\pm2.2$\,mas with respect to the field.
Including its line-of-sight motion, the 3D velocity is $232\pm30$\,km\,s$^{-1}$ with a trajectory that rules out suggestions that it could be a rejuvenated globular cluster.
%If it is typical of clusters in this region, would tend to argue against the in-spiral of young clusters to the Galactic Centre.
Instead, the Arches Cluster may be transitioning between the x1 and x2 orbit families associated with the Galaxy's barred potential.

\subsection{The Galactic Center}
\label{sec:gc}

\begin{figure}
\psfig{file=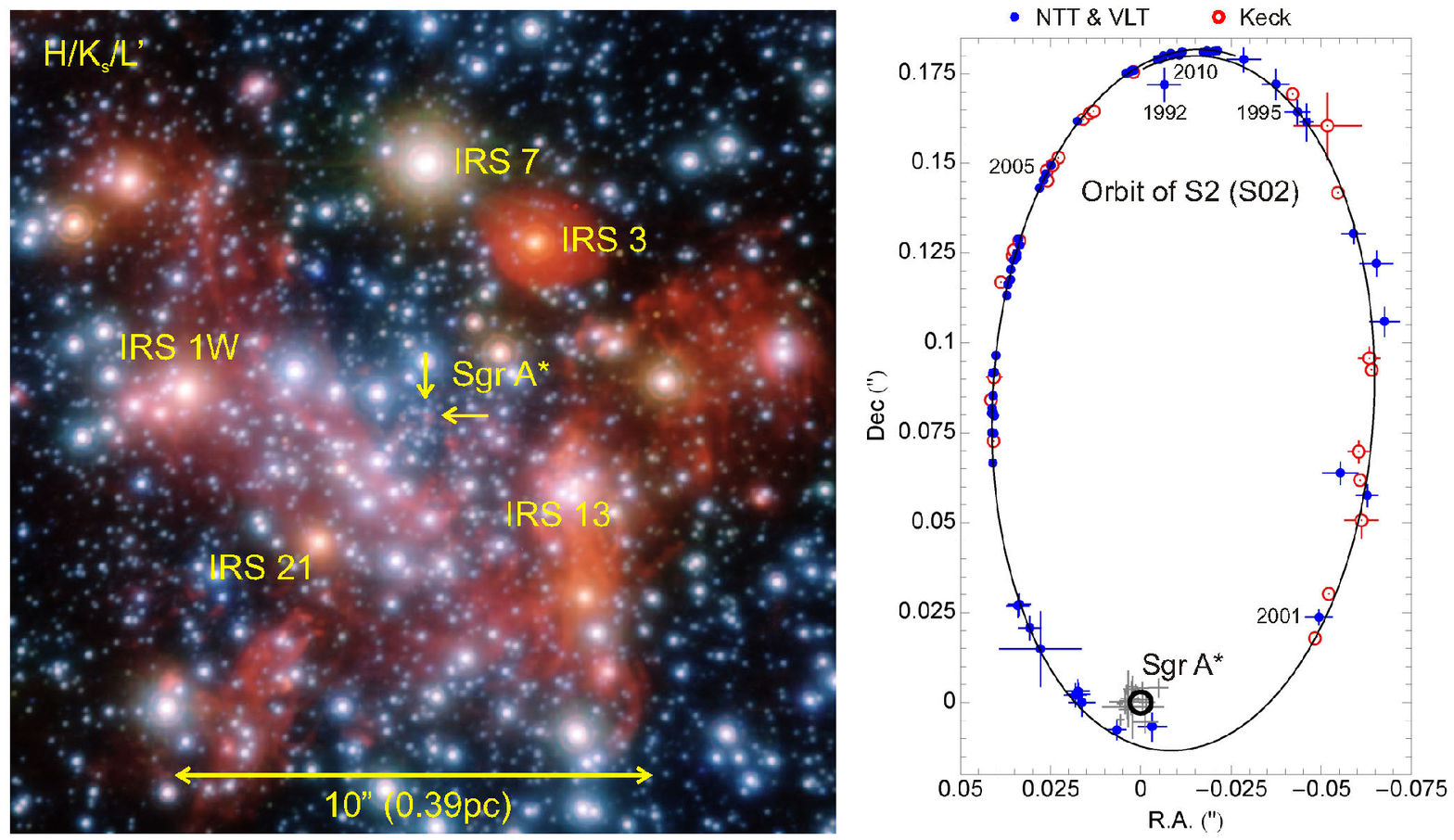,width=\textwidth}
\caption{Left: adaptive optics imaging of the Galactic Center in H, K$_s$ and L bands (blue, green, and red respectively). The location of Sgr\,A* is marked by the 2 yellow arrows.
Right: astrometric measurements of the star S2 (or S02) over a period of nearly 20 years has tracked more than one orbit around Sgr\,A* (the grey crosses mark the location of the near-infrared flares, very likely coming from within ten times the Schwarzschild radius of the massive black hole ).
In the same coordinate system the Keck and VLT astrometry is consistent.
Adapted from \cite{gen10} (courtesy of S.~Gillessen).}
\label{fig:gc}
\end{figure}

The Galactic Center (GC) provides an exquisitely detailed view of the physical processes occuring in the nucleus of our Galaxy and around its central massive black hole (BH), which can be directly applied to the nuclei of other galaxies.
It is a rapidly advancing field in which spatially resolving the stellar populations is of particular interest.
The severe crowding means that AO plays a central role despite the fact that the GC is highly obscured and has no optically bright guide stars.
Instead, because a $K\sim6$\,mag star lies just 5.5\arcsec\ away, it is a primary science driver for infrared wavefront sensing.
Currently, NaCo on the VLT is the only camera with an infrared WFS \citep{rou03}.
%, and it can easily achieve 50\% Strehl ratio in the H-band on this guide star.
In contrast, when observing the GC from Keck~II, a laser guide star must be used;
that the GC never rises above 41$^\circ$ from that latitude adds to the difficulty of obtaining high resolution data.
But observations from both observatories have been highly successful.
A complete review of recent observational and theoretical progress on the central parsec is presented by \cite{gen10}.
The main advances that have been made using AO are summarised below.

The $\sim40$\,mas resolution that can be achieved in the H-band on 8--10-m class telescopes (Fig.~\ref{fig:gc}, left) has enabled the stellar distribution to be traced to scales well below $0.04$\,pc ($1\arcsec$ at the distance of 8\,kpc) from Sgr\,A*.
Although the cusp maximum is centered on Sgr\,A*, the radial distribution of different stellar types varies considerably; and much of the increase in the central 1\arcsec\ is due to a high concentration of B stars \citep{bar10}.

Analysis of the stellar motions (Fig.~\ref{fig:gc}, right), afforded by 150--300\,$\mu$as astrometry \citep{fri10,ghe08}, shows that while the vast majority of stars in the central parsec are old and have randomly oriented orbits \citep{yel10}, about half of the young stars in the central 10--15\arcsec\ are confined to a warped clockwise disk, while many of the remainder may be in a second counter-clockwise disk \citep{lu09,bar10}.
This has put strong constraints on the mechanism of the star formation event(s) 6\,Myr ago that gave rise to these stars.

A combination of astrometry and line-of-sight velocities has enabled the full 3D orbits of about 30 stars to be determined, showing beyond reasonable doubt that Sgr\,A* is a massive BH \citep{ghe08,gil09}.
The combined Keck~II and VLT data sets are yielding ever more precise measurements of the distance to the GC and mass of the BH, for both of which the systematics associated with the distance is now the dominant error term.
The current best values are 8.3\,kpc and $4.3\times10^6$\,M$_\odot$ \citep{gen10}.

Light curves of near-IR flares from the accretion flow towards Sgr\,A* are now regularly measured.
Statistical analysis of their frequency, brightness, and variations indicate that there may be two types: a power-law distribution of occasional bright flares showing substructure, superimposed on a lognormal distribution of faint continuous variability characterised by red noise \citep{do09,dod11}.
Such data, combined with spectral index and polarisation measurements as well as simultaneous multi-wavelength observations are providing detailed constraints on the physical models of flare emission \citep{gen10}.

\subsection{Galaxy Nuclei and Active Galaxies}
\label{sec:galaxies}

\subsubsection{Black Hole Masses}

Our understanding that the growth of supermassive black holes is tied to that of their host galaxies is founded on measurements of BH masses.
These have led to the emergence of relations between the BH mass and the velocity dispersion, mass, and luminosity of the stellar spheroid around it \citep{geb00,hae04,fer05}.
AO is increasingly dominating what was once the domain of optical longslit spectroscopy with HST, 
because the high resolution it affords is coupled with a large collecting area (allowing one to study faint galaxies) and integral field spectroscopy (the 2D coverage of which enables a more robust recovery of the orbital distribution), at near-IR wavelengths (to probe dust obscured nuclei).
\cite{dav06} showed that it is possible to measure BH masses in type~1 active galactic nuclei (AGN) using spatially resolved stellar kinematics, providing a complementary method to reverberation mapping which relies on tracking the temporal variability of the broad lines.
For NGC\,3227 they argued that $M_{BH}$ was lower than previous estimates, lying in the range (7--20)$\times10^6$\,M$_\odot$.
The most recent reverberation mass of $(7.6\pm1.7)\times10^6$\,M$_\odot$ derived from a well sampled light curve \citep{den10} is consistent with this range.
Nevertheless, uncertainties for measuring $M_{BH}$ in quiescent or active spiral galaxies are likely to remain high due to data quality \citep{kra09} as well as systematics and degeneracies associated with the distribution function of the stellar orbits, the co-existence of multiple stellar populations, and the presence of significant nuclear gas masses \citep{dav08a,gul09}.

Coupled to this is the question of pseudo-bulges.
Because they are built through secular disk processes rather than merger events (and therefore have different stellar populations, mass distributions, and kinematics), it is not clear whether the black hole properties should correlate with them in the same way as for classical bulges \citep{orb11,kor11}.
It is now realised that many local disk galaxies have at least a pseudo-bulge component to their central regions \citep{wei09,fis11}, and so this may contribute to the scatter of disk galaxies on the $M_{BH}-\sigma_*$ plane.
Using AO to derive black hole masses in such galaxies, Nowak et al. (\citeyear{now09,now10}) suggest that to really understand the co-evolution of BHs and bulges one does need to tease apart the pseudo- and classical bulge components.

For elliptical galaxies, \cite{geb11} argue that the main sources of uncertainty on $M_{BH}$ are the treatment of the dark matter halo, an incomplete orbit library, and triaxiality.
To overcome them for M\,87, these authors combine AO integral field spectroscopy with wider field data to find a mass of $(6.6\pm0.4)\times10^9$\,M$_\odot$.
That this exceeds the mass expected from the $M_{BH}-\sigma_*$ relation by twice its uncertainty suggests that the high mass end of the relation is poorly constrained and/or its scatter is larger than previously thought.
%, and that there is much still to learn about the physics underlying the relation.

\subsubsection{Gas Inflow and Outflow}

\begin{figure}
\psfig{file=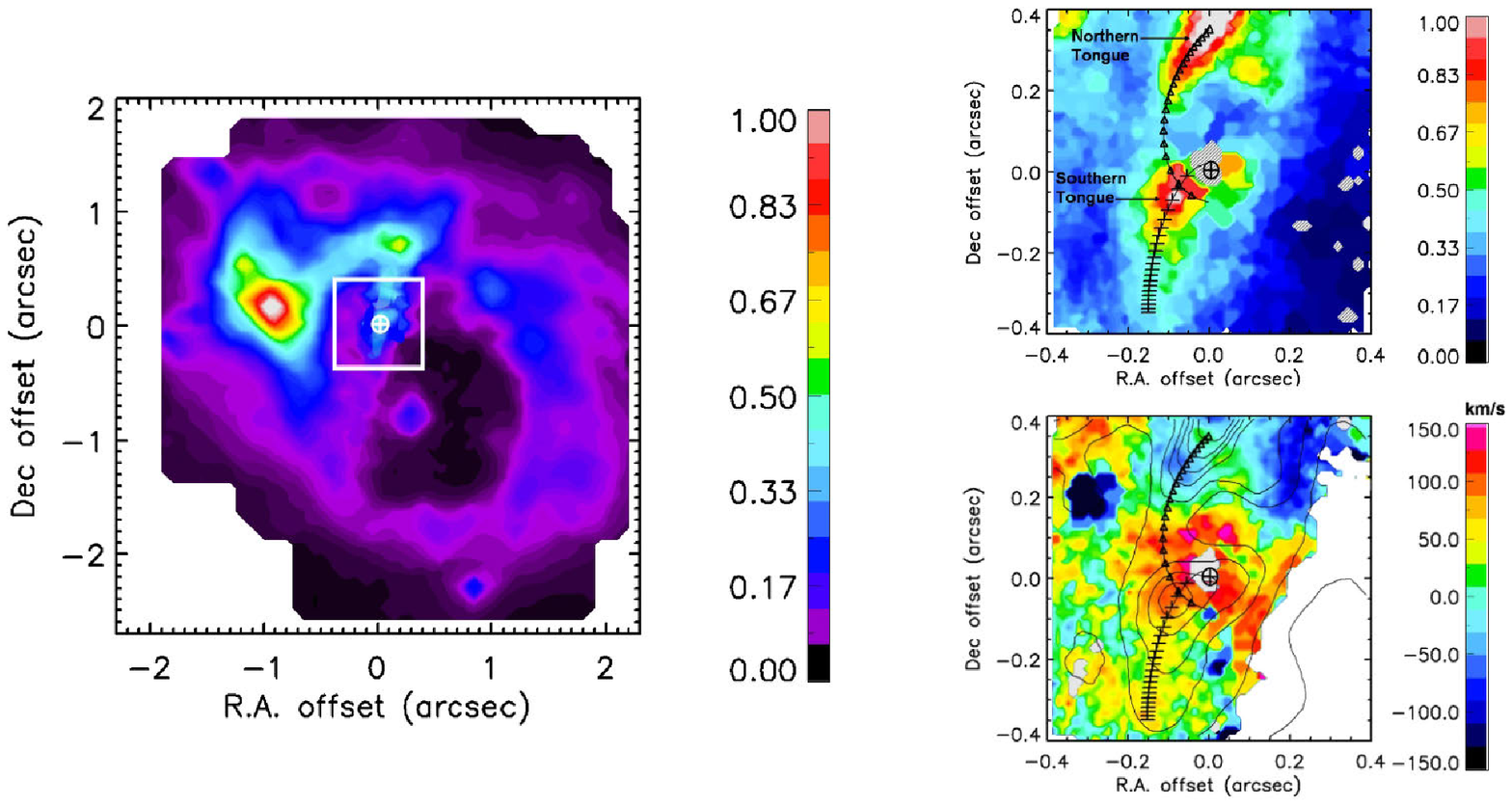,width=\textwidth}
\caption{AO integral field observations of the type~2 AGN NGC1068 (1\arcsec = 70\,pc) in the 2.12\,$\mu$m H$_2$ line.
Left: the central few arcsec show an expanding ring of molecular gas with the brightest emission on the northeast side. The location of the AGN is marked. The square denotes the region viewed in more detail on the right where gas is rapidly inflowing almost directly towards the AGN.
Right: two tongues of emission are revealed in the central arcsec (line emission (top) and projected velocity (bottom); the line properties could not be extracted at the location of the AGN itself).
The trajectories of the best fitting models of inflow are shown, and follow both the emission and its velocity -- which on the north side is approaching (blue colours), and on the south side is receeding (yellow colours), accelerating as it nears the AGN (red colours).
Adapted from \cite{mue09}.}
\label{fig:n1068}
\end{figure}

Nearby active galactic nuclei (AGN) provide an excellent opportunity to study the mechanisms that drive gas towards the central BH.
Their proximity means that with AO, scales down to a few parsecs can be resolved.
However, it is the combination of AO with integral field spectroscopy that is leading the way forward, by probing the full 2D distribution and kinematics of stars, molecular gas, and ionised gas.
Such techniques, in the optical or in the near-IR using AO, have revealed inward flows of gas at relatively low rates along circumnuclear spiral arms in a number of galaxies \citep{fat06,sto07,rif08,dav09,schm11}.
Although there can be exceptions to this, such as the dramatic case of NGC\,1068 where the gas appears to be streaming
almost directly towards the AGN \citep[][Fig.~\ref{fig:n1068}]{mue09}, such inflow is in principle sustainable for Gyr timescales \citep{dav09}.
It is believed to be associated with the circumnuclear dust
structures that have been mapped in many active and inactive galaxies.
The implication is that gas streaming ought to be common, but its relation to AGN accretion is still far from clear.

Part of the reason for this is that such processes bring gas to the central tens of parsecs where one might expect a starburst to ensue;
but the role of nuclear star formation in fuelling AGN is still open to debate.
Stellar population synthesis of optical \citep{fer04,sar06} and infrared \citep{rif09a} data suggest that while some AGN appear to be associated with recent  ($<100$\,Myr old) starbursts, the nuclear stellar spectra of others are dominated by intermediate age or very old populations.
A possible solution to this puzzle was proposed by \cite[][see also \citealt{wil10}]{dav07}.
Using AO to probe scales below tens of parsecs, they presented evidence that the youngest starbursts are associated with only weak AGN accretion, and that AGN fuelling is much more efficient in a post-starburst after the early phases of turbulent stellar feedback have ceased.
Hydrodynamical simulations \citep{scha10} tend to support this view, and have shown how stellar ejecta from a post-starburst may generate a turbulent 1-pc scale gas disk consistent with those inferred from maser observations.
If this gas is related to the obscuring torus, the simulations imply that the torus may be a dynamically evolving structure with different components responsible for the various observed phenomena,
including small scales structures seen with VLTI \citep{tri07,rab09,bur09,bur10} as well as larger scale components \citep{hic09}.
The use of AO integral field spectroscopy to map the locations of different age stellar populations \citep{rif10,rif11b} is a new aspect that should provide important clues in the emerging picture of how nuclear starbursts are related to the torus and to AGN fuelling.

Seyfert nuclei require little gas to fuel their moderate luminosities, so one might ask whether the far larger inflowing gas masses are piling up or being expelled.
Here also integral field spectroscopy coupled to AO is revealing details about AGN outflows via their ionised and coronal line emission \citep{rif09b,sto10,mue11,rif11a}, in a way that is highly complementary to longslit spectroscopy with HST \citep{cren11}.
Since the outflow rates are 100 or more times greater than the accretion rate onto the BH, it seems that AGN driven winds are entraining a significant amount of gas from the local ISM.
Clues about the complex interaction between the outflow and the ambient ISM are also appearing from models of the outflow geometry.
These are providing hints that the more molecular gas there is in the vicinity, the faster and narrower the outflow \citep{mue11}.
These first results illustrate how AO, by enabling one to probe the detailed physics of outflows in local AGN, is a key technology in understanding a phenomenon that has shaped galaxy evolution.

\subsubsection{Quasars and Mergers}
\label{sec:qsohost}

While it is possible to probe to small scales in nearby Seyfert galaxies, the more luminous QSOs provide a much greater observational challenge, due to their distance and the often overwhelming brightness of the AGN with respect to the host galaxy.
As emphasized by \cite{guy06} in their survey of 32 nearby QSOs using Gemini North and Subaru, the key issue is knowing the PSF so that it can be accurately subtracted (see Sec.~\ref{sec:psf}) in order to reveal the underlying host galaxy.
At high redshift, one has to contend with the $(1+z)^4$ surface brightness dimming and so, even with AO, detecting the host galaxy can itself be difficult \citep{cro04,fal05}.

The specific role of mergers in fuelling QSOs can be assessed by targeting the $\sim1$\% of QSOs that exhibit double peaked [O{\sc iii}] lines, which could trace pre-coalescent dual AGN.
Remarkably, AO imaging of such candidates shows that only 30--40\% have double nuclei on kiloparsec scales \citep{fu11,ros11}.
Instead, \cite{fu11} argue that the origin of the double peaked line profiles may lie in other processes such as outflows or jet-cloud interactions;
and \cite{ros11} suggest that most QSOs may not be associated with gas rich major mergers, an idea that is gaining weight from other observations \citep{cis11}.
As discussed in Sec.~\ref{sec:highz}, the role of secular processes in galaxy evolution at early cosmic times is now also receiving increasing attention.

Nearby luminous merging systems provide an alternative approach to this issue.
One of the best known dual AGN, NGC\,6240, has been the focus of several AO studies.
These have highlighted numerous massive young star clusters around the nuclei which are undetected in optical HST images \citep{max05,pol07}.
Comparison of AO images from K-band (2.1\,$\mu$m) to L-band (3.8\,$\mu$m) with X-ray and radio continuum data have revealed the location of the two AGN \citep{max07}.
Although the AGN themselves are highly obscured, across most of the merger the extinction is rather less \citep{eng10}: 
combining AO data with HST images shows it to be consistent with the \cite{cal00} reddening law, derived for isolated star forming galaxies rather than mergers.
Spatially resolved stellar kinematics afforded by AO now provide, in addition to the tidal arm morphology, independent constraints on the merger geometry \citep{eng10}; and have yielded the intriguing result that the BH in the southern nucleus also lies above the high mass end of the $M_{BH}-\sigma_*$ relation \citep{med11}.

\subsection{The High Redshift Universe}
\label{sec:highz}

One of the most remarkable applications of adaptive optics has been in spatially resolving the internal structure and kinematics of star forming galaxies at $z \sim 1.5$--3, the epoch of peak mass assembly.
These galaxies are usually no more than 1--2\arcsec\ across, and the optical emission lines that constitute prime diagnostics of their dynamical and physical properties are redshifted to the near-IR.
They would make ideal AO targets, except that most deep fields where such galaxies are identified are specifically chosen to avoid bright stars.
At the same time the low surface brightness means the resolution is often limited to 0.1--0.2\arcsec\ by instrumental and observational constraints (e.g. seeing variability over long integrations).
Making use of AO in this regime is difficult.
For their large spectroscopic imaging survey of 63 high redshift galaxies, \cite{for09} used AO on only 12 targets (a number that has doubled in an extension to the survey, \citealt{man11}).
Similarly \cite{law09b} detected only 13 of their targets using AO, \cite{wri09} observed only 6 galaxies and \cite{wis11} 13, and \cite{man09} selected 10.
Quantitatively, the severity of the selection criteria means that even when using a laser guide star, only about 10\% of high redshift galaxies have a suitable guide star within 30--40\arcsec; and when one takes into account other criteria (e.g. ensuring that the faint emission lines fall in regions of high atmospheric transmission but not too close to bright OH airglow lines), the fraction goes down to $\sim1$\% or less \citep{man09,man11}.

\begin{figure}
\psfig{file=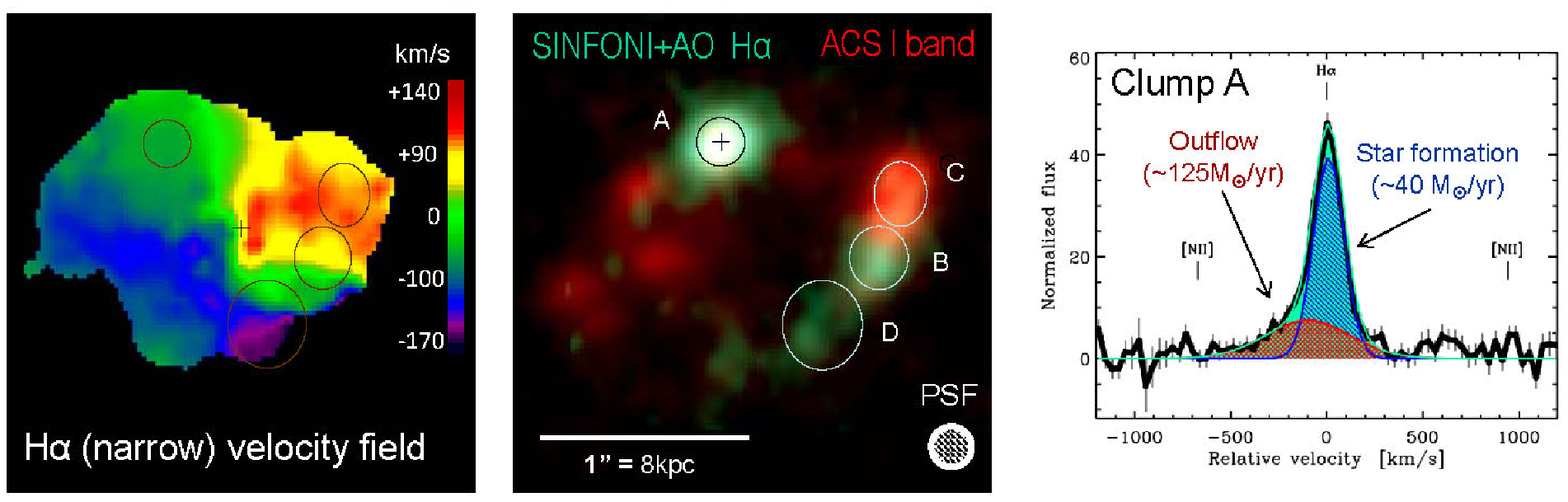,width=\textwidth}
\caption{AO integral field spectroscopy is now an important tool in probing the physical structure and dynamical state of high redshift galaxies.
These data of the $z\sim2$ galaxy ZC406690 reveal a coherent velocity field indicative of disk rotation (left), the individual star-forming clumps (center), and a blue wing on the spectral line profile of one of these clumps that traces a powerful outflow (right).
In the right panel, the blue profile tracing the star formation and the red profile tracing the outflow sum to give the green profile which is a fit to the observed spectrum.
Adapted from \cite{genz11}.}
\label{fig:zc406690}
\end{figure}

Following the kinematical confirmation with AO that massive star forming galaxies at $z\sim2$ are not necessarily mergers but can also be gas rich disks \citep{gen06}, it is now thought that about 1/3 of such galaxies are in fact disks \citep{for09}.
As shown in Fig.~\ref{fig:zc406690} for one such object, these disks are unlike those of local galaxies.
%, which are dynamically cool, have $\sim10$\% gas fractions form stars at a few $M_\odot$\,yr$^{-1}$.
They are rapidly forming stars, often in giant star-forming complexes or clumps.
Fitting beam-smeared disk models to the kinematics shows that they have high intrinsic velocity dispersions of 20--100\,km\,s$^{-1}$ \citep{wri09,cre09,law09b}.
Many studies now suggest that such dispersions are characteristic of normal massive high-redshift disks, and are connected to the high gas accretion rates through cold flows at early cosmic times, high gas fractions, and global instability to star formation.
AO is a key capability because it separates the $\sim1$\,kpc (0.1--0.2\arcsec) sized clumps from the inter-clump regions, which are blurred together in seeing-limited conditions.
Even higher linear resolutions, corresponding to as little as $\sim100$\,pc, can be reached if AO is applied to gravitationally lensed objects \citep{sta08,jon10}.
Using AO with integral field spectroscopy is now shedding light on the detailed properties of the individual clumps and showing that they can drive powerful winds, with outflow rates that may exceed their star formation rates \citep{genz11}.
This has important consequences for the clump lifetimes, and hence the evolution of the galaxies, for example whether they survive long enough to migrate inwards and build a nascent bulge \citep{gen08} or fuel an AGN \citep{amm09,wri10}.

Through the use of AO and integral field spectroscopy, and despite the difficulties of finding suitable targets, this field is now thriving.
The observations, although still relatively few, have already led to fundamental developments in our understanding of galaxy evolution.
They are now driving the requirements for multi-object AO systems that can operate simultaneously on many integral field units deployed across a wide field (see Sec.~\ref{sec:moao}).

%%%%%%%%%%%%%%%%%%%%%%%%%%%%%%%%%%%%%%%%%%%%%%%%%%

\section{Point Spread Function}
\label{sec:psf}

Knowledge of the point spread function (PSF) is of obvious importance for analyzing AO data sets, yet the PSF delivered by adaptive optics systems has an unfavourable reputation, based on its complex shape and its spatial and temporal variability.
While the traditional advice to cope with this has been to observe a PSF reference star and use it to deconvolve the data, there are in fact a variety of options available to both measure and compensate for the PSF.
Despite this, the limited applicability of all these methods means that the issue of recovering the AO PSF remains an unsolved problem.
As such, our aim in this section is to set the scene for future developments that might assist in quantifying spatial and spectral PSF variability, and ultimately in motivating efforts towards PSF reconstruction.

We begin by discussing how the PSF is used to extract the desired information from the data.
Despite the variety of scientific analyses described in Sec.~\ref{sec:science}, the PSF is typically used in only a few different ways.
We outline the aims of these methods, and look at the accuracy with which they require one to measure the PSF.
In the second part of the section, we turn to how one might achieve this.
There are a number of methods to empirically measure the PSF, either with additional observations or from the science data itself, but none of them are globally applicable.
The alternative of reconstructing the PSF from wavefront data would be preferred by astronomers, but implementing it robustly is still a major challenge.

\subsection{The Role of the PSF}

The main methods of PSF-dependent post-processing can be broadly grouped into four classes:
(i) deconvolution,
(ii) model convolution,
(iii) point source photometry and astrometry, or
(iv) speckle suppression.
These regimes, which are discussed in the following sections, disentangle the scientific information from the effects of the PSF in very different ways depending on the aim of the analysis.
%In myopic deconvolution, approximate initial estimates of the PSF allow it to be recovered at the same time as enhancing the contrast of faint structure in the data.
%Alternatively, when a model convolved with the PSF is matched to the data, the limitation is the simplicity of the model and so it is often sufficient to approximate the PSF with bivariate functions.
%The accuracy of photometric and astrometric measurements from crowded stellar fields is directly related to the precision with which one can eextract the PSF.
%This is taken to extremes in high contrast imaging, but the approach is to use careful optical design and data processing techniques to suppress, rather than measure, the speckles in the PSF halo.

\subsubsection{Deconvolution}
\label{sec:deconv}

The intensity distribution across an image $I$ corresponds to the convolution of the true intensity distribution of an object $O$ with the PSF $P$ together with some additive noise $N$: $I = P \otimes O + N$.
The aim of deconvolution is to infer the unknown $O$.
%This is most straightforwardly done in Fourier space where the convolution becomes a multiplication.
%Inverting the problem then yields an estimate of the Fourier transform of the object $\hat{\tilde{O}}$ via simple division
%\[
%\hat{\tilde{O}} = \hat{I}/\hat{P} = \hat{O} + \hat{N}/\hat{P}.
%\]
During post-processing, it
%deconvolution (albeit rather more sophisticated than the most basic form above) 
is commonly applied in solar astronomy, asteroid studies, and imaging of planetary surfaces and atmospheres, targets which all contain structure that does not easily lend itself to characterisation with a simple analytical model.
Generally, deconvolution does not lead to super-resolution \citep{sta02}.
Instead, the aim is to enhance low contrast features in the data, in effect by removing the characteristic broad AO PSF halo that dilutes them.
Such techniques are particularly important for generating dopplergrams (i.e. difference images of the red and blue wings of an emission line) of solar features, which are sensitive to variations in AO performance and hence PSF core/halo contrast \citep{rim11}.

\begin{figure}
\psfig{file=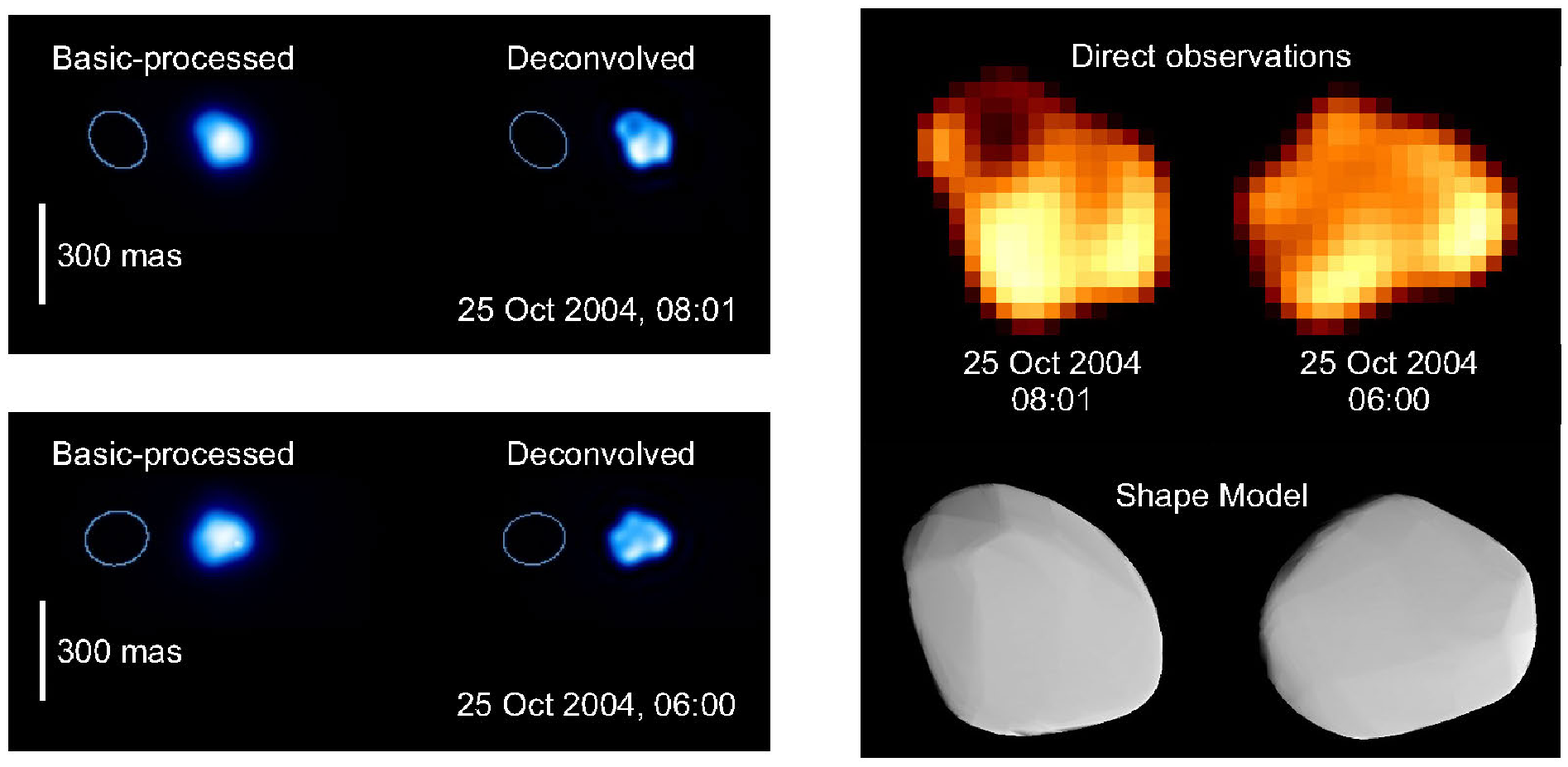,width=12cm}
\caption{Left: Keck~II AO images, taken 2 hours apart, of the asteroid 9\,Metis in 2 orientations. The deconvolution with AIDA has enhanced the contrast of the surface features.
Right: the features across the deconvolved images match the shape inferred from the independently derived lightcurve inversion model at the appropriate orientation, and hence remove the pole ambiguity from that model.
Adapted from \cite{mar06b} (courtesy of F.~Marchis).}
\label{fig:deconv}
\end{figure}

\cite{sta02} describe a number of methods to perform deconvolution based on different mathematical approaches, each of which is suited to specific circumstances, for example whether the astronomical object has well defined edges (e.g. planets, asteroids) or not (e.g. galaxies).
However, there are many issues that must be carefully controlled.
Chief among these is noise amplification, to which deconvolution, as an inverse process, is inherently prone.
Another key issue is the limb brightness of planetary surfaces, which can be artifically enhanced due to Gibbs oscillations resulting from a high luminosity gradient \citep{hir06}.
Such problems can be reduced by explicitly including terms in the algorithm to preserve the edge properties, as was done in the AIDA code \citep{hom07} which was developed for processing asteroid images (Fig.\ref{fig:deconv}).
Its implementation is a specific extension of MISTRAL \citep{mug04}, a myopic restoration tool.
The strength of myopic deconvolution is that it does not assume the PSF is perfectly known.
Instead, it estimates the PSF and object simultaneously, using a number of PSF reference images in addition to the object frames, and applying soft constraints to its properties.
% (e.g. non-negativity, the band limit of its optical transfer function).
By using priors in this way, it is more robust than blind convolution, which makes no assumptions about the PSF.
On the other hand, blind deconvolution is applicable to solar images where no image of a PSF can be taken.
Instead, prior constraints are provided by a careful optical design that yields simultaneous imaging of 2 or more narrow bands, together with defocussed data that enable one to make use of phase diversity methods.
Using such data it is possible correct high order aberrations post-facto \citep{noo05}.

\subsubsection{Model Convolution}
\label{sec:conv}

The requirement to disentangle object data from the PSF does not necessarily mean that a classical deconvolution must be performed.
If the target can be reasonably described by an analytic expression, then this can be convolved with the PSF and the result compared directly to the observed data.
Thus, given $I = P \otimes O + N$,
one defines a model $M$ that provides an estimate of the observed intensity distribution
$\tilde{I} = P \otimes M$.
Iterative improvements to the model $M$ lead to $\tilde{I} \sim I$, at which point it is reasonable to assume that $M \sim O$.
Because typically only a few parameters are needed to specify a model, the problem is highly constrained and does not lead to noise amplification inherent in deconvolution.
Perhaps the most important advantage of this method is that it deconstructs the intrinsic source properties so that the final product is a set of parameters (rather than another image) which can directly provide insights into the source itself.
The methodology also enables one to make reliable estimates of the uncertainties of the parameters, because it is straightforward to quantify the impact of changing them.
Since it is often the inevitable simplicity of the model rather than the PSF that limits how well the convolved model matches the observed data, one does not require a highly detailed measurement of the PSF.
It can be sufficient just to know the strength and extent of the wings, which enables one to represent the core and halo of the PSF in a simple way as a pair of bivariate Gaussian or Moffat functions.

Being well suited to data with low signal-to-noise, the method is frequently applied to deep images of distant galaxies, the light intensity profiles of which are commonly parametrized using a S\'ersic function.
%This has been shown to be a first order Taylor expansion of any real light profile and so, in the limit of low resolution and low signal-to-noise, can approximate the light profile of any galaxy \citep{and11}.
But while fitting a simple function to a galaxy profile is the strength of model convolution, it is also a weakness: one only knows how well that particular family of models matches the data, and it is not always clear whether (or how) the fit may be biassed by complex structure that is not accounted for properly.
It is to overcome this limitation that perturbations in the form of Fourier modes have been included to the azimuthal shapes in the popular galaxy profile fitting code GALFIT3 \citep{pen10}.

Kinematics are a special case where deconvolution is not an option because of the inter-dependency imposed by the PSF between the observed luminosity, velocity, and dispersion (velocity width).
In this case, the only alternative to model convolution is to deconvolve each spectral plane of the original datacube, noting that the PSF that varies gradually from one plane to the next.
%Although it is standard procedure to clean datacubes produced by millimetre and radio interferometers, such techniques have seldom been applied to optical or near-infrared datacubes.
Early work by \cite{fer00} used an independent higher resolution image to guide such a deconvolution, leading to a factor 3 gain in spatial resolution for the kinematics of the [O\,{\sc III}] outflow in Mkn\,573.
More recent efforts have focussed on the specific context of extracting the spectrum of a supernova (i.e. point source) superimposed on a galaxy \citep{bon11}, but the method has more general applications.

\subsubsection{Point Source Photometry and Astrometry}
\label{sec:photastrom}

The information contained in images of stellar fields (from binary stars to entire star clusters) comprises the flux and position of each source.
In densely populated fields, because the sensitivity is limited by the crowding rather than photon and detector noise, the key to extracting accurate photometry and astrometry is in obtaining a good estimate of the PSF with which to fit the sources.
In addition it is important to quantify the flux level at which one can no longer distinguish between faint sources and residual speckles or other noise features, since this directly relates to the spatially variable completeness correction that must be applied to the derived number counts.
Various codes have been developed explicitly for processing stellar fields, most notably DAOPHOT \citep{ste87} and StarFinder \citep{dio00}.
Within DAOPHOT, the reference PSF is based on analytical fits to several stars (with a look-up table for small empirical corrections), and so is able to handle undersampled data as well as accommodate some degree of spatial variation.
On the other hand, StarFinder was designed to work on AO data with complex PSF shapes, and so uses an empirical PSF that is extracted from a combination of various stars in the field.
Although it currently does not account for anisoplanatism, a local PSF can be extracted for each distinct isoplanatic region.
Crucially, even though the core of the PSF may be very narrow, with a FWHM of $\sim50$\,mas, high photometric precision requires that the reference PSF is determined out to at least 1\arcsec\ so that the faint wings are properly characterised.
It is because of this that \cite{scho10} argues that, if one is careful, the process can be improved by first applying a linear (e.g. Wiener) deconvolution, which reduces the strength of the PSF wings and makes the core brighter and narrower.

\subsubsection{Speckle Suppression}
\label{sec:speckle}

High contrast imaging takes the requirements on PSF knowledge to extremes, and has led to the instrumental and observational requirement that the complex speckle pattern in the wings of the PSF cancels out when data are combined.
This requires very detailed characterisation of the PSF, which can be done under specific conditions.
Here we outline several ways that have been developed to achieve it:

\paragraph{Chromatic} methods include simultaneous differential imaging (SDI) and spectral deconvolution (SD).
SDI makes use of the different spectral energy distributions of the star and its companion.
One can then define two narrow bands close to each other in wavelength so that in a difference image, the speckles will subtract out but the companion does not (e.g. because methane absorption longwards of 1.62\,$\mu$m in the cool atmosphere of a late type brown dwarf makes it almost disappear in an image taken at this wavelength).
SD \citep{spa02} makes use of the fact that the radial location of the speckles from the centre of the PSF is proportional to wavelength, while the location of a companion with respect to the star is fixed. Re-scaling all individual images of an IFS cube proportional to its wavelength aligns the speckles but makes the planet move inwards with increasing wavelength. Speckles are now well fitted by a smooth (e.g. a low-order polynomial) function to each pixel along the cube's dispersion axis while the planet produces a narrow bump while traveling through the pixel at a certain wavelength range. Since this bump is badly fitted by the smooth function, the subtraction of the fit removes most of the speckles and leaves the planet. This latter method has been used to reach 9\,mag contrast at 0.2\arcsec\ separation without a coronagraph \citep{tha07}.
%also P1640 test on Alcor at Palomar by \cite{crep11}

\paragraph{Polarimetric} differential imaging (PDI) is based on the fact that at small angular radii, speckles remain unpolarised.
Thus, if two images at orthogonal polarisations are taken simultaneously, the speckles will subtract out.
Only the polarised light is left, which might trace, for example, scattered light from a circumstellar disk or even the atmosphere of an exoplanet. The potential of this technique was first demonstrated on the T~Tauri star TW~Hya \citep{apa04}, and a highly sensitive differential polarimeter will be used for the VLT exoplanet imaging instrument SPHERE \citep{beu06}.
%e.g. also  \cite{mur08}

\paragraph{Temporal} methods exploit rapid variations in the speckle pattern.
The `dark speckle' method proposed by \cite{lab95} and demonstrated by \cite{boc01} makes use of the fact that the rapidly changing speckle pattern will never reach its minimum level (which is dependent on the static long-exposure pattern) in locations where there is emission from a faint companion.
However, while this technique enables one to detect faint companions, it excludes performing photometry.
A parallel method also based on using statistical techniques to distinguish speckles from companions, but enabling one to estimate the intensity of the companion, has been proposed by \cite{gla10}.

\paragraph{Spatial} methods such as angular differential imaging (ADI) 
%make use of the different coherence properties of speckles and companions.
rely on the fact that the major sources of quasi-static speckles arise within the telescope and instrument. ADI can be used if the field of view, and hence the image of the exoplanet, is allowed to
rotate with respect to the telescope/instrument configuration and hence the quasi-static speckle pattern. 
With suitable processing of such a sequence of images \citep[e.g.][]{laf07b}, one can cancel the speckles while enhancing the signal from a companion.
ADI is rather simple to implement on alt-azimuth telescopes, because field and pupil rotate with respect to each other. 
It was first demonstrated by \cite{mar06c} and is now widely used for high-contrast imaging.

\paragraph{Deep Suppression} of speckles over a limited part of the PSF is possible through phase correction by the DM. The Fraunhofer diffraction formula states that the complex amplitude $A(x)e^{i\phi(x)}$ of light is essentially propagated between aperture and image plane by a Fourier transform. Absorbing the amplitude into a complex phase $\phi_c(x)$ and considering small errors $e^x \sim 1+x$, we see that sinusoidal phase or amplitude errors across the aperture produce two speckles of light that are symmetric to the image center $FT(\sin(ax)) \propto \delta(f-a) - \delta(f+a)$. The angular separation of these speckles from the center grows with spatial frequency of the ripple. Since the DM has a spatial frequency cut-off due to its finite number of actuators, it can only produce or correct speckles up to a maximum angular separation $\theta_{AO}$ called the correction radius. 
In order to suppress a coherent patch of light created by phase and/or amplitude aberrations inside the correction radius, the DM just needs to create an anti-speckle of the same amplitude but with a $\pi$ phase-shift at the same location. Obviously this scheme would work perfectly for pure phase aberrations in the aperture since the DM would simply have to flatten the phase. Unfortunately, the phases of the symmetric speckles created by a phase sinusoid $\cos(kx+\theta)$, e.g. created by the DM, are $\pi/2-\theta$ and $\pi/2+\theta$, while the phases of speckles created by an amplitude sinusoid are $\theta$ and $-\theta$ \citep{guy05}. Hence, only one of the amplitude speckles can be suppressed by an aperture plane DM at the expense of an amplification of the other one. Exploiting the Talbot effect, i.e. the transformation of a phase sinusoid into an amplitude sinusoid by Fresnel propagation, it is however possible to correct all image plane speckles over a certain wavelength range using a second DM. Algorithms that describe in more detail how to suppress speckles and create a dark hole with a single DM were presented by, for example, \cite{giv07} and \cite{guy10}; and a suppression of the diffracted and scattered light near a star-like source to a level of $6 \times 10^{-10}$ at $4 \lambda/D$ using such methods has been demonstrated in the laboratory \citep{tra07}.

\subsection{Estimating the PSF}

We have seen that there are various ways in which a PSF might be used for a scientific analysis.
Similarly, there are a variety of options for estimating the PSF.
However, each of them has limited applicability, and so we consider also PSF reconstruction based on the wavefront sensor (and ancilliary) data, and discuss whether this is a viable alternative.

\subsubsection{Empirical Extraction}
\label{sec:psfextract}

The four methods described here illustrate that there are often circumstances which make it possible to derive at least some information about the PSF from observations.
However, they also emphasize that this is usually at best a crude approximation, and that independent guidance and confirmation about the PSF shape is much needed.

\paragraph{Targeted Reference Star}

Observers are often recommended to take images of an isolated reference star in order to estimate the PSF.
This can work well if a number of implicit assumptions are fulfilled:
(i) the atmospheric conditions are sufficiently stable that a short and temporally distinct exposure on the PSF star is a good representation of the PSF in a long science exposure (a seeing monitor can help to verify whether this condition is met);
(ii) the intensity and distribution of flux on the WFS is the same for both the PSF reference and the science target's guide star (which is not necessarily the case for non-stellar guide stars such as galaxy nuclei, which may be extended and/or superimposed on a bright background);
(iii) if an off-axis guide star is used for the science target, the anisoplanatism for the PSF reference should be the same (i.e. one may need a pair of stars at a specific separation, one matching the guide star magnitude and the other for the science camera).
Clearly, there are situations where taking targeted images of a reference star can be successful, particularly if several such images are used for myopic deconvolution as described in Sec.~\ref{sec:deconv}.
However, even if one accepts the reduced observing efficiency such observations imply, it is often hard -- or even impossible -- to meet all 3 conditions.

\paragraph{Unresolved Sources in the Science Field}

For observations of stellar fields, we have seen that the standard procedure is to extract the PSF using several stars in the field.
This approach can also be applied to other science targets if there are unresolved sources in the vicinity.
More often, such sources will be distributed over a wide field and so it
will be necessary to account for anisoplanatism. 
Methods have been developed to estimate an off-axis PSF by making use of knowledge of the $C_n^2(z)$ distribution through the atmosphere; 
and in principle these could be turned around to derive an on-axis PSF
from off-axis stars.
As an alternative, there has been some effort to show one can use calibration frames containing many stars to quantify empirically how the PSF varies across a field \citep{ste02,ste05,cre05}.
Applying this to a science field where there are at least a few stars then enables one to interpolate the PSF at any point in the field.
The approximate nature of the inferred PSF makes this method suitable for model convolution.
However, it is impractical for spectroscopy because the field of view is simply too small.

\paragraph{Within the Science Target}

Rather than extract the PSF from stars in the field or from observations of other stars, any feature in the science target that is spatially unresolved can be used to estimate the PSF.
Chromatic, polarimetric, and angular differential imaging (Sec.~\ref{sec:speckle}) are special cases where the instrumentation and observing technique are designed to provide a PSF reference image.
Similarly, there are a number of different components contributing to the total spatial and spectroscopic emission from AGN.
In Seyfert nuclei, the broad line region is only a few lightdays across
and is always unresolved (in QSOs, its size can be measured in
lightyears but, because of their greater distance, is still unresolved).
Alternatively, the near-IR non-stellar continuum, which usually arises from hot dust close around the AGN, is only 0.1--1\,pc across and hence unresolved on 8--10-m class telescopes at distances of $\sim20$\,Mpc or more.
The spatial distribution of both these quantities can be extracted
using the spectral information in a datacube.
% (perhaps using principle components analysis, as proposed by \citealt{ste09}).
This method of deriving the PSF can be successful at yielding its overall shape without fine detail, and so is suited to spectroscopic studies where one uses a model to derive the intrinsic source properties;
but it is applicable to only a limited number of the science targets.

\paragraph{Comparison to Higher Resolution Data}

In some circumstances it may be possible to make use of a higher resolution broad-band image to infer the PSF (for example when analysing kinematics from undersampled integral field spectroscopy).
The rationale is that the observed intensity distribution $I$ is the convolution of the intrinsic object $O$ with a
PSF $P$ so that $I = P \otimes O$.
One can find a broadening function $F$ which, when convolved with
the higher resolution observation $I_h$, reproduces the lower
resolution observation $I_l = F \otimes I_h$.
%Putting these together leads to 
%$P_l \otimes O = F \otimes P_h \otimes O$.
%And since convolution is associative, 
The lower resolution PSF is then $P_l = F \otimes P_h$.
While this can be successful, the point of AO is to yield the highest resolution data, making this a niche method.

\subsubsection{Reconstruction}
\label{sec:recon}

None of the empirical approaches described above is applicable under a wide range of observation conditions and astronomical targets. An alternative method was therefore introduced by \citet{ver97} and successfully applied to the low-order curvature AO system PUEO. This method models the optical transfer function (OTF, the Fourier transform of the PSF) as the product of three contributors, the OTFs of i) telescope and instrument, ii) high spatial frequency turbulence that cannot be corrected by the DM, and iii) the low spatial frequency AO corrected residual wavefront.  While the contribution of telescope and instrument is supposed to be quasi-static and can be measured, and the uncorrected turbulence is supposed to follow Kolmogorov statistics described solely by the Fried parameter $r_0$, the determination of the AO residual wavefront requires a deep understanding of the AO control dynamics and WFS characteristics as well as {\it a priori} modeling assumptions on atmospheric turbulence and noise characteristics.

The expected difficulties with making PSF reconstruction work in more complex cases (low signal-to-noise, off-axis guide star, or when using a LGS), made \citet{jol10} speculate about alternative ways to retrieve a reasonable PSF estimate that could still be useful for astronomical applications. One approach could be to derive the PSF using analytical models which are fed with some basic parameters (turbulence profile, AO system parameters, guide star parameters) and heuristically tuning it to the actual system PSF. Another approach to reconstruct PSFs would be to make use of point sources in the field and estimate the PSFs lying in between either by a generalization of the method presented by \citet{bri06} to derive the anisoplanatic degradation, or by interpolating the few parameters of an empirically determined analytical PSF profile (e.g. Moffat distribution). 
Certainly, formulation of a proper metric and required accuracy for PSF reconstruction, widely accepted in the astronomical community, is a necessary prerequisite for the development of an appropriate algorithm.
The new work on PSF reconstruction started by \cite{jol11} may provide a basis from which a suitable metric can be developed.

%%%%%%%%%%%%%%%%%%%%%%%%%%%%%%%%%%%%%%%%%%%%%%%%%%
\section{Novel Techniques}
\label{sec:novel}

The science described in Sec.~\ref{sec:science} was performed using SCAO systems. These have only a single deformable mirror to correct the wavefront perturbations, which are measured using a single reference source (or with a laser guide star, a single source for each of the high order and tip-tilt components). 
The scientific applications of such systems are constrained by the limited performance, particularly at shorter wavelengths, the small field of view, and poor sky coverage (see Sec.~\ref{sec:basicao}).
In this section we discuss the motivation for novel alternative schemes for wavefront sensing and correction that promise to widen the parameter space of adaptive optics.
%and we outline the techniques they employ. 
We focus on concepts that have already been implemented and tested on sky, or are currently being built.

\begin{figure}
\psfig{file=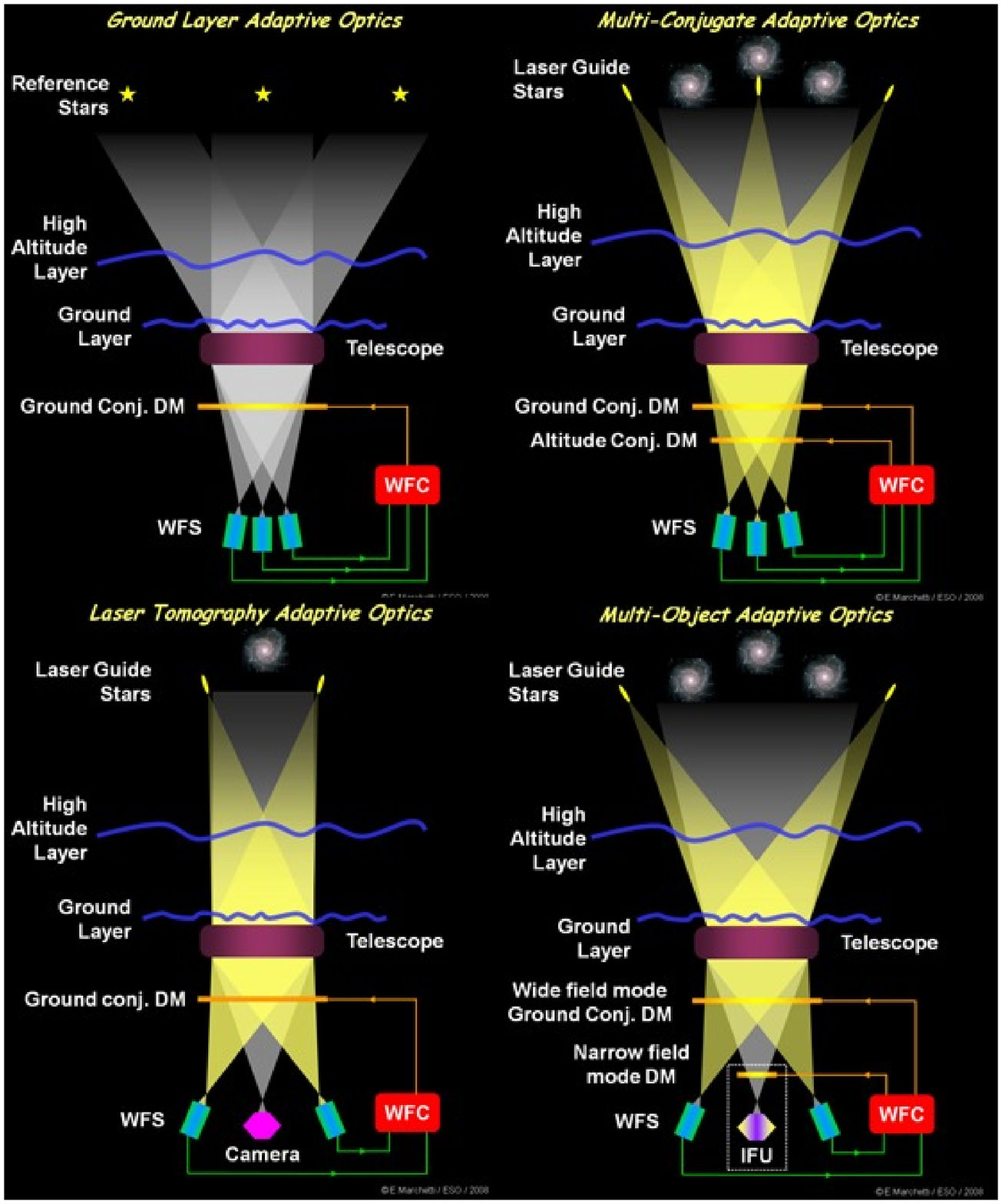,width=\textwidth}
\caption{Conceptual drawings of various novel AO concepts (courtesy of E.~Marchetti, ESO). Clockwise from top left: Ground Layer AO, Multi-Conjugate AO, Multi-Object AO, and Laser Tomography.}
\label{fig:ao_concepts}
\end{figure}

\subsection{Laser Tomography AO}
\label{sec:ltao}

We have seen in Sec.~\ref{sec:lgs} that while using a LGS improves the sky coverage with respect to NGS, there are still limitations, notably the cone effect.
For a sodium LGS on an 8--10-m class telescope, the cone effect reduces the Strehl ratio obtainable by a factor 0.85 in K-band and 0.6 in J-band.
This moderate impact would become devastating for the next generation ELTs, limiting the maximum achievable K-band Strehl to only 15\%, and much less at shorter wavelengths.
In order to overcome the cone effect, measurements from several LGSs can be combined to fully reconstruct the turbulence column in the direction of the astronomical target (lower left panel in Fig.~\ref{fig:ao_concepts}). 
The separation of the LGSs can be significantly larger than the isoplanatic angle $\theta_0$, but their beams should still overlap at the highest turbulent layer in order to prevent there being unsampled turbulence \citep{fus10}. The reconstructed volume is then collapsed along the third dimension and finally projected onto a single DM. This technique is called laser tomography AO (LTAO) and has been tested in the laboratory \citep{cos10}.

By involving a tomographic wavefront reconstruction, LTAO provides both a scientific and a technical stepping stone towards the (more complex) capabilities described later.
It also permits a degree of flexibility in how the wavefront correction is done.
If the science requires a high Strehl on a single target, the wavefront correction can fully overcome the cone effect.
Calculations for ATLAS on the European ELT suggest that, optimised on-axis, one can reach about 50\% Strehl in the K-band \citep{fus10}.
For extended sources, one can adjust the weighting of the wavefront errors as they are projected onto the DM to provide a more uniform, but more moderate, performance over a larger field.
In the most extreme case, one can achieve a modest performance of 10--15\% Strehl over a field as wide as 120\arcsec.
At the same time, the area over which one can search for a NGS to compensate the lack of tip-tilt signal from each LGS is widened, improving sky coverage.

\subsection{Seeing Enhancement}

In the preceeding sections we have seen that sometimes, due to the limitations of objects (e.g. surface brightness or size) and instrumentation (e.g. affordable detector area), one must perform a trade-off between signal-to-noise, resolution, and field of view.
It means that there are many situations where one cannot make full use of the diffraction limit, but can still benefit from enhanced resolution.
On the other hand, if the tip-tilt star is very faint, one may find there is significant residual jitter, augmented by anisoplanatic effects from the unavoidable offset to the science target.
\cite{dav08b} suggested that, extrapolating this regime to the limit, it is conceptually only a small step to dispense with tip-tilt completely when using a laser guide star;
and that by doing so, one can achieve spatial resolutions down to 0.2\arcsec\ with 100\% sky coverage.
Such a mode of operation has been commissioned at the VLT, where the fast tracking at 30\,Hz frame rate performed by the telescope itself compensates for vibrations such as wind shake, and is effective for observations of single sources.

\subsubsection{Ground Layer Correction}
\label{sec:glao}

Enhanced resolution over a wide field -- in principle up to $\sim10$\arcmin\ -- would bring many advantages. An obvious application is to stellar fields: one can perform the analysis simultaneously over the entire field, rather than splitting it into isoplanatic pieces.
This has two immediate benefits: the completeness correction is no longer dependent on distance from the guide star, 
%(which is also an issue for imaging extragalactic fields and deriving number counts, and is equivalent to a significant multiplex gain)
and it increases the potential for deriving a better PSF reference. Even if the science target is a single source, uniform correction over a wide field vastly improves the ability to empirically measure the PSF.

Multi-object spectroscopy would also benefit enormously from such a capability, especially for extragalactic studies.
As we have seen in Sec.~\ref{sec:highz}, the morphologies, kinematics, star formation rates, and metallicities of high redshift galaxies provide crucial insights into galaxy formation and evolution.
Key issues still to be resolved include the respective roles of mergers and secular evolution in galaxy growth;
how and when the various elements of the Hubble sequence arose;
how the galaxies acquired their angular momentum; 
and the roles played by AGN and stellar feedback in regulating BH and galaxy growth.
These all require deep, spatially resolved spectroscopy of faint sources with sizes of 0.1--1\arcsec.
On 8--10-m class telescopes, seeing limited conditions do not provide the necessary spatial resolution, while at the diffraction limit signal-to-noise is a major constraint due to the faint surface brightness of the spatially extended regions.
Intermediate resolution offers a reasonable compromise, and having it over a wide field adds a very important multiplex advantage -- a combination that is likely to be competitive also for the ELTs.

Such gains can be achieved using ground layer adaptive optics (GLAO, upper left panel in Fig.~\ref{fig:ao_concepts}), where only the atmospheric turbulence close to the ground is corrected. The key issue is that light rays from all objects, regardless of their angular offset from the field centre, pass through the same low layer of turbulence because it lies directly in front of the telescope mirror. Correcting it, by combining the wavefronts from several well separated reference sources so that signal from the ground layer is strengthened while the differing contributions from high altitudes cancel out, therefore enhances the resolution over a wide field.
An intriguing alternative is use a single low altitude Rayleigh LGS, as demonstrated by the GLAS system \citep{ben08}.
Here the cone effect is beneficial because high altitude turbulence must not enter the GLAO measurement \citep{tok04}.

Since at most astronomical sites, much of the turbulence does exist close to the ground, the GLAO gain can be substantial (perhaps most dramatically in Antarctica; \citealt{tra09}).
\citet{rig02} and \citet{tok04} have shown there is a trade-off between field of view and performance, which is roughly independent of telescope size.
The height $H$ (m) to which turbulence can be fully compensated depends on the projected actuator spacing $d$ (m), which is matched to $r_0$, and the field of view $\phi_0$ (arcsec) as $H \sim 2\times10^5 d / \phi_0$.
Thus for wider fields of view, only turbulence closer to the ground can be corrected, limiting the resolution gain.
Detailed simulations for the ARGOS system, which is being installed on the LBT, suggest that GLAO should yield a factor 2--3 improvement in FWHM across a 240\arcsec\ field of view, with a median resolution of 0.35\arcsec\ \citep{rab10}.
This level of performance has already been validated on the MMT, where the first system to average the wavefront from multiple laser guide stars yielded a gain in K-band resolution from 0.70\arcsec\ to 0.33\arcsec\ uniformly across a 120\arcsec\ field \citep{bar09,har10}.
One important aspect of GLAO is that it can function even in poorer atmospheric conditions where high order AO struggles. It is this that will make it such an important scientific tool in the future -- especially for the ELTs -- making poor seeing useful and good seeing excellent.

\subsection{Wide Fields at the Diffraction Limit}

The scientific motivation for reaching the diffraction limit across a wide field is the same as for GLAO but, for 8--10-m class telescopes, with an additional factor 4--6 improvement in resolution and point source sensitivity.
Here we describe the rationale for two contrasting methods that achieve this. The first provides diffraction limited performance across a contiguous field, the second at a number of selected points within a large patrol field.
%For both of these, the higher AO performance is a critical trade-off with sky coverage, and we discuss how this might be overcome.

\subsubsection{Multi-Conjugate AO}
\label{sec:mcao}

In solar astronomy, the short wavelengths and daytime observations yield a usable field of view with SCAO of order only 10\arcsec, which is not enough to encompass most sunspots or active regions, nor to efficiently monitor flare trigger mechanisms.
This then is a strong motivation to develop wide field AO correction.
It can be achieved using multiple reference sources in different directions so that the columns of atmospheric turbulence they probe overlap partially at high altitude and fully at low altitude.
In principle it is then possible to infer the full 3D structure of the turbulence.
In practice one attributes the measured perturbations to 2 or 3 layers at specific heights and uses separate deformable mirrors, one conjugated to each layer, to correct them -- hence multi-conjugate adaptive optics (MCAO, upper right panel in Fig.~\ref{fig:ao_concepts}).
MCAO was first discussed by \cite{dic75} and \cite{bec88}; and the first systems, developed in the context of solar astronomy, were tested at the Vacuum Tower Telescope %at Teide Observatory 
\citep{ber05} and at the Dunn Solar Telescope 
%at the National Solar Observatory 
\citep{lan04}, increasing the diameter of the corrected field to as much as 60\arcsec.

\begin{figure}
\psfig{file=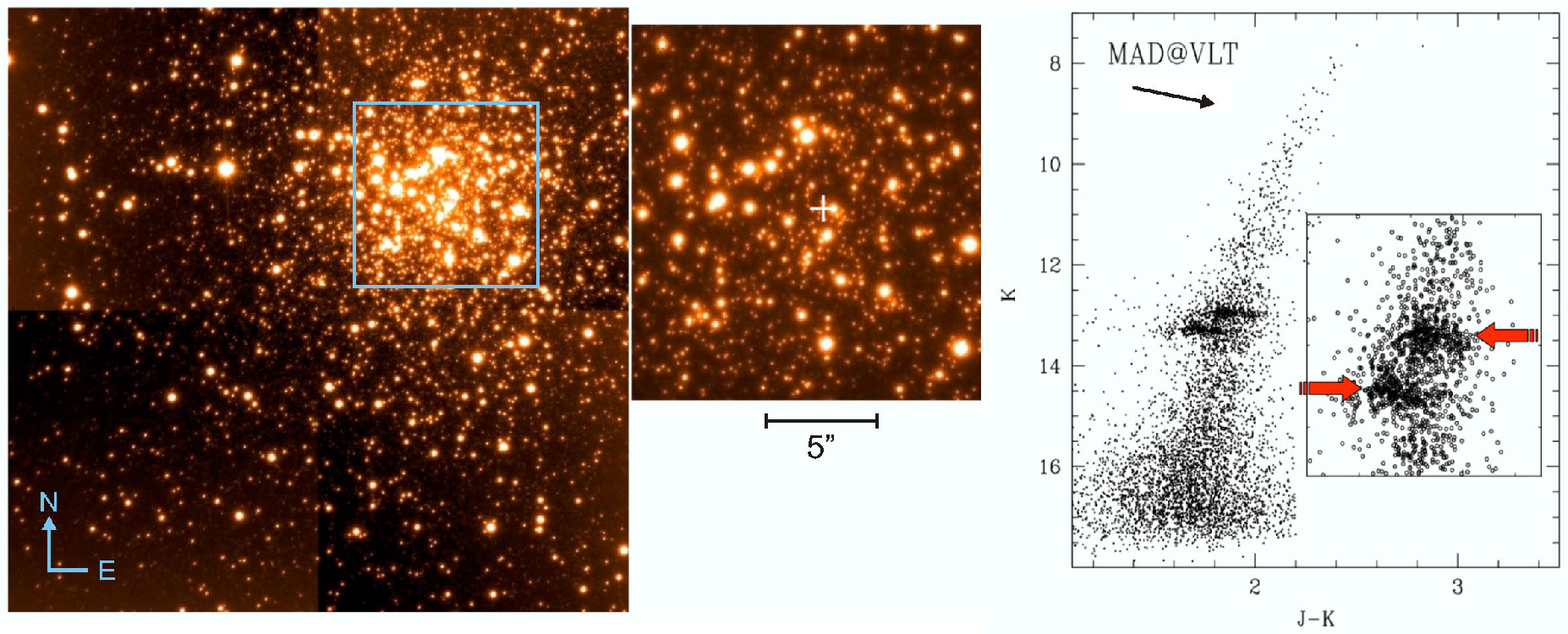,width=\textwidth}
\caption{Left: K-band image of Terzan\,5 taken with MAD, which provided 0.1\arcsec\ resolution over the entire 60\arcsec\ field. Right: Colour magnitude diagram of the cluster, with the two horizontal branch populations indicated. At top left, the arrow indicates the impact of reddening; uncertainties are indicated to the right.
From \cite{fer09} (courtesy of F.~Ferraro).}
\label{fig:terzan}
\end{figure}

Planetary, galactic, and extragalactic science had to wait until the MCAO demonstrator MAD was commissioned at the VLT \citep{marc08}.
The system used natural guide stars to deliver near-IR resolution down to 0.1\arcsec\ across a 120\arcsec\ field.
Despite the limited sky coverage, numerous results have been published.
One that typifies the potential of MCAO is discovery of two stellar populations in the star cluster Terzan\,5 \citep{fer09}.
These are clearly separated in the $K,J-K$ colour magnitude diagram (Fig~\ref{fig:terzan}) because the near-IR data are only mildly affected by the differential extinction.
Both populations contain several hundred stars but the brighter horizontal branch clump is redder, more compact, more metal rich ([Fe/H] $\sim 0.3$), and younger ($\sim 6$\,Gyr) than the fainter clump (iron abundance [Fe/H] $\sim -0.2$ and age $\sim 12$\,Gyr).
The authors argued that these characteristics are what one might expect if Terzan\,5 was the remnant of a dwarf galaxy that had been disrupted as it merged with the Galactic bulge;
with the implication that galactic spheroids may have been assembled hierarchically from pre-formed, evolved stellar systems.

MAD tested two different ways to achieve the wide field AO correction.
For classical `star oriented' MCAO, the wavefront signals from several guide stars (laser or natural) are measured on separate detectors, and combined afterwards during the computation of how much turbulence originates from the different layers.
An alternative `layer oriented' concept \citep{rag00} measures each layer separately by optically combining the guide star signals for that layer onto a single detector.
Each layer is then corrected independently, making the approach computationally simpler.
In both cases, the quality and uniformity of the correction depend on the brightness and locations of the guide stars -- which can be controlled far better using laser guide stars.
The first laser MCAO system will operate in a star oriented mode using 5 LGS and typically 3 tip-tilt stars:
GeMS is currently being commissioned on the Gemini South telescope \citep{nei11}.
A layer oriented NGS system is being built for the LBT \citep{her08}.

An additional advantage of MCAO is that the sky coverage is higher than for SCAO, but this can be improved still further by making use of `image sharpening' at near-IR wavelengths for the tip-tilt stars.
The design of the Keck Next Generation AO system includes DMs specifically to correct the wavefront from the tip-tilt stars \citep{wiz10}.
For future ELTs, tip-tilt stars can be selected within the wide region beyond the science field where there is still a substantial correction, leading to a sky coverage exceeding 50\% at the Galactic Pole \citep{dio10,her10}.

Technically the performance of an MCAO system is limited by a number of practical considerations.
These include the finite number of guide stars, and the fact that the insensitivity of LGS to tip-tilt propagates into the quadratic modes for MCAO so that several NGS are required.
From a scientific perspective, arguably the most important issue is the field of view that can be corrected.
\cite{rig00} have shown that this is limited by the number $N_{DM}$ of DMs in the AO system to $\phi_0 \sim 10^6 d N_{DM} / H$ where
$H$ (m) is the height to which the turbulence is corrected, $d$ (m) is the actuator spacing (matched to $r_0$) and $\phi_0$ (arcsec) is the field of view.
For example (and part of the rationale behind the design of GeMS), if the turbulence extends up to 12\,km, the condition above means that one needs 3~DMs to achieve a K-band Strehl of 80\% uniformly across a 60\arcsec\ field.
That MAD was able to provide 25\% Strehl over a 120\arcsec\ field with only 2~DMs is related to the distribution of turbulence through the atmosphere.
\cite{marc08} showed that on these occasions more than half of the turbulence was in the ground layer at altitudes $<500$\,m.
To achieve high Strehl ratios over fields wider than this, MCAO is impractical.

\subsubsection{Multi-Object AO}
\label{sec:moao}

Surveys of high redshift galaxies require multi-object (integral field) spectroscopy to probe the evolution of their star formation histories, metallicities, and internal dynamics.
The number density of galaxies for which this can be done is at most a few per square arcmin, so patrol fields of tens of square arcminutes are required to make such programmes observationally efficient.
In this respect, MCAO is of limited use: the corrected field is too small, and neither the instrumentation nor the science aims make use of the contiguous field.
Instead, multi-object adaptive optics (MOAO) has been developed with the purpose of targeting specific objects within a large patrol field.

This is achieved by having a number of wavefront reference sources in and around the patrol field which are used to reconstruct the 3D structure of the turbulence within it.
The interpolated wavefront towards each science target is then corrected in open loop using a deformable mirror that has been allocated to that target (lower right panel in Fig.~\ref{fig:ao_concepts}; see also Sec.~\ref{sec:lgs}).
To validate this unusual approach, several MOAO developments are currently under way.
These have led to on-sky tests of open loop adaptive optics correction in the I-band \citep{and08}, as the first step to an MOAO system that will be commissioned on Subaru \citep{con10};
as well as of a true MOAO correction in the H-band using several guide stars in the field \citep{gend11}.
%The results are encouraging, but also demonstrate the scope of the work required to build a scientifically productive MOAO system.

\subsection{Extreme AO}
\label{sec:extremeao}

At the opposite end of the AO parameter space is the aim to provide exceptionally high performance on bright ($<10$\,mag) stars.
Reaching a Strehl ratio in excess of 90\% in the H-band is just the first step in a chain of complementary optical and data processing techniques optimised to probe deep within the PSF \citep{opp09}, but one that is absolutely necessary for the rest to work well.
The primary scientific goal of this effort is to study young and evolved giant exoplanets orbiting nearby stars in wide 1--100\,AU orbits that are inaccessible to current Doppler techniques \citep{mac07}.
This can be done by direct imaging of the circumstellar environment, at a contrast ratio of $10^5$ within 0.1\arcsec\ (and to $10^9$ at 0.04\arcsec\ with ELTs), and several orders of magnitude more at 1\arcsec.
Targeting the nearest stars will probe the smallest orbits, and hence provide the opportunity to detect planets via reflected stellar light.
In contrast, planets around young (0.1--1\,Gyr) stars -- particularly stars within nearby young associations -- will be self-luminous and can be detected to low masses through their own emission.
Focussing on known planets with systematic residuals in their radial velocity curves will provide the best chance of detecting additional planets in wider orbits.
Combining these approaches will lead to a clearer understanding of the regime in which planets can form via core accretion, the role of migration in the evolution of planetary systems, and whether our solar system, with giant planets at 5--10\,AU, is unique.

While the AO concept is no different from conventional SCAO (see Fig.~\ref{fig:ao_prince}), the technical implementation is highly challenging.
The wavefront sensing must be performed with a small $\sim20$\,cm aperture spacing to spatially sample the turbulence.
This requires a large format detector, which must be able to read out the measurements with a low $<1$\,e$^-$ readnoise and a fast $>1$\,kHz frame rate in order to sample the turbulence temporally. 
The deformable mirror must match the WFS, and have a high stroke so that both atmospheric and instrumental perturbations can be fully corrected (as noted in Sec.~\ref{sec:dm}, one way to achieve this is to use 2 DMs corresponding to a woofer and tweeter). 
Image and pupil instability due to thermo-mechanical effects, and non-common path errors must be actively minimised using additional internal control loops.
On top of these provisions, highly efficient reconstruction and filtering algorithms and needed to reduce to $<1\,$ms the delay between measuring and applying a wavefront correction.

The challenges and rewards of high contrast imaging have led to a massive effort in the development of extreme AO systems.
The 672 actuator system at the LBT \citep{esp10} could be considered a forerunner to extreme AO.
The first truly high order system, with more than 3000 actuators on the DM, is PALM-3000 for the 5-m telescope at Palomar Observatory, which had its first light in June 2011 \citep{dek11}.
Two dedicated extreme AO instruments will be SPHERE \citep{beu06} and GPI \citep{mac08}, which are close to commissioning at the VLT and Gemini observatories.
And the AO system on Subaru will soon receive the SCExAO upgrade \citep{mar09}, which includes an additional higher order ($>1000$ actuator) DM as well as coronographs to better serve its high contrast instrumentation.

\subsection{Towards the Visible}
\label{sec:visao}

The high performance of extreme AO systems will lead to improvements at visible wavelengths, but only in the few arcsec around relatively bright stars.
Currently, the only way to achieve Strehl ratios as high as 20\% at 700\,nm over a wider field using fainter stars is through a combination of low order AO and image selection, as demonstrated by the LuckyCam experiment \citep{law06,law09a}.
The key to this technique is to keep only data obtained during the (perhaps brief and intermittent) periods when the AO performance was at its best, tradeing depth for resolution.
The enabling technology is a fast (10\,Mpix\,s$^{-1}$ giving frame rates of 20--50\,Hz) essentially noiseless read-out of a large format detector \citep{mac03}.
The role of AO is to increase the likelihood that the PSF is dominated by a single bright speckle at any given time, and hence the fraction of frames above a quality threshold (and with a wide isoplanatic angle).
The selected frames are then combined using a shift-and-add alignment.
\cite{law09a} found that when using 20\% of the frames, the co-added image reached the diffraction limited of the Palomar 5-m telescope, and the FWHM and Strehl ratio were both a factor 2 better than with AO alone.
Both shorter ($<10$\,ms) exposure times \citep{smi09} and the use of Fourier selection \citep{gar10} may lead to greater gains.
A different approach is required to apply lucky imaging to pre-existing instruments, as planned for the Magellan Telescope DSM, which recent tests have shown can yield significant gains in the visible with bright guide stars (L.~Close, priv. comm.).
By including a fast shutter which enables real-time `frame' selection at a rate of $\sim100$\,Hz, and by using an additional 2\,kHz `clean-up' tip-tilt system (as a substitute for shift-and-add), the aim is to feed an integral field spectrometer with $>20$\% Strehl ratio at 0.6--1.0\,$\mu$m \citep{mal10}.
The attraction of achieving the diffraction limit at visible wavelengths has also led to rapid growth in the number of AO lucky imaging cameras in operation at 2--4\,m class telescopes, opening a regime in which they can compete effectively with larger telescopes.

%%%%%%%%%%%%%%%%%%%%%%%%%%%%%%%%%%%%%%%%%%%%%%%%%%
\section{Lessons Learned and Future Outlook}
\label{sec:future}

Since the first astronomical AO systems were opened to the community in the early 1990s, numerous technical achievements have been accomplished, a multitude of novel techniques have been established, and it is now inconceivable to consider building a large telescope without AO.
Yet many of the advanced concepts still exist only on paper or are just being demonstrated in the laboratory. While many of these innovations emerged during the first half of the last decade, recent years were mostly dedicated to raising the technological readiness level and approaching on-sky demonstration. It is becoming more evident that AO is conceptually mature, and it appears that few paths remain which have not yet been explored. The largest progress in AO is now to be expected from technological developments mainly in the areas of high-density deformable mirrors and powerful real-time computers. Also fast and low-noise near-IR detectors have a great potential, while optical detectors with sub-electron read-noise and very high quantum efficiency are already close to perfection.

The demand for AO, and particularly for LGS-AO, is high. 
For example, the time for which SINFONI is scheduled on the VLT is divided roughly equally between LGS-AO, NGS-AO, and seeing limited observations (ESO User Support Department).
Similarly, for the last 5~years nearly half of Keck~II science nights have been allocated to AO programmes; and of the AO science published in 2010, most extragalactic, and about half of galactic, papers had made use of the LGS \citep{wiz11}.
In contrast, although scientific exploitation of AO is beginning to flourish, it struggled to get started, as reflected by the relative citation rates of AO papers in the main peer-reviewed journals.
%This delayed growth can be quantified if one takes the mean number of citations per paper in the main peer-reviewed journals as a measure of scientific impact.
%Observational AO papers from the period 1998-2000, after 10 years of development, are less cited than other papers from the same period.
%A decade on, the situation has changed: AO science papers from 2008-2010 are on average more highly cited than other observational papers.
To some extent, this growth of scientific productivity reflects the level of investment into AO \citep{fro06}, which is a concern if the scientific impact of AO is to continue growing.
But perhaps a more important lesson is that there is a `threshold' in the science domain both for case studies (does the spatial resolution give access to new understanding of a physical process?) and for the statistical approach (are there sufficient well-selected targets?).
In either case, for the increasingly complex systems that are now being developed, it is important to bear in mind that

\smallskip
\noindent
{\it there is a vast gulf between technological demonstration and scientific productivity}.
\smallskip

This can be seen in the expectations of early AO systems, which were raised very high and led to disappointment within the community and a loss of credibility.
Estimates of the number of targets that could benefit from AO, based on statistical distributions of classes of objects, had been misleading in their optimism.
Instead, analysis of telescope pointings (e.g. to assess the impact of `LGS collisions' where scattered light from a laser at one telescope falls within the field of view of another) show that astronomers frequently return to the same targets and fields.
The reason is that observational progress relies heavily on objects for which there is ancilliary data and prior knowledge.
So far, the science impact of AO has benefitted greatly from case studies of relatively few intensively studied objects.
In the future, the use of AO on large representative samples will become increasingly important.
Astronomers need AO to function for those targets, even though they may be demanding (e.g. low airmass or faint and far off-axis guide stars).
Thus:

\smallskip
\noindent
{\it AO ought to be accessible to targets for which the primary selection criteria are astrophysical rather than technical}.
\smallskip

One way to achieve this is to better understand the requirements imposed on the AO system by the targets (rather than vice-versa) using appropriate performance metrics.
Strehl ratio directly conflicts with sky coverage, particularly in fields that have faint or sparse guide/tip-tilt stars;
and encircled energy leaves the issue of spatial resolution completely open.
\cite{dav10} suggested encircled energy should be used as a criterion in terms of the fractional energy within the core of the PSF, in conjunction with an independent specification on the maximum acceptable width of the core.
Similar criteria (e.g. encircled energy within a pixel) are already used for a few instruments, and they will remain important for the ELTs since a number of the prime science drivers require high ($<100$\,mas) but not diffraction limited resolution.
%Another aspect is the wavelength at which the metrics apply, and the scientific benefit of improved performance at shorter wavelengths should not be overlooked.
To fully embrace the top level scientific requirements,

\smallskip
\noindent
{\it A simple set of AO performance metrics ought to be widely applicable to all types of astrophysical targets and science cases}.
\smallskip

AO performance is strongly dependent on atmospheric conditions, which has meant that for much of the time it is scientifically ineffective.
To address this, the operating parameters of the current generation of AO systems can be (automatically) adjusted to match the ambient conditions.
However, their design specifications still generally only cover the better part of the seeing distribution.
Queue scheduling, while beneficial, can only provide part of the solution to dealing with different atmospheric conditions.
Instead:

\smallskip
\noindent
{\it AO ought to provide useful performance in moderate to poor atmospheric conditions}.
\smallskip

Coupled to this, especially for deep observations of distant galaxies, is the issue of long integrations. These can exceed several hours, and 80\,hr integrations are planned for MUSE \citep{bac09}.
To accumulate this in a reasonable time means the AO system must maintain its performance during long sequences of exposures despite rapid, large amplitude variations in the seeing on short timescales.
Analysing turbulence profiles from 7 sites over 83 nights, \cite{rac09} found that the variability is much stronger on timescales of minutes than between nights.
And at 10\,Hz timescales, seeing can change by as much as a factor of 2 within a second (C.~Mackay, priv. comm.).
Following strong variability is demanding for AO systems, but the requirement for long integrations means that:

\smallskip
\noindent
{\it AO performance ought to mitigate the effects of a highly and rapidly variable atmosphere}.
\smallskip

Even for well designed AO systems, functioning optimally is hard.
This is partly due to the large number of expected and unexpected opto-mechanical and operational interfaces to the telescope and instrument.
In the past, some of these (e.g. vibrations) have had a detrimental impact on AO performance.
To avoid such situations, the telescope, AO system, and instrument need to be considered as a joint facility.
System engineering of the entire instrument-AO-telescope system is therefore a key ingredient, and

\smallskip
\noindent
{\it an AO system ought to be designed together with its telescope and instrument}.
\smallskip

Optimal functionality also implies a significant workload to manage and maintain a complex system at operational readiness.
This is particularly so for LGS systems because of the additional control loops required, the beam relay, and the complexity of the lasers.
Finally, performance depends very much on calibration and characterisation.
This has a major impact on how well one can perform wavefront reconstruction, remove non-common path errors, and compensate optical distortions.
These can, for example, directly affect astrometric precision; but the impact of AO on astrometry, and how best to calibrate it, is still poorly understood \citep{tri10}.
It is clear that, as the requirements become more demanding and AO systems more complex, 

\smallskip
\noindent
{\it the operational effort to continually achieve optimal performance may be considerable}.
\smallskip
       
Extracting the best science requires PSF calibration.
This is a crucial ingredient in the post-processing of AO data, and one that remains a major issue because it has not received the attention it needs.
Because of the difficulties in empirically estimating the PSF, it is important to make use of the knowledge we have of the AO system and atmospheric conditions -- ideally to provide a full reconstruction of the PSF, but minimally to provide some information about its shape and spatial variability.
In this way, 

\smallskip
\noindent
{\it AO systems ought to provide post-processing support by enabling the PSF to be derived}.
\smallskip

The rate of technical progress in AO over the last 2 decades has been astounding;
and that it can lead to important insights in astrophysical processes cannot be disputed.
But if we wish to rely increasingly on the ability of AO to boost the performance of current and future telescopes, it is crucial that the astronomical and AO communities come together so that they can understand their respective requirements and constraints.
Only this way can we ensure that the scientific productivity and impact of AO continue to grow.

%%%%%%%%%%%%%%%%%%%%%%%%%%%%%%%%%%%%%%%%%%%%%%%%%%

\subsection*{Acknowledgements}

The authors are grateful to all those who have provided figures, comments, suggestions, or ideas, and extend special thanks to: Carmelo Arcidiacono, Yann Cl\'enet, Laird Close, Steven Cornelissen, Alison Davies, Celine D'Orgeville, Gaspard Duch\^ene, Frank Eisenhauer, Simone Esposito, Francesco Ferraro, Pierre Ferruit, Natascha F\"orster Schreiber, Reinhard Genzel, Stefan Gillessen, Greg Herczeg, Stefan Hippler, Miska Le Louarn, Craig Mackay, Vincenzo Mainieri, Enrico Marchetti, Frank Marchis, Francisco M\"uller S\'anchez, Richard Myers, Imke de Pater, J\'er\^ome Paufique, Sebastian Rabien, David Rosario, Jean-Pierre V\'eran, Elise Vernet, and Peter Wizinowich.

%%%%%%%%%%%%%%%%%%%%%%%%%%%%%%%%%%%%%%%%%%%%%%%%%%

% too many acronyms to list

\subsection*{Key Terms}

{\bf Strehl Ratio}
the ratio of the observed peak intensity in a real point source image, compared to the theoretical maximum peak intensity of a perfect imaging system working at the diffraction limit.

%%%%%%%%%%%%%%%%%%%%%%%%%%%%%%%%%%%%%%%%%%%%%%%%%%


\begin{thebibliography}{}

\bibitem[\'Ad\'amkovics et al.(2007)]{ada07}
\'Ad\'amkovics M., Wong M., Laver C., de Pater I. 2007.
Science, 318, 962

\bibitem[Ammons et al.(2009)]{amm09}
Ammons S.M. et al. 2009.
%Melbourne J., Max C., Koo D., Rosario D., 2009,
AJ, 137, 470

\bibitem[Andersen et al.(2008)]{and08}
Andersen D. et al. 2008.
in {\it Adaptive Optics Systems},
eds Hubin N., Max C., Wizinowich P., 
Proc. SPIE, 7015, 70150H

%\bibitem[Andrae, Melchior \& Jahnke(2011)]{and11}
%Andrae R., Melchior P., Jahnke K. 2011.
%MNRAS, accepted

\bibitem[Apai et al.(2004)]{apa04}
Apai D. et al. 2004.
A\&A, 415, 671

%\bibitem[Arcidiacono et al.(2004)]{arc04}
%Arcidiacono C. et al. 2004.
%in {\it Advancements in Adaptive Optics},
%eds Bonaccini Calia D., Ellerbroek B., Ragazzoni R., 
%Proc. SPIE, 5490, 563

\bibitem[Arsenault et al.(2010)]{ars10}
Arsenault R. et al. 2010.
The Messenger 142, 12

\bibitem[Babcock(1953)]{bab53}
Babcock, H.~W. 1953
PASP, 65, 229

\bibitem[Bacon et al.(2009)]{bac09}
Bacon R. et al. 2009.
in {\it Science with the VLT in the ELT Era},
ed Moorwood A.,
(Springer), pp.331

\bibitem[Baranec et al.(2009)]{bar09}
Baranec C. et al. 2009.
ApJ, 693, 1814

\bibitem[Bartko et al.(2010)]{bar10}
Bartko H. et al. 2010.
ApJ, 708, 834

\bibitem[Bastian, Covey \& Meyer(2010)]{bas10}
Bastian N., Covey K., Meyer M. 2010.
ARA\&A, 48, 339

\bibitem[Bate(2009)]{bat09}
Bate M. 2009.
MNRAS, 392, 590

\bibitem[Beckers(1988)]{bec88}
Beckers J. 1988.
in {\it Very Large Telescopes and their Instrumentation},
ed Ulrich M.-H.,(ESO) p.693

\bibitem[Beckers(1993)]{bec93}
Beckers J. 1993.
ARA\&A, 31, 13

\bibitem[Benn(2008)]{ben08}
Benn C. et al. 2008.
in {\it Adaptive Optics Systems},
eds Hubin N., Max C., Wizinowich P., 
Proc. SPIE, 7015, 701523

\bibitem[Berkefeld, Soltau \& von der L\"uhe(2005)]{ber05}
Berkefeld T., Soltau D., von der L\"uhe O. 2005.
in {\it Astronomical Adaptive Optics Systems and Applications II},
eds Tyson R., Lloyd-Hart M., 
Proc. SPIE, 5903, 219

\bibitem[Beuzit et al.(2006)]{beu06}
Beuzit J.-L., 2006 et al. 2006.
The Messenger, 125, 29

\bibitem[Biller et al.(2007)]{bil07}
Biller B. et al. 2007.
ApJSS, 173, 143

\bibitem[Biller et al.(2011)]{bil11}
Biller B. et al. 2011.
%Allers K., Liu M., Close L., Dupuy T., 2011
ApJ, 730, 39

\bibitem[Boccaletti et al.(2001)]{boc01}
Boccaletti A. et al. 2001.
%Moutou C., Mouillet D., Lagrange A.-M., Augereau J.-C., 2001,
A\&A, 367, 371

\bibitem[Boccas et al.(2006)]{boc06}
Boccas M. et al. 2006.
in {\it Advances in Adaptive Optics II},
eds Ellerbroek B.L., Bonaccini Calia D.
Proc. SPIE, 6272, 62723L

\bibitem[Bonaccini Calia et al.(2006)]{bon06}
Bonaccini Calia D. et aol. 2006.
in {\it Advances in Adaptive Optics II},
eds Ellerbroek B., Bonaccini Calia D., 
Proc. SPIE, 6272, 627207

\bibitem[Bonaccini Calia et al.(2010)]{bona10}
Bonaccini Calia D. et al. 2010.
The Messenger 139, 12

\bibitem[Bongard et al.(2011)]{bon11}
Bongard S., Soulez F., Thiebaut E., P\'econtal E. 2011.
MNRAS in press

\bibitem[Bono et al.(2010)]{bon10}
Bono G. et al. 2010.
ApJL, 708, L74

%\bibitem[Bouchez et al.(2003)]{bou03}
%Bouchez A. et al. 2003.
%%Brown M., Troy M., Burruss R., Dekany R., West R., 2003
%in {\it Adaptive Optical System Technologies II.}
%eds Wizinowich P., Bonaccini D.,
%Proc. SPIE, 4839, 1045

%\bibitem[Bouchez et al.(2010)]{bou10}
%Bouchez A. et al. 2010.
%in {\it Adaptive Optics Systems II},
%eds Ellerbroek B., Hart M., Hubin N., Wizinowich P., 
%Proc. SPIE, 7736, 77361Q

\bibitem[Britton(2006)]{bri06}
Britton M.C. 2006.
PASP, 118, 490, 885

\bibitem[Brusa et al.(2003)]{bru03}
Brusa G. et al. 2003.
in {\it Astronomical Adaptive Optics Systems and Applications},
eds Tyson R.K., Lloyd-Hart M., 
Proc. SPIE, 5169, 26

\bibitem[Burtscher et al.(2009)]{bur09}
Burtscher L. et al. 2010.
%Meisenheimer K., Jaffe W., Tristram K., R\"ottgering H., 2010,
PASA, 27, 490

\bibitem[Burtscher et al.(2010)]{bur10}
Burtscher L. et al. 2009.
%Jaffe W., Raban D., meisenheimer K., Tristram K., R\"ottgering H., 2009,
ApJ, 705, L53

\bibitem[Calzetti et al.(2000)]{cal00}
Calzetti D. et al. 2000.
ApJ, 533, 682

%\bibitem[Campbell et al.(2010)]{cam10}
%Campbell M. et al. 2010.
%MNRAS, 405, 421

\bibitem[Chabrier et al.(2000)]{cha00}
Chabrier G., Baraffe I., Allard F., Hauschildt P., 2000.
ApJ, 542, 464

\bibitem[Chauvin et al.(2005)]{cha05}
Chauvin G. et al. 2005
A\&A, 438, L25

\bibitem[Cid Fernandes et al.(2004)]{fer04}
Cid Fernandes R. et al. 2004.
%Gu Q., Melnick J., Terlevich E., Terlevich R., Kunth D., Rodrigues Lacerda R., Joguet B., 2004,
MNRAS, 355, 273

\bibitem[Cisternas et al.(2011)]{cis11}
Cisternas M. et al. 2011.
ApJ, 726, 57

\bibitem[Close et al.(2003)]{clo03}
Close L., Siegler N. Freed M., Biller B., 2003.
ApJ, 587, 507

\bibitem[Close et al.(2007a)]{clo07a}
Close L. et al. 2007a.
ApJ, 660, 1492

\bibitem[Close et al.(2007)]{clo07b}
Close L. et al. 2007b.
ApJ, 665, 736

\bibitem[Close(2010)]{clo10}
Close L. 2010.
Nature, 468, 1048

\bibitem[Close et al.(2010)]{clo10b}
Close L et al. 2010.
in {\it Adaptive Optics Systems II},
eds Ellerbroek B., Hart M., Hubin N., Wizinowich P., 
Proc. SPIE, 7736, 77360T

\bibitem[Conan et al.(2010)]{con10}
Conan R et al. 2010.
in {\it Adaptive Optics Systems II},
eds Ellerbroek B., Hart M., Hubin N., Wizinowich P., 
Proc. SPIE, 7736, 77360T

\bibitem[Connelley, Reipurth \& Tokunaga(2009)]{con09}
Connelley M., Reipurth B., Tokunaga A. 2009.
AJ, 135, 2496

\bibitem[Costille et al.(2010)]{cos10}
Costille A. et al. 2010.
JOSA A, 27, 469

\bibitem[Cox (2001)]{cox01}
Cox A.N. 2001.
in {\it Allen's Astrophysical Quantities}, Springer New York

\bibitem[Crenshaw et al.(2011)]{cren11}
Crenshaw M., Fischer T., Kraemer S., Schmitt H. 2011.
in {\it Narrow Line Seyfert 1s and their Place in the Universe},
eds Foschini L.

%\bibitem[Crepp et al.(2011)]{crep11}
%Crepp J. et al. 2011.
%ApJ, 729, 132

\bibitem[Cresci et al.(2005)]{cre05}
Cresci G., Davies R., Baker A., Lehnert M. 2005.
A\&A, 438, 757

\bibitem[Cresci et al.(2009)]{cre09}
Cresci G. et al. 2009.
ApJ, 697, 115

\bibitem[Croom et al.(2004)]{cro04}
Croom S. et al. 2004.
%Schade D., Boyle B., Shanks T., Miller L, Smith R., 2004,
ApJ, 606, 126

\bibitem[Currie et al.(2011)]{cur11}
Currie T. et al. 2011.
ApJ, 729, 128

%\bibitem[Dasyra et al.(2007)]{das07}
%Dasyra K. et al. 2007.
%ApJ, 657, 102

\bibitem[Davidge et al.(2005)]{dav05}
Davidge T. et al. 2005.
%Olsen K., Blum R., Stephens A., Rigault F., 2005, 
AJ, 129, 201

\bibitem[Davies et al.(2006)]{dav06}
Davies R. et al. 2006.
ApJ, 646, 754

\bibitem[Davies et al.(2007)]{dav07}
Davies R. et al. 2007.
%M\"uller S\'anchez F., Genzel R., Tacconi L., Hicks E., Friedrich S., Sternberg A., 2007
ApJ, 671, 1388

\bibitem[Davies(2008a)]{dav08a}
Davies R. 2008a.
NewAR, 52, 307

\bibitem[Davies et al.(2008b)]{dav08b}
Davies R. et al. 2008b.
The Messenger, 131, 7

\bibitem[Davies et al.(2009)]{dav09}
Davies R. et al. 2009.
%Maciejewski W., Hicks E., Tacconi L., Genzel R., Engel H., 2009,
ApJ, 702, 114

\bibitem[Davies et al.(2010)]{dav10}
Davies R. et al. 2010.
in {\it Adaptive Optics Systems II},
eds Ellerbroek B., Hart M., Hubin N., Wizinowich P., 
Proc. SPIE, 7736, 77361G

\bibitem[Dicke(1975)]{dic75}
Dicke R. 1975.
ApJ, 198, 605

\bibitem[de Pater et al.(2004)]{pat04}
de Pater I. et al. 2004.
%Marchis F., Macintosh B., Roe H., Le Mignant D., Graham J., Davies A., 2004,
Icarus, 169, 250

\bibitem[de Pater et al.(2010)]{pat10}
de Pater I. et al. 2010.
%Wong M., Marcus P., Luszcz-Cook S., \'Ad\'amkovics M., Conrad A., Asay-Davies X., Go C., 2010,
Icarus, 210, 742

\bibitem[Deep et al.(2011)]{dee11}
Deep A. et al. 2011.
%Fiorentino G., Tolstoy E., Diolaiti E., Bellazzini M., Ciliegi P., Davies R., Conan J.-M., 2011,
A\&A, in press

\bibitem[Dekany et al.(2011)]{dek11}
Dekany R. et al. 2011.
in the {\it 2nd international conference on Adaptive Optics for Extremely Large Telescope}, Sept 2011.

\bibitem[Denney et al.(2010)]{den10}
Denney K. et al. 2010.
ApJ, 721, 715

\bibitem[Descamps \& Marchis(2008)]{des08}
Descamps P., Marchis F. 2008.
Icarus, 193, 74

\bibitem[Descamps et al.(2011)]{des11}
Descamps P. et al. 2011.
Icarus, 211, 1022

\bibitem[Diolaiti et al.(2000)]{dio00}
Diolaiti E. et al. 2000.
%Bendinelli O., Bonaccini D., Close L., Currie D., Parmeggiani G., 2000,
A\&AS, 147, 335

\bibitem[Diolaiti et al.(2010)]{dio10}
Diolaiti E. et al. 2010.
in {\it Adaptive Optics Systems II},
eds Ellerbroek B., Hart M., Hubin N., Wizinowich P., 
Proc. SPIE, 7736, 77360R

\bibitem[Do et al.(2009)]{do09}
Do T., et al. 2009.
ApJ, 691, 1021

\bibitem[Dodds-Eden et al.(2011)]{dod11}
Dodds-Eden K. et al. 2011. 
ApJ, 728, 37

\bibitem[D'Orgeville et al.(2011)]{org11}
D'Orgeville C. et al.  2011.
in the {\it 2nd international conference on Adaptive Optics for Extremely Large Telescope}, Sept 2011.

\bibitem[Duch\^ene et al.(2004)]{duc04}
Duch\^ene G., McCabe C., Ghez A., Macintosh B. 2004.
ApJ, 606, 960

\bibitem[Duch\^ene et al.(2007)]{duc07}
Duch\^ene G. et al. 2007.
%Bontemps S., Bouvier J., Andr\'e P., Djupvik A., Chez A., 2007,
A\&A, 476, 229

%\bibitem[Duch\^ene(2008)]{duc08}
%Duch\^ene G.
%NewAR, 52, 117

\bibitem[Duch\^ene et al.(2010)]{duc10}
Duch\^ene G. et al. 2010.
ApJ, 712, 112

%\bibitem[Duch\^ene(2012)]{duc12}
%Duch\^ene G.
%ARAA, in prep

\bibitem[Dupuy \& Liu(2011)]{dup11}
Dupuy T., Liu M. 2011.
ApJ, 733, 122

%\bibitem[Dutrey et al.(2007)]{dut07}
%Dutrey A., Guilloteau S., Ho P. 2007.
%in {\it Protostars and Planets V}
%eds Reipurth B., Jewitt D., Keil K.,
%University of Arizona Press, Tucson, p.495--506

\bibitem[Eckart et al.(2000)]{eck00}
Eckart A. et al. 2000.
Experimental Astronomy, 10, 1

\bibitem[Eisenhauer et al.(1998)]{eis98}
Eisenhauer F., Quirrenbach A., Zinnecker H., Genzel R. 1998.
ApJ, 498, 278

\bibitem[Ellerbroek \& Tyler(1998)]{ell98}
Ellerbroek B.L., Tyler D.W. 1998.
PASP, 110, 165

\bibitem[Engel et al.(2010)]{eng10}
Engel H. et al. 2010.
A\&A, 524, A56

%\bibitem[Engel et al.(2011)]{eng11}
%Engel H. et al. 2011.
%Davies R., Genzel R., Tacconi L., Sturm E., Downes D., 2011
%ApJ, 729, 58

\bibitem[Esposito et al.(2010)]{esp10}
Esposito S. et al. 2010.
in {\it Adaptive Optics Systems II},
eds Ellerbroek B., Hart M., Hubin N., Wizinowich P., 
Proc. SPIE, 7736, 773609

\bibitem[Falomo et al.(2005)]{fal05}
Falomo R., Kotilainen J., Scarpa R., Treves A. 2005.
A\&A, 434, 469

%\bibitem[Falomo et al.(2009)]{fal09}
%Falomo R et al. 2009.
%A\&A, 501, 907

\bibitem[Fathi et al.(2006)]{fat06}
Fathi K. et al. 2006.
%Storchi-Bergmann T., Riffel R.A., Winge C., Axon D., Robinson A., Capetti A., Marconi A., 2006,
ApJ, 641, L25

\bibitem[Ferrarese \& Ford(2005)]{fer05}
Ferrarese L., Ford H. 2005.
SSRv, 116, 523

\bibitem[Ferraro et al.(2009)]{fer09}
Ferraro F. et al. 2009.
Nature, 426, 483

\bibitem[Ferruit, P\'econtal \& Wilson(2000)]{fer00}
Ferruit P., P'econtal E., Wilson A., 2000,
in {\it Imaging the Universe in Three Dimensions},
eds van Breugel W., Bland-Hawthorn J.,
ASP Conf Ser. vol. 195, p.289

\bibitem[Fisher \& Drory(2011)]{fis11}
Fisher D., Drory N. 2011.
ApJL, 733, L47

\bibitem[F\"orster Schreiber et al.(2009)]{for09}
F\"orster Schreiber N. et al. 2009. 
ApJ, 706, 1364

\bibitem[Foy \& Labeyrie(1985)]{foy85}
Foy R., Labeyrie A. 1985.
A\&A 152, L29

\bibitem[Fritz et al.(2010)]{fri10}
Fritz T. et al. 2010.
MNRAS, 401, 1177

\bibitem[Frogel(2006)]{fro06}
Frogel J. 2006.
Gemini Focus, 33, 82

\bibitem[Fu et al.(2011)]{fu11}
Fu H., Myers A., Djorgovski S., Yan L. 2011.
ApJ, 733, 103

\bibitem[Fusco et al.(2010)]{fus10}
Fusco T. et al. 2010.
in {\it Adaptive Optics Systems II},
eds Ellerbroek B., Hart M., Hubin N., Wizinowich P.,
Proc. SPIE, 7736, 77360D

\bibitem[Fugate(1992)]{fug92}
Fugate, R.Q. 1992.
in {\it Progress in Telescope and Instrumentation Technologies, ESO Conference and Workshop Proceedings}
edited by Marie-Helene Ulrich

\bibitem[Garrel et al.(2010)]{gar10}
Garrel V. et al. 2010.
in {\it Adaptive Optics Systems II},
eds Ellerbroek B., Hart M., Hubin N., Wizinowich P.,
Proc. SPIE, 7736, 77365V

\bibitem[Gebhardt et al.(2000)]{geb00}
Gebhardt K. et al. 2000.
%Adams J., Richstone D., Lauer T., Faber S., G\"ultekin K., Murphy J., Tremaine S., 2011,
ApJ, 539, L13

\bibitem[Gebhardt et al.(2011)]{geb11}
Gebhardt K. et al. 2011.
%Adams J., Richstone D., Lauer T., Faber S., G\"ultekin K., Murphy J., Tremaine S., 2011,
ApJ, 729, 119

\bibitem[Gendron \& Lena(1994)]{gend94}
Gendron E., Lena P. 1994.
A\&A, 291, 337

\bibitem[Gendron et al.(2011)]{gend11}
Gendron E. et al. 2011.
A\&A, 529, L2

\bibitem[Genzel et al.(2006)]{gen06}
Genzel R. et al. 2006.
Nature, 442, 786

\bibitem[Genzel et al.(2008)]{gen08}
Genzel R. et al. 2008.
ApJ, 687, 59

\bibitem[Genzel et al.(2010)]{gen10}
Genzel R., Eisenhauer F., Gillessen S. 2010.
Rev. Mod. Phys., 82, 3121

\bibitem[Genzel et al.(2011)]{genz11}
Genzel R. et al. 2011.
ApJ, 733, 101

\bibitem[Ghez et al.(1993)]{ghe93}
Ghez A., Neugebauer G., Matthews K. 1993.
AJ, 106, 2005

\bibitem[Ghez et al.(2008)]{ghe08}
Ghez A. et al. 2008.
ApJ, 689, 1044

\bibitem[Gillessen et al.(2009)]{gil09}
Gillessen S. et al. 2009.
%Eisenhauer F., Trippe S., Alexander T., Genzel R., Martins F., Ott T., 2009,
ApJ, 692, 1075

\bibitem[Give'on et al.(2007)]{giv07}
Give'on A. et al. 2007.
in {\it Astronomical Adaptive Optics Systems and Applications III},
eds Tyson R.K., Lloyd-Hart M.
Proc. SPIE, 6691

\bibitem[Gladysz et al.(2010)]{gla10}
Gladysz S., Yaitskova N., Christou J. 2010.
JOSA A, 27, A65

\bibitem[Glenar et al.(1997)]{gle97}
Glenar D., Hillman J., Le Louarn M., Fugate R., Drummond J. 1997.
PASP, 109, 326

\bibitem[Golimowski et al.(1993)]{gol93}
Golimowski D., Durrance S., Clampin M. 1993.
ApJ, 411, L41

\bibitem[Goode et al.(2010)]{goo10}
Goode P. et al. 2010.
%Yurchyshyn V., Cao W., Abramenko V., Andic A., Ahn K., Chae J., 2010,
ApJL, 714, 31

%\bibitem[Grier et al.(2009)]{gri09}
%Grier C. et al. 2009.
%in {\it Co-evolution of Central Black Holes and Galaxies}, 
%eds Peterson B., Somerville R., Storchi-Bergmann T., 2009,
%Proc. IAU Symposium 267, pp.204

\bibitem[G\"ultekin et al.(2009)]{gul09}
G\"ultekin K. et al. 2009.
ApJ, 698, 198

\bibitem[Guyon(2005)]{guy05}
Guyon O. 2005.
ApJ, 629, 592

\bibitem[Guyon et al.(2006)]{guy06}
Guyon O., Sanders D., Stockton A. 2006.
ApJS, 166, 89

\bibitem[Guyon(2007)]{guy07}
Guyon O. 2007.
in {\it Astronomical Adaptive Optics Systems and Applications III},
eds Tyson R., Lloyd-Hart M.,
Proc. SPIE, 6691, 66910G

\bibitem[Guyon et al.(2010)]{guy10}
Guyon O. et al. 2010.
PASP, 122, 71

\bibitem[H\"aring \& Rix(2004)]{hae04}
H\"aring N., Rix H.-W. 2004.
ApJ, 604, L89

\bibitem[Hampton et al.(2006)]{ham06}
Hampton P.J. et al. 2006.
in {\it Advanced Software and Control for Astronomy},
eds Lewis H., Bridger A.,
Proc. SPIE, 6274, 62741Z

\bibitem[Harayama, Eisenhauer \& Martins(2008)]{har08}
Harayama Y., Eisenhauer F., Martins F., 2008.
ApJ, 675, 1319

\bibitem[Hart et al.(2010)]{har10}
Hart M. et al. 2010.
Nature, 466, 727

\bibitem[Hartung et al.(2004)]{har04}
Hartung M. et al. 2004.
A\&A, 421, 17

\bibitem[Hayano et al.(2010)]{hay10}
Hayano Y. et al. 2010.
in {\it Adaptive Optics Systems II},
eds Ellerbroek B., Hart M., Hubin N., Wizinowich P.,
Proc. SPIE, 7736, 77360N

\bibitem[Herbst et al.(2008)]{her08}
Herbst T. et al. 2008.
in {\it Ground-based and Airborne Instrumentation for Astronomy II},
eds McLean I., Casali M.,
Proc. SPIE, 7014, 70141A

\bibitem[Herriot et al.(2010)]{her10}
Herriot G. et al. 2010.
in {\it Adaptive Optics Systems II},
eds Ellerbroek B., Hart M., Hubin N., Wizinowich P., 
Proc. SPIE, 7736, 77360B

\bibitem[Hicks et al.(2009)]{hic09}
Hicks E. et al. 2009.
%Davies R., Malkan M., Genzel R., Tacconi L., M\"uller S\'anchez F., Sternberg A., 2009,
ApJ, 696, 448

\bibitem[Hirtzig et al.(2006)]{hir06}
Hirtzig M. et al. 2006. 
A\&A, 456, 761

\bibitem[Hom et al.(2007)]{hom07}
Hom E. et al. 2007.
%Marchis F., Lee T., Haase S., Agard D., Sedat J., 2007,
JOSA A, 24, 1580

\bibitem[Honda et al.(2009)]{hon09}
Honda M. et al. 2009.
ApJ, 690, L110

%\bibitem[Hubin et al.(2005)]{hub05}
%Hubin N. et al. 2005.
%C.~R.~Physique, 6, 1099

\bibitem[Hu\'elamo et al.(2011)]{hue11}
Hu\'elamo N. et al. 2011.
A\&A, 528, L7

\bibitem[Jones et al.(2010)]{jon10}
Jones T. et al. 2010.
%Swinkbank A.M., Elis R., Richard J., Stark D., 2010,
MNRAS, 404, 1247

\bibitem[Jolissaint et al.(2010)]{jol10}
Jolissaint L. et al. 2010.
in {\it Adaptive Optics Systems II},
eds Ellerbroek B., Hart M., Hubin N., Wizinowich P., 
Proc. SPIE, 7736, 773621

\bibitem[Jolissaint(2011)]{jol11}
Jolissaint L., Neyman C., Wizinowich P., Christou J. 2011.
in the {\it 2nd international conference on Adaptive Optics for Extremely Large Telescope}, Sept 2011.

%\bibitem[Kaenders et al.(2010)]{kae10}
%Kaenders W. et al. 2010.
%in {\it Adaptive Optics Systems II},
%eds Ellerbroek B., Hart M., Hubin N., Wizinowich P., 
%Proc. SPIE, 7736, 773621

\bibitem[Kasper et al.(2007)]{kas07}
Kasper M., Apai D., Janson M., Brandner W. 2007.
A\&A, 472, 321

%\bibitem[King(2003)]{kin03}
%King A. 2003.
%ApJ, 596, L27

\bibitem[K\"ohler(2011)]{koh11}
K\"ohler R. 2011. 
A\&A, 530, 126

%\bibitem[Komossa et al.(2003)]{kom03}
%Komossa S., Burwitz V., Hasinger G., Predehl P., Kaastra J., Ikebe Y., 2003,
%ApJ, 582, L15

\bibitem[Konopacky et al.(2010)]{kon10}
Konopacky Q. et al. 2010.
%Ghez A., Barman T., Rice E., Bailey J., White R., McLean I., Duch\^ene G., 2010,
ApJ, 711, 1087

\bibitem[Kormendy \& Bender(2011)]{kor11}
Kormendy J., Bender R. 2011.
Nature, 469, 374

\bibitem[Krajnovi\'c et al.(2009)]{kra09}
Krajnovi\'c D., McDermind R., Cappellari M., Davies R. 2009.
MNRAS, 390, 1839

\bibitem[Kraus et al.(2012)]{kra11a}
Kraus A., Ireland M., Hillenbrand L., Martinache F. 2012.
ApJ, 745, 19

\bibitem[Kraus \& Ireland(2012)]{kra11b}
Kraus A., Ireland M. 2012.
ApJ, 745, 5

\bibitem[Labeyrie(1995)]{lab95}
Labeyrie A. 1995.
A\&A, 298, 544

\bibitem[Lafreni\`ere et al.(2007a)]{laf07}
Lafreni\`ere D. et al. 2007a.
ApJ, 670, 1367

\bibitem[Lafreni\`ere et al.(2007b)]{laf07b}
Lafreni\`ere D. et al. 2007b.
ApJ, 660, 770

%\bibitem[Lagrange et al.(1996)]{lag96}
%Lagrange A.-M., Beuzit J.-L., Mouillet D., 1996,
%JGR, 101, 14831

\bibitem[Lagrange et al.(2010)]{lag10}
Lagrange A.-M. et al. 2010.
Science, 329, 57

\bibitem[Langlois et al.(2004)]{lan04}
Langlois M. et al. 2004.
in {\it Advancements in Adaptive Optics.},
eds Bonaccini Calia D., Ellerbroek B., Ragazzoni R., 
Proc. SPIE, 5490, 59

\bibitem[Law, Mackay \& Baldwin(2006)]{law06} % LuckyCam
Law N., Mackay C., Baldwin J. 2006.
A\&A, 446, 739

\bibitem[Law et al.(2009a)]{law09a} % LuckyCam
Law N. et al. 2009a.
ApJ, 692, 924

\bibitem[Law et al.(2009b)]{law09b}
Law D. et al. 2009b.
%Steidel C., Erb D., Larkin J., Pettini M., Shapley A., Wright S., 2009b,
ApJ, 697, 2057

\bibitem[Le Roux et al.(2004)]{ler04}
Le Roux B. et al. 2004.
JOSAA, 21, 1261

\bibitem[Liu(2008)]{liu08}
Liu M. 2008.
in {\it Adaptive Optics Systems},
eds Hubin N., Max C., Wizinowich P., 
Proc. SPIE, 7015, 701508

%\bibitem[Loose(2011)]{loo11}
%Loose C., 2011
%Diploma Thesis

\bibitem[Lu et al.(2009)]{lu09}
Lu J. et al. 2009.
%Ghez A., Hornstein S., Morris M., Becklin E., Matthews K., 2009,
ApJ, 690, 1463

\bibitem[Macintosh et al.(2007)]{mac07}
Macintosh B. et al. 2007. 
C.~R.~Physique, 8, 365

\bibitem[Macintosh et al.(2008)]{mac08}
Macintosh B. et al. 2008.
in {\it Adaptive Optics Systems},
eds Hubin N., Max C., Wizinowich P., 
Proc. SPIE, 7015, 701518

\bibitem[Mackay, Baldwin \& Tubbs(2003)]{mac03}
Mackay C., Baldwin J., Tubbs R. 2003.
in {\it Future Giant Telescopes},
eds Angel R., Gilmozzi R., 
Proc. SPIE, 4840, 436

\bibitem[Males et al.(2010)]{mal10}
Males J et al. 2010.
in {\it Adaptive Optics Systems II},
eds Ellerbroek B., Hart M., Hubin N., Wizinowich P.,
Proc. SPIE, 7736, 773660

\bibitem[Mancini et al.(2011)]{man11}
Mancini C. et al. 2011.
ApJ, 743, 86

\bibitem[Mannucci et al.(2009)]{man09}
Mannucci F. et al. 2009.
MNRAS, 398, 1915

\bibitem[Marchetti et al.(2008)]{marc08}
Marchetti E. et al. 2008.
in {\it Adaptive Optics Systems},
eds Hubin N., Max C., Wizinowich P., 
Proc. SPIE, 7015, 70150F

\bibitem[Marchis et al.(2005)]{mar05} % silvia
Marchis F., Descamps P., Hestroffer D., Berthier J. 2005.
Nature, 436, 822

\bibitem[Marchis et al.(2006a)]{mar06a} % patroclus
Marchis F. et al. 2006a.
Nature, 439, 565

\bibitem[Marchis et al.(2006b)]{mar06b} % survey
Marchis F. et al. 2006b.
%Kaasalainen M., Hom E., Berthier J., Enriquez J., Hestroffer D., Le Mignant D., de Pater I., 2006b,
Icarus, 185, 39

%\bibitem[Marchis et al.(2011)]{mar11}
%Marchis F. et al. 2011.
%Enriquez J., Emery J., Berthier J., Descamps P., Vachier F., 2011,
%Icarus, 213, 252

\bibitem[Marois et al.(2006)]{mar06c}
Marois C. et al. 2006.
ApJ, 641, 556

\bibitem[Marois et al.(2008)]{mar08}
Marois C. et al. 2008.
%Macintosh B., Barman T., Zuckerman B., Song I., Patience J., Lafreni\`ere D., Doyon R., 2008,
Science, 322, 1348

\bibitem[Marois et al.(2010)]{mar10}
Marois C. et al. 2010.
%Zuckerman B., Konopacky Q., Macintosh B., Barman T., 2010,
Nature, 468, 1080

\bibitem[Martinache \& Guyon(2009)]{mar09}
Martinache F., Guyon O. 2009.
in {\it Techniques and Instrumentation for Detection of Exoplanets IV},
ed Shaklan S.,
Proc. SPIE, 7440, 74400O

\bibitem[Max et al.(2005)]{max05}
Max C. et al. 2005.
%Canalizo G., Macintosh B., Raschke L., Whysong D., Antonucci R., Schneider G., 2005,
ApJ, 621, 738

\bibitem[Max et al.(1997)]{max97}
Max C.E. et al. 1997.
Science, 277, 1649

\bibitem[Max et al.(2007)]{max07}
Max C., Canalizo G., de Vries W. 2007.
Science, 316, 1877

\bibitem[McCabe et al.(2011)]{mcc11}
McCabe C. et al. 2011.
%Duch\^ene G., Pinte C., Stapelfeldt K., Ghez A., M\'enard F., 2011, 
ApJ, 727, 90

%\bibitem[McGurk et al.(2011)]{mcg11}
%McGurk R. et al.
%ApJ, submitted

\bibitem[Medling et al.(2011)]{med11}
Medling A. et al. 2011.
%Ammons S.M., Max C., Davies R., Engel H., Canalizo G., 2011,
ApJ, submitted

\bibitem[Melbourne et al.(2010)]{mel10}
Melbourne J. et al. 2010.
%Williams B., Dalcanton J., Ammons S.M., Max C., Koo D., Girardi L., Dolphin A., 2010, 
ApJ, 712, 469

\bibitem[Merkle et al.(1989)]{mer89}
Merkle F. et al. 1989.
The Messenger, 58, 1

\bibitem[Merline et al.(1999)]{mer99}
Merline W. et al. 1999.
Nature, 401, 565

\bibitem[Merline et al.(2001)]{mer01}
Merline W. et al. S/2001(617) 1. IAU Circ. 7741

\bibitem[Metchev \& Hillenbrand(2009)]{met09}
Metchev S., Hillenbrand L. 2009.
ApJS, 181, 62

%\bibitem[Minowa et al.(2010)]{min10}
%Minowa Y. et al. 2010.
%in {\it Adaptive Optics Systems II},
%eds Ellerbroek B., Hart M., Hubin N., Wizinowich P.,
%Proc. SPIE, 7736, 77363N

%\bibitem[Momany et al.(2008)]{mom08}
%Momany Y et al. 2008.
%MNRAS, 391, 1650

%\bibitem[Moretti et al.(2009)]{mor09}
%Moretti A. et al. 2009.
%A\&A, 493, 539

\bibitem[M\"uller S\'anchez et al.(2009)]{mue09}
M\"uller S\'anchez F. et al. 2009.
%Davies R., Genzel R., Tacconi L., Eisenhauer F., Hicks E., Friedrich S., Sternberg A., 2009,
ApJ, 691, 749

\bibitem[M\"uller S\'anchez et al.(2011)]{mue11}
M\"uller S\'anchez F. et al. 2011.
%Prieto M.A., Hicks E., Vives-Arias H., Davies R., Malkan M., Tacconi L., Genzel R., 2011,
ApJ, 739, 69

\bibitem[Mugnier et al.(2004)]{mug04}
Mugnier L., Fusco T., Conan J.-M. 2004.
JOSA A, 21, 1841

%\bibitem[Murakawa et al.(2008)]{mur08}
%Murakawa K. et al. 2008.
%A\&A, 488, L75

\bibitem[Neichel \& Rigaut(2011)]{nei11}
Neichel B., Rigaut F. 2011.
Gemini Focus, 42, 29

%\bibitem[Nielsen \& Close(2010)]{nie10}
%Nielsen E., Close L., 2010.
%ApJ, 717, 878

\bibitem[Nowak(2009)]{now09}
Nowak N. 2009.
PhD thesis, Ludwig-Maximilians-Universit\"at M\"unchen

\bibitem[Nowak et al.(2010)]{now10}
Nowak N. et al. 2010.
MNRAS, 403, 646

\bibitem[Olsen, Blum \& Rigaut(2003)]{ols03}
Olsen K., Blum R., Rigaut F. 2003.
AJ, 126, 452

\bibitem[Olsen et al.(2006)]{ols06}
Olsen K. et al. 2006.
%Blum R., Stephens A., Davidge T., Massey P., Strom S., Rigault F., 2006,
AJ, 132, 271

\bibitem[Oppenheimer \& Hinkley(2009)]{opp09}
Oppenheimer B., Hinkley S. 2009.
ARA\&A, 47, 253

\bibitem[Orban de Xivry et al.(2011)]{orb11}
Orban de Xivry G. et al. 2011.
MNRAS, 417, 2721

%\bibitem[Ortolani et al.(2011)]{ort11}
%Ortolani S. et al. 2011.
%Barbuy B., Momany Y., Saviane I., Bica E., Jilkova L., Malta Salerno G., Jungwiert B., 2011,
%arXiv:1106.2725

\bibitem[Peng et al.(2010)]{pen10}
Peng C., Ho L., Impey C., Rix H.-W. 2010.
AJ, 139, 2097

\bibitem[Perrin et al.(2003)]{per03}
Perrin M.D. et al. 2003.
ApJ, 596, 702

\bibitem[Petit et al.(2008)]{pet08}
Petit C. et al. 2008.
Optics Express, 16, 87

\bibitem[Poyneer, Macintosh \& Veran(2007)]{poy07}
Poyneer L.A., Macintosh B., Veran J.-P. 2007.
JOSAA, 24, 2645

\bibitem[Poyneer et al.(2008)]{poy08}
Poyneer L.A. et al. 2008.
Applied Optics, 47, 1317

\bibitem[Poyneer, Gavel \& Brase(2002)]{poy02}
Poyneer L.A., Gavel D.T., Brase J.M. 2002.
JOSAA, 19, 2100

\bibitem[Pollack, Max \& Schneider(2007)]{pol07}
Pollack L., Max C., Schneider G. 2007.
ApJ, 660, 288

\bibitem[Raban et al.(2009)]{rab09}
Raban D. et al. 2009.
%Jaffe W., R\"ottgering H., Meisenheimer K., Tristram K., 2009,
MNRAS, 394, 1325

\bibitem[Rabien et al.(2002)]{rab02}
Rabien S. et al. 2002.
in {\it Adaptive Optical System Technologies II},
eds Tyson R.K., Bonaccini Calia D., Roggemann M.C.
Proc. SPIE, 4494, 325

\bibitem[Rabien et al.(2010)]{rab10}
Rabien S. et al. 2010.
in {\it Adaptive Optics Systems II},
eds Ellerbroek B., Hart M., Hubin N., Wizinowich P., 
Proc. SPIE, 7736, 77360E

\bibitem[Racine(2009)]{rac09}
Racine R. 2009.
in {\it Optical Turbulence: Astronomy Meets Meteorology}
eds Masciadri E., Sarazin M., 
(Imperial College Press), pp.13-22

\bibitem[Ragazzoni(1996)]{rag96}
Ragazzoni R. 1996.
J. Modern Opt., 43, 289

\bibitem[Ragazzoni(2000)]{rag00}
Ragazzoni R. 2000.
in {\it Proceedings of the Backaskog workshop on extremely large telescopes},
eds Andersen T., Ardeberg A., Gilmozzi R.,
ESO Conference and Workshop Proceedings, 57, p.175

\bibitem[Ragazzoni(2011)]{rag11}
Ragazzoni R. et al. 2011.
in the {\it 2nd international conference on Adaptive Optics for Extremely Large Telescope}, Sept 2011.

\bibitem[Riffel et al.(2008)]{rif08}
Riffel R.A. et al. 2008.
%Storchi-Bergmann T., Winge C., McGregor P., Beck T., Schmitt H., 2008,
MNRAS, 385, 1129

\bibitem[Riffel et al.(2009a)]{rif09a}
Riffel R., Pastoriza M., Rodr\'iguez-Ardila A., Bonatto C. 2009a.
MNRAS, 400, 273

\bibitem[Riffel et al.(2009b)]{rif09b}
Riffel R.A., Storchi-Bergmann T., Dors O., Winge C. 2009b.
MNRAS, 393, 783

\bibitem[Riffel et al.(2010)]{rif10}
Riffel R.A., Storchi-Bergmann T., Riffel R., Pastoriza M. 2010. 
ApJ, 713, 469

\bibitem[Riffel \& Storchi-Bergmann(2011)]{rif11a}
Riffel R.A., Storchi-Bergmann T. 2011.
MNRAS, 411, 469

\bibitem[Riffel et al.(2011)]{rif11b}
Riffel R., Riffel R.A., Ferrari F., Storchi-Bergmann T. 2011.
MNRAS, 416, 493

\bibitem[Rigaut et al.(1991)]{rig91}
Rigaut F. et al. 1991.
A\&A, 250, 280

\bibitem[Rigaut, Ellerbroek \& Flicker(2000)]{rig00}
Rigaut F., Ellerbroek B., Flicker R. 2000.
in {\it Adaptive Optical Systems Technology},
ed Wizinowich P., 
Proc. SPIE, 4007, 1022

\bibitem[Rigaut et al.(2002)]{rig02}
Rigaut F. 2002.
in {\it Beyond Conventional Adaptive Optics},
eds Vernet E., Ragazzoni R., Esposito S., Hubin N.,
ESO Conference and Workshop Proceedings, 58, p.11

\bibitem[Rimmele(2000)]{rim00}
Rimmele T. 2000.
in {\it Adaptive Optical Systems Technology}, 
ed. Wizinowich P., 
Proc. SPIE, 4007, 218

%\bibitem[Rimmele(2008)]{rim08}
%Rimmele T., Hegwer S., Richards K., Woeger F. 2008.
%Proc. Advanced Maui Optical and Space Surveillance Technologies Conference,
%ed. Ryan S.,

\bibitem[Rimmele \& Marino(2011)]{rim11}
Rimmele T., Marino J. 2011.
Living Rev. Solar Phys., 8, 2, 
http://www.livingreviews.org/lrsp-2011-2 (cited on 14 June 2011)

\bibitem[Roddier(1988)]{rod88}
Roddier F. 1988.
Applied Optics, 27, 1223

\bibitem[Roddier et al.(1996)]{rod96}
Roddier C. et al. 1996.
%Roddier F., Northcott M., Graves J., Jim K., 1996
ApJ, 463, 326

\bibitem[Rosario et al.(2011)]{ros11}
Rosario D. et al. 2011.
%McGurk R., Max C., Shields G., Smith K., 2011,
ApJ, 739, 44

\bibitem[Rosensteiner(2011)]{rose11}
Rosensteiner M. 2011.
submitted to JOSAA

\bibitem[Ross(2009)]{ros09}
Ross T.S. 2009.
Applied Optics, 48, 1812

\bibitem[Rousset et al.(1990)]{rou90}
Rousset G. et al. 1990.
A\&A, 230, L29

\bibitem[Rousset et al.(2003)]{rou03}
Rousset G. et al. 2003.
in {\it Adaptive Optical System Technologies II},
eds Wizinowich P., Bonaccini D., 
Proc. SPIE, 4839, 140

%\bibitem[Sanders et al.(1988)]{san88}
%Sanders D. et al. 1988.
%Soifer T., Elias J., Madore B., Matthews K., Neugebauer G., Scoville N., 1988, ApJ, 325, 74

\bibitem[Sankarasubramanian \& Rimmele(2008)]{san08}
Sankarasubramanian K., Rimmele T. 2008.
JApA, 29, 329

\bibitem[Sarzi et al.(2006)]{sar06}
Sarzi M., Shields J., Pogge R., Martini P. 2006.
in {\it Stellar Populations as Building Blocks of Galaxies}, 
eds Vazdekis A., Peletier R., 
2006,
Proc. IAU Symposium 241, pp.489-492 (CUP)

\bibitem[Scharmer et al.(2011)]{sch11}
Scharmer G., Henriques V., Kiselman D., de la Cruz Rodr\'iguez J. 2011.
Science, 15 July, p.316

\bibitem[Schartmann et al.(2010)]{scha10}
Schartmann M. et al. 2010.
%Burkert A., Krause M., Camenzind M., Meisenheimer K., Davies R., 2010,
MNRAS, 403, 1801

\bibitem[Schnorr M\"uller et al.(2011)]{schm11}
Schnorr M\"uller A. et al. 2011.
%Storchi-Bergmann T., Riffel R.A., Ferrari F., Steiner J., Axon D., Robinson A., 2011,
MNRAS, 413, 149

\bibitem[Sch\"odel(2010)]{scho10}
Sch\"odel R. 2010.
A\&A, 509, A58

%\bibitem[Sch\"odel et al.(2011)]{scho11}
%Sch\"odel R. et al. 2011.
%arXiv:1110.2261

\bibitem[Seo et al.(2009)]{seo09}
Seo, B.-J. et al. 2009.
Applied Optics, 48, 5997

\bibitem[Shatsky \& Tokovinin(2002)]{sha02}
Shatsky N., Tokovinin A. 2002. 
A\&A, 382, 92

\bibitem[Sicardy et al.(1999)]{sic99}
Sicardy B. et al. 1999.
Nature, 400, 731

\bibitem[Siegler et al.(2003)]{sie03}
Siegler N., Close L., Mamajek E., Freed M. 2003.
ApJ, 598, 1265
	
\bibitem[Sinquin et al.(2008)]{sin08}
Sinquin J.-C., Lurcon J.-M., Guillemard C. 2008.
in {\it Adaptive Optics Systems.},
eds Hubin N., Max C.E., Wizinowich P.L. 
Proc. SPIE, 7015, 70150O

\bibitem[Skemer et al.(2011a)]{ske11a}
Skemer A. et al. 2011a.
%Close L, Sz\"ucs L., Apai D., Pascucci I., Biller B., 2011,
ApJ, 732, 107

\bibitem[Skemer et al.(2011b)]{ske11b}
Skemer A. et al. 2011b.
%Close L, Sz\"ucs L., Apai D., Pascucci I., Biller B., 2011,
ApJ, in press

\bibitem[Smith et al.(2009)]{smi09}
Smith A., Bailey J., Hough J., Lee S. 2009.
MNRAS, 398, 2069

\bibitem[Sparks \& Ford(2002)]{spa02}
Sparks W.B., Ford H.C. 2002.
ApJ, 578, 543

\bibitem[Starck et al.(2002)]{sta02}
Starck J., Pantin E., Murtagh F. 2002.
PASP, 114, 1051

\bibitem[Stark et al.(2008)]{sta08}
Stark D. et al. 2008.
Nature, 455, 775

\bibitem[Steinbring et al.(2002)]{ste02}
Steinbring E. et al. 2002.
PASP, 114, 1267

\bibitem[Steinbring et al.(2005)]{ste05}
Steinbring E. et al. 2005.
%Faber S., Macintosh B., Gavel D., Gates E., 2005,
PASP, 117, 847

%\bibitem[Steiner et al.(2009)]{ste09}
%Steiner J., Menezes R., Ricci T., Oliveira A. 2009.
%MNRAS, 395, 64

\bibitem[Stetson(1987)]{ste87}
Stetson P. 1987.
PASP, 99, 191

\bibitem[Stolte et al.(2008)]{sto08}
Stolte A. et al. 2008.
%Ghez A., Morris M., Lu J., Brandner W., Matthews K., 2008,
ApJ, 675, 1278

\bibitem[Storchi-Bergmann et al.(2007)]{sto07}
Storchi-Bergmann T. et al. 2007.
%Dors O., Riffel R.A., Fathi K., Axon D., Robinson A., Marconi A., \"Ostlin G., 2007,
ApJ, 670, 959

\bibitem[Storchi-Bergmann et al.(2010)]{sto10}
Storchi-Bergmann T. et al. 2010.
%Sim\~oes Lopes R., McGregor P., Riffel R.A., Beck T., Martini P., 2010,
MNRAS, 402, 819

\bibitem[Thatte et al.(2007)]{tha07}
Thatte N. et al. 2007.
%Abuter R., Tecza M., Nielsen E., Clarke F., Close L., 2007,
MNRAS, 378, 1229

\bibitem[Thiebaut \& Tallon(2010)]{thi10}
Thiebaut E., Tallon M. 2010.
JOSAA, 27, 1046

\bibitem[Tokovinin(2004)]{tok04}
Tokovinin A. 2004.
PASP, 116, 941

\bibitem[Trauger \& Traub(2007)]{tra07}
Trauger J.T, Traub W.A. 2009.
Nature, 446, 771

\bibitem[Travouillon et al.(2009)]{tra09}
Travouillon T. et al. 2009.
PASP, 121, 668

\bibitem[Trippe et al.(2010)]{tri10}
Trippe S. et al. 2010.
%Davies R., Eisenhauer F., F\"orster Schreiber N., Fritz T., Genzel R., 2010,
MNRAS, 402, 1126

\bibitem[Tristram et al.(2007)]{tri07}
Tristram K. et al. 2007.
A\&A, 474, 837

\bibitem[van Noort, Rouppe van der Voort \& L\"ofdahl(2005)]{noo05}
van Noort M., Rouppe van der Voort L., L\"ofdahl M. 2005.
Solar Phys., 228, 191

\bibitem[Veran et al.(1997)]{ver97}
Veran J.-P. et al. 1997.
JOSA A, 14, 3057

\bibitem[Verinaud \& Cassaing(2001)]{ver01}
Verinaud C., Cassaing F. 2001.
A\&A, 365, 314

\bibitem[Verinaud et al.(2005)]{ver05}
Verinaud, C. et al. 2005.
MNRAS, 357, L26

\bibitem[Wallner(1983)]{wal83}
Wallner E.P. 1983.
JOSA, 73, 1771

%\bibitem[Watson et al.(2007)]{wat07}
%Watson A., Stapelfeldt K., Wood K., M\'enard F. 2007.
%in {\it Protostars and Planets V},
%eds Reipurth B., Jewitt D., Keil K.,
%University of Arizona Press, Tucson, p.523--538

%\bibitem[Watson et al.(2008)]{wat08}
%Watson L. et al. 2008.
%Martini P., Dasyra K., Bentz M., Ferrarese L., Peterson B., Pogge R., Tacconi L., 2008,
%ApJ, 681, L21

\bibitem[Weinzirl et al.(2009)]{wei09}
Weinzirl T. et al. 2009.
ApJ, 696, 411

\bibitem[Wild et al.(2010)]{wil10}
Wild V., Heckman T., Charlot S. 2010.
MNRAS, 405, 933

\bibitem[Wisnioski et al.(2011)]{wis11}
Wisnioski E. et al. 2011.
MNRAS, in press

\bibitem[Wizinowich et al.(2000)]{wiz00}
Wizinowich P. et al. 2000.
PASP, 112, 315

\bibitem[Wizinowich et al.(2006)]{wiz06}
Wizinowich P.L. et al. 2006.
PASP, 118, 297

\bibitem[Wizinowich et al.(2010)]{wiz10}
Wizinowich P. et al. 2010.
in {\it Adaptive Optics Systems II},
eds Ellerbroek B., Hart M., Hubin N., Wizinowich P., 
Proc. SPIE, 7736, 77360K

\bibitem[Wizinowich(2011)]{wiz11}
Wizinowich P. 2011.
in the Keck Newsletter, vol. 11., http://www2.keck.hawaii.edu/inst/newsletters/Vol11/index.html

%\bibitem[Wong et al.(2009)]{won09}
%Wong M. et al. 2009.
%in {\it 40th Annual Meeting of the Division for Planetary Sciences of the American Astronomical Society}

\bibitem[Wright et al.(2009)]{wri09}
Wright S. et al. 2009.
%Larkin J., Law D., Steidel C., Shapley A., Erb D., 2009,
ApJ, 699, 421

\bibitem[Wright et al.(2010)]{wri10}
Wright S., Larkin J., Graham J., Ma C.-P. 2010.
ApJ, 711, 1291

\bibitem[Wyatt(2008)]{wya08}
Wyatt M. 2008.
ARA\&A, 46, 339

\bibitem[Yelda et al.(2010)]{yel10}
Yelda S. et al. 2010.
ApJ, 725, 331

%\bibitem[Yu et al.(2011)]{yu11}
%Yu Q., Lu Y., Mohayaee R., Colin J. 2011.
%arXiv:1105.1963

\bibitem[Zapatero Osorio et al.(2004)]{zap04}
Zapatero Osorio M. et al. 2004.
ApJ, 615, 958

\end{thebibliography}
\end{document}